\documentstyle[aps,prl,eqsecnum,preprint,tighten]{revtex}
\begin{document}   
\title{Mesoscopic conductance and its fluctuations\\
at non-zero Hall angle}
\author{Shanhui Xiong,\footnote{Present Address: International Center for
Theoretical Physics, 34100 Trieste, Italy} N. Read, and A. Douglas Stone}
\address{Departments of Physics and Applied Physics, P.O. Box 
208284\\
Yale University, New Haven, CT 06520-8284}
\date{December 15, 1996}
\maketitle
\newcommand{\und}{\underline}
\newcommand{\ov}{\overline}
\newcommand{\be}{\begin{equation}}
\newcommand{\ee}{\end{equation}}
\newcommand{\bea}{\begin{eqnarray}}
\newcommand{\eea}{\end{eqnarray}}
\newcommand{\br}{{\bf r }}
\newcommand{\bA}{{\bf A }}
\newcommand{\brp}{{\bf r}^{\prime} }
\newcommand{\bna}{{\nabla}}
\newcommand{\DD}{\stackrel{\leftrightarrow}{D}}
\newcommand{\DDs}{\stackrel{\leftrightarrow}{D^*}}
\newcommand{\non}{\nonumber}
\newcommand{\dl}{\delta}
\newcommand{\al}{\alpha}
\newcommand{\om}{\omega}
\newcommand{\gm}{\gamma}
\newcommand{\la}{\lambda}
\newcommand{\wc}{\omega_c}
\newcommand{\Gprrp}{G^+(\br,\brp)}
\newcommand{\Gmrrp}{G^-(\br,\brp)}
\newcommand{\Gprpr}{G^+(\brp,\br)}
\newcommand{\Gmrpr}{G^-(\brp,\br)}
\newcommand{\rarrow}{\rightarrow}
\newcommand{\gpm}{G^{\pm}}
\newcommand{\gnpm}{G_0^{\pm}}
\newcommand{\Gb}{\ov{G}}
\newcommand{\Gbp}{\ov{G}^+}
\newcommand{\Gbm}{\ov{G}^-}
\newcommand{\Gbrrp}{\ov{G}(\br,\brp;E)}
\newcommand{\Gnrrp}{G_0(\br,\brp;E)}
\newcommand{\dlA}{\Delta \bA}
\newcommand{\sx}{\sigma_{xx}}
\newcommand{\sy}{\sigma_{xy}}
\newcommand{\ea}{\end{array}}
\newcommand{\ba}{\begin{array}}
\newcommand{\fba}{\ov{\varphi}}
\newcommand{\zd}{z^{\dagger}}

\begin{abstract}
We consider the bilocal conductivity tensor, the two-probe conductance and 
its fluctuations for a disordered phase-coherent two-dimensional system of 
non-interacting electrons in the presence of a magnetic field, including 
correctly the edge effects. Analytical results are obtained by perturbation 
theory in the limit $\sx \gg 1$. For mesoscopic systems the conduction 
process is dominated by diffusion but we show that, due to the lack of 
time-reversal symmetry, the boundary condition for diffusion
is altered at the reflecting edges. Instead of the usual condition, that the 
derivative along the direction normal to the wall of the diffusing variable 
vanishes, the derivative at the Hall angle to the normal 
vanishes. We demonstrate the origin of this boundary condition {}from 
different starting points, using (i) a simplified Chalker-Coddington network 
model, (ii) the standard diagrammatic perturbation expansion, and 
(iii) the nonlinear $\sigma$-model with the topological term, thus 
establishing connections between the different approaches. Further boundary
effects are found in quantum interference phenomena.
We evaluate the mean bilocal conductivity tensor $\sigma_{\mu\nu}(\br,\brp)$, 
and the mean and variance of the conductance, to leading order in $1/\sx$ and 
to order $(\sy/\sx)^2$, and find that the variance of the conductance 
increases with the Hall ratio. Thus the conductance fluctuations are no 
longer simply described by the unitary universality class of the
$\sigma_{xy}=0$ case, but instead there is a one-parameter family of
probability distributions. Our results differ {}from previous calculations, 
which neglected $\sy$-dependent vertices beyond the change in boundary 
conditions. In the quasi-one-dimensional limit, the usual universal result 
for the conductance fluctuations of the unitary ensemble is recovered, 
in contrast to results of previous authors. We also give an extensive 
discussion of current conservation conditions in the nonlinear sigma model.

\end{abstract}

\tableofcontents

\section{Introduction}
 
In the past decade much attention has been given to the statistical 
properties of quantum  conductors with complete phase coherence 
(with size $L$ smaller than the phase coherence length $L_{in}$).
A notable feature of such systems is the lack of self-averaging in their
transport properties. In mesoscopic systems for which $L$ is less than the 
localization length $\xi$, the conductance fluctuation amplitude
(the standard deviation of the conductance), is found, at low magnetic field, 
to be of order 1 and independent of system size and the degree of disorder, 
(but dependent on the dimensionality, shape, and overall symmetry properties of 
the system) \cite{Webb,Stone2,LS,Alsh2,Skocpol,Oded,LSF}. (Note that, 
in the present paper, all conductivities 
$\sigma$ and conductances $g$ have a factor $e^2/h$ removed, 
so they are dimensionless in two dimensions; their 
dimensionful analogues are recovered by multiplying by $e^2/h$.)
These universal conductance fluctuations (UCF) 
have been well understood within the framework of perturbation
theory \cite{LS,Alsh2,LSF} and the one parameter scaling theory of 
quantum conductance \cite{gang4,AKL}. The physics underlying the UCF is the
long-ranged spatial correlation among the wave functions of the 
conduction electrons in the diffusive regime. The universality of the 
phenomenon has also stimulated the formulation of a random matrix theory 
description of quasi-one-dimensional (quasi-1D) 
\cite{Imry,Mello,Random,Beenakker} and quasi-zero-dimensional
(quasi-0D) conductors \cite{Baranger2,Jalabert}, which reproduces 
quantitatively the results of diagrammatic perturbation theory. In the earlier 
perturbative work the effect of a magnetic field was only 
included through the introduction of an appropriate Aharonov-Bohm 
phase in the zero-field propagator. This leads to the elimination of the
Cooperon contributions to the conductance fluctuations, causing a cross-over
{}from the so-called orthogonal to the unitary ensemble, and a 
consequent reduction of the variance by a factor of two \cite{Alsh2,LSF}. 
In the quasi-1D case, this result is also recovered by the random-matrix 
approaches \cite{Random,Beenakker}.

In two dimensions (2D), the interplay of quantum interference effects and 
magnetic field leads to the quantum Hall effect in high magnetic field $B$
\cite{Klitzing}. In mesoscopic samples, conductance fluctuations persist in 
fairly high magnetic field $\wc\tau_0> 1$, where $\wc$ is 
the cyclotron frequency, $\tau_0$ is the elastic scattering time in 
zero magnetic field, and the fluctuation amplitude remains comparable
to the low field limit \cite{Timp,Geim,Kinaret,Bhatt2,Ando2,Wang,Cho}.
For $B$ fields sufficiently high that quantization of the Hall conductance sets 
in, the fluctuations are strongly suppressed in the plateau regions, but 
reappear in both longitudinal and Hall resistance in the transition regions
between plateaus.
It is therefore of theoretical interest to generalize the theory for 
conductance fluctuations to all fields. At $\wc\tau_0>1$, the trajectories of 
the electrons are significantly influenced by the Lorentz force between 
successive scattering events and the dynamical effect of 
the magnetic field must be treated. The diffusion at high field occurs 
by a different mechanism {}from the low field regime. For the short-ranged 
random potential model, the center of the cyclotron orbit hops a length of 
order the cyclotron radius $R_c$ whenever it encounters a scattering center, 
therefore $R_c$ plays the role of the mean free path $l$. The bare 
conductivity $\sx^0$ in the middle of the $N$th Landau level is of order 
$(2N+1)/\pi$ \cite{Ando1} ($N=0$, $1$, \ldots).  Despite the altered nature of 
the microscopic diffusion, a unified treatment of mesoscopic conductance 
fluctuations in relatively high fields is possible because, even at 
$\wc\tau_0>1$, there exists a perturbative regime where the transport process 
is dominated by diffusion. As long as there are many Landau levels occupied,
$1/\sx^0$ or $1/k_Fl$ serves as a small parameter, and perturbation theory is
still useful. Previous perturbation theories have shown that the weak 
localization correction to $\sx$ for the unitary class 
is of order $-(\sx^0)^{-1}\log(L/l)$ \cite{Brezin,Hikami,Chalker2}. 
The localization correction is relatively small for systems with $L$ less 
than the crossover length $\xi_{pert}=le^{(\sx^0)^2}$ (for $\sx^0$ large). 
For $l< L < \xi_{pert}$, the conductance fluctuations are expected to be 
similar to the UCF\@. At $L>\xi_{pert}$, the renormalization group flows 
for the system, driven by non-perturbative effects
\cite{Pruisken1,LLP,Pruiskenrev,Pruisken2}, 
carry it either to one of the localization fixed points where $\sx\rightarrow 
0$ and $\sy$ becomes quantized, or to one of the non-trivial fixed points 
where $\sy$ is half integer and $\sx$ approaches a universal value. 
Numerical work finds that the conductance fluctuations in the critical regimes 
\cite{Bhatt2,Ando2,Wang,Cho} have a different distribution {}from the UCF and 
one would expect this distribution to be beyond the scope of a perturbative 
treatment. We comment further on the critical regime in the conclusion, Sec.\
VI.
    
In this paper we study the conductance and its variance for
a two-probe geometry in two dimensions in the presence of disorder 
and magnetic field and in the perturbative regime ($l\ll L\ll\xi_{pert}$). 
As shown in Figure~\ref{two probe geometry}, the edges of the sample are 
defined by a hard-wall confinement potential and the two ends of the sample are 
connected to highly conducting leads. The first analytic work on this problem 
was by two of the authors (Xiong and Stone \cite{Xiong}) in which they 
generalized the previous diagrammatic perturbative techniques \cite{Chalker2} 
to treat the conductance fluctuations. At the level of the self-consistent 
Born approximation (SCBA), the only effect they found of the magnetic field 
$B$ was a field-dependent diffusion constant $D(B)$. Since the value of the
diffusion constant cancels from the conductance fluctuations, they found no 
effect of the magnetic field on the amplitude of the two-probe conductance 
fluctuations (other than the well-known factor of two reduction associated 
with the crossover to the unitary ensemble), although the correlation field, 
$B_c$, which determines the spacing of the fluctuations in magnetic field, 
was found to increase with increasing field in a manner consistent with 
experiment \cite{Geim}. The reason for the increase is that for systems 
with $L> L_{in}$, $B_c\sim \phi_0/L_{in}^2$, where $L_{in}^2=D(B)\tau_{in}$ 
($\phi_0=h/e$, $\tau_{in}$ is the inelastic scattering time). Since the 
diffusion constant $D(B)$ decreases with increasing magnetic field (a result 
reviewed in Sec.\ II), $B_c$ increases. Although this conclusion about the 
correlation field is basically correct, the conclusion concerning the 
fluctuation amplitude is now understood to be correct only for a periodic 
boundary condition in the transverse direction and must be revised for the 
case of a system with reflecting edges. As discovered independently by 
Khmel'nitskii and Yosefin (KY), Maslov and Loss (ML), and one of the present 
authors (Read) \cite{Yosef,Maslov,Read}, the boundary condition is modified 
{}from the vanishing of the normal derivative of the diffusing variable to the 
vanishing of the derivative at an angle to the normal. KY and Read further 
showed that this angle is the Hall angle $\theta_{H}=\tan^{-1}\sy^0/\sx^0$, 
where $\sy^0$ is the bare Hall conductivity. These authors pointed out the 
possibility that the mesoscopic conductance fluctuations may depend on 
magnetic field due to the boundary condition. KY and ML attempted to evaluate 
this dependence both numerically and analytically.  Simulations performed for 
the two-probe conductance of small systems in the non-quantized regime 
\cite{Maslov} show that the maximum fluctuation amplitude appears towards the 
bottom of the Landau levels where the Hall ratio is large, indicating some 
dependence on the Hall ratio. However, the analytic calculations by KY and ML 
\cite{Yosef,Maslov} do not agree with our present results since these authors 
have merely modified the diffusion propagator in previous expressions for UCF 
diagrams. Like KY, we find that $\sy^0$ enters not only the boundary condition 
but also the current vertex.  Moreover the altered boundary condition permits 
new diagrams to occur which roughly speaking describe interference effects 
associated with a $\sigma_{xy}^0$ dependent ``interaction" of the diffusion 
modes. These diagrams, which were not considered by ML and KY, must be 
included when edges are present. We evaluate the two-probe conductance and 
its variance to leading order in $1/\sx^0$ and to order $\gamma^2$, where 
$\gamma=\sigma_{xy}^0/\sigma_{xx}^0$. For wide samples with $W\sim L$, where 
$L$ and $W$ are now the length and the width of the sample, we find that the 
variance does not depend on $\sigma_{xx}^0$ and $\sigma_{xy}^0$ individually,
but does depend on $\gamma$. The variance increases as $\gamma^2$ for small 
$\gamma$ and hence is no longer independent of magnetic field (although it is 
still independent of size in this order in $1/\sigma_{xx}^0$, and has no 
direct dependence on the mean-free path).  Interestingly, however, in the 
quasi-1D limit ($W\ll L$), the Hall ratio is absorbed into an effective 1D 
conductivity which cancels in all diagrams.  Therefore in this limit
the UCF result of the unitary class is recovered (in contrast 
to the claims by KY and ML that this is modified for $\gamma \neq 0$). 
The implication is that quasi-1D conductance fluctuations are still
described by the standard random matrix theory of disordered conductors,
but that the 2D fluctuations, even in perturbation theory, define a
family of random-matrix ensembles parametrized by the Hall ratio $\gamma$.

In this article, we use the disorder-averaged diagrammatic 
approach and the field theoretical approach in a complementary
way. To study transport properties of a system with phase coherence, 
the appropriate starting point is the bilocal conductivity tensor 
$\sigma_{\mu\nu}(\br,\brp)$. In Sec.\ II, we evaluate the mean
$\sigma_{\mu\nu}(\br,\brp)$ to leading order in $1/\sx^0$ using the 
diagrammatic approach and demonstrate the microscopic origin of the edge 
contributions. In Sec.\ III, we set up the field-theoretical formalism for 
the evaluation of linear response functions. We discuss the connection between 
the tilted boundary condition and the nonlinear $\sigma$-model action with a 
topological term proportional to $\sy^0$ \cite{Pruisken1,Pruiskenrev}. 
Previously it was known that this 
topological term is crucial to the critical transition of the quantum Hall 
effect at large length scale ($L\gg \xi_{pert}$) \cite{LLP,Pruiskenrev}. 
For a system with reflecting edges, this term is a non-vanishing surface term,
which {\it does} influence transport properties in perturbation theory, 
valid when 
$L\ll \xi_{pert}$, through various $\sy^0$-dependent boundary contributions. 
In Sec.\ IV and Sec.\ V, the conductance and its variance are calculated by 
expanding in power series in $1/\sx^0$ and $\gamma$. In the remainder of 
the introduction we begin the discussion of the main ideas and summarize our
results. The details of the calculations, and further discussions, are 
given in later sections.

\subsection{Local conductivity parameters, two-probe 
conductance, and edge states}

In this article we focus on calculations of the two-probe conductance
in a magnetic field. It is however possible to generalize our calculations
to treat the conductance matrix of a multi-probe conductor as has been
done previously for zero (or weak) field \cite{Kane2,Hershfield}.  
Two-probe conductance describes an experimental set-up in which voltage 
measurements are made only between the current source and sink and not between 
distinct voltage probes as in a typical Hall measurement (two-probe 
measurements are not uncommon for mesoscopic conductors, because of the 
difficulty of making multiple contacts). Therefore such a measurement
cannot separately determine $\sx(B)$ and $\sy(B)$.  In this subsection, we show
how the assumption of a local form for the conductivity in a system with edges 
leads to the result that the two probe conductance is approximately 
proportional to $\sx$ when $\sx \gg |\sy|$ and to $|\sy|$ when $|\sy|\gg\sx$.   
Interestingly, such an
argument already indicates the appearance of the tilted boundary condition
which affects the conductance fluctuations as well.

We wish to find the current produced in linear response to an applied 
electric field.
We will consider the two-probe conductance which results from assuming
a local form for the conductivity: 
\be
 j_{\mu}(\br)=\frac{e^2}{h}\sigma^0_{\mu\nu}{\cal E}_{\nu}(\br),
\label{Ohms}
\ee  
where due to the macroscopic homogeneity of the sample and Onsager relations
the conductivity parameters obey $\sx^0=\sigma^{0}_{yy}$ and $\sy^0 = 
-\sigma^{0}_{yx}$. It is a common misconception that {\boldmath$\cal E$} in
this formula is the same as the applied electric field $\bf E$. 
In fact, in general, {\boldmath $\cal E$} in this formula should be interpeted 
as the {\em electromotive} field, that is, as minus the gradient of the
electrochemical potential $V=\phi-\mu/e$, where $\phi$ is the electric
potential, and $\mu$ the chemical potential (more generally, $\mbox{\boldmath
$\cal E$}={\bf E}+\nabla\mu/e$). In this paper, we will neglect 
electron-electron interactions of all kinds, so $\phi$ can be viewed as the 
externally-applied electric potential. (If it is desired to include Coulomb 
interactions through a self-consistent potential, then $\phi$ is the total
electric potential, the sum of the externally-applied potential and the
potential produced by the electrons in the system; the latter potential is 
determined by the change in the expectation value of the density, 
in response to the external field, through the 3D Poisson equation. This is 
distinct {}from similar-looking equations below which have the form of the 2D 
Laplace equation. Other short-range interaction effects would contribute to 
$\mu$.) The chemical potential in the above expressions is defined in terms
of a local quasi-equilibrium (in the presence of a nonzero response 
current) which must be established by inelastic effects. Hence, this
formulation is only valid on scales greater than the inelastic 
length $L_{in}$ (this of course requires that there be some interaction 
between the electrons). It is only in this sense, which implies the absence of 
effects due to single-particle phase coherence, that Eq.\ (\ref{Ohms}) is a 
classical formula; it does not require that all effects of quantum mechanics be 
neglected. Eq.\ (\ref{Ohms}) (when valid) is the most convenient form for 
expressing the linear response, since a voltage measurement determines 
electrochemical potential differences. It does not imply that a local relation 
exists between the electromotive and electric fields. 
As the chemical potential is determined by the local conditions, in particular 
by the local density, and that density is affected by the transport, which 
in our case will be diffusive, this relation is not local. Thus the current 
response to the {\em electric} field is actually nonlocal, as in Ref.\  
\cite{Kane}, even in this ``classical'' case. We will return to this in 
Sec.\ ID.

We will now calculate the two-probe conductance of a rectangular sample with 
insulating edges connected to conducting leads at each end. The potential 
in the leads is assumed to be held at constant values, $V_1$ and $V_2$.  
At every point on the insulating edges the normal current must vanish.  
Using Eq.\ (\ref{Ohms}) and $\mbox{\boldmath$\cal E$}=-\nabla V$ it follows 
that $(\partial_y -\gamma\partial_x) V(\br)=0$, i.e. the electromotive field 
at the Hall angle to the normal must vanish. Thus in this case the appearance 
of a tilted boundary condition on the electrochemical potential follows simply 
{}from the fact that the field is tilted {}from the current by the Hall angle. 
{}From the continuity equation, $\nabla \cdot {\bf j}= 0$, and one finds that 
$ \nabla^2 V=0$ in bulk. Solving the Laplace equation for $V(\br)$ with fixed 
voltages at the two ends and the tilted boundary conditions at the edges is not
a simple exercise, but has been done for this 2D rectangular geometry
by conformal mapping \cite{Girvin} (we give a solution in another form in Sec.\
ID\@). {}From this solution one can obtain the two-probe conductance for 
arbitrary Hall ratio, however here we analyze only its limiting behavior.
The conductance $g^0$ can be found by integrating the current over a 
cross-section perpendicular to the current flow:
\begin{equation}
g^0= -\sx^0\int_0^W dy\, (\partial_x+\gamma\partial_y) V(\br)/(V_2-V_1). 
\label{gclassical}
\end{equation}
First consider the case $\gamma \rightarrow 0$; it is then convenient to
integrate over all transverse cross-sections in the sample and divide by
the sample length $L$. The integral then just evaluates the voltage 
at the two ends where it is fixed, yielding the familiar ohmic result
\be
 \lim_{\gamma\rightarrow 0} g^0= \sx^0 W/L.
\ee
In the opposite limit
$\sx^0\rightarrow 0$, $|\gamma|\rightarrow \infty$, the boundary condition
implies that everywhere along the edge ${\cal E}_x = 0$, i.e.\ there is no
voltage drop along the edge and the voltage must remain equal to $V_1$
along one edge and $V_2$ at the other edge except at singularities 
at diagonally opposite corners (where it jumps between $V_1$ and $V_2$).
The transverse potential drop is equal to $(V_1-V_2){\rm sgn}\, \gamma$, 
therefore
\be
\lim_{|\gamma|\rightarrow \infty} g^0= \mbox{}-\lim_{|\gamma|\rightarrow 
\infty} \sx^0\int_0^W dy\, 
\gamma\partial_y V(\br)/(V_2-V_1)=|\sy^0|.
\ee
So, using the classical local conductivity, the two-probe conductance changes 
{}from being dominated by the longitudinal conductance at small $\gamma$ 
to being dominated by the Hall conductance at large $\gamma$. 
We can also solve the limit $L/W\rightarrow\infty$ with $\gamma$ fixed; for a
fixed current, the voltage drop is dominated by the part of the geometry a
distance greater than $W$ {}from the ends, in which the current distribution is
essentially independent of $y$, and one finds that
$g^0\rightarrow[\{(\sigma_{xx}^0)^2+(\sigma_{xy}^0)^2\}/\sigma_{xx}^0]W/L
\equiv(1/\rho_{xx}^0)W/L$. Note that the crossover to this behavior occurs at a
value of $L/W$ that depends on $\gamma$. In contrast to the above results for a
rectangular sample with edges, for periodic boundary conditions in the
transverse direction, which is equivalent to transport along a cylinder, the
conductance is always $\sigma_{xx}^0W/L$, for any value of $\gamma$. We note
that this geometry is equivalent, through a conformal mapping, to the Corbino
disk geometry, in which the voltage drop is radial, and since the equations 
are conformally invariant, results for the cylinder apply to the disk also.
Although the local formulation cannot be used in the fully phase-coherent
(``quantum'') case where $L_{in}\gg L$, we expect that the physics illustrated 
by this argument will be relevant to the average quantum conductance.

Indeed, in the {\em quantized} Hall regime $L\gg \xi$, previous arguments 
based on the Landauer formula for two-probe conductance in terms of 
transmission coefficients have noted the relation between two-probe 
conductance and Hall conductance. These approaches assume that the incident 
and outgoing channels are $N$ edge states \cite{Halperin,Buttiker}.  These 
edge states are analogous to classical skipping orbits advancing in one sense 
along each edge and occasionally being scattered into the bulk.  Any actual 
calculation of these transmission coefficients will be equivalent to 
evaluating the bilocal conductivity between the two ends \cite{Baranger}, 
however physical arguments are made that in high field the backscattering of 
edge states will be suppressed giving perfect edge transmission and 
$g=N=|\sy|$; i.e.\ the two-probe conductance is equal to the Hall conductance 
which takes its quantized value. (Since $\sx=0$ this is consistent with the 
classical result above). Although physically appealing, such arguments assume 
that the interaction of bulk and edge states can be ignored.  However in a 
bulk 2D sample it is known that it is only localization effects which prevent 
edge states {}from backscattering through the bulk states; they thus require 
that $W\gg\xi$, the localization length. We note that away {}from the critical 
values $E_{cN}$ ($N=0$, $1$, \ldots) of the Fermi energy, which lie near the 
center of the $N$-th Landau level, $\xi\simeq\xi_{pert}$, while 
$\xi\rightarrow\infty$ as $E_F\rightarrow E_{cN}$ for each $N$. We cannot 
calculate analytically the conductance in this non-perturbative regime $L$, 
$W>\xi_{pert}$; however our results below do describe samples for which $L$, 
$W\ll\xi_{pert}$ and localization effects have not inhibited backscattering of 
edge states. The key ingredient to describe the edge-bulk coupling in the 
perturbative regime is the tilted boundary condition for diffusion which we 
derive in the next subsections. In the fully phase-coherent case, the boundary
condition is however modified further.

\subsection{Classical network model, edge states, and tilted boundary 
condition on diffusion}

Khmel'nitskii and Yosefin \cite{Yosef} (KY), Maslov and Loss \cite{Maslov} 
(ML), and Read \cite{Read} have previously obtained the boundary condition on 
the diffusion process in high magnetic field. KY's argument appears to be
related in part to the classical conductivity formulas reviewed in Sec.\ IA\@. 
ML considered the effect of the edge in high field on
the microscopic diffusion process using a Boltzmann equation approach.
They found that the tendency to skip in only one direction when colliding
with the edge does lead to the tilted boundary condition on the diffusion
equation.  They expressed the tilt angle in terms of the
ratio of the mean free path along the edge and the bulk mean free path. 
Read \cite{Read} used the non-linear sigma model approach, which will be
described later in this paper. KY and Read were able to identify the tilt 
angle as the Hall angle.  In this subsection
we rederive the boundary condition in a particularly transparent manner using
a classical version of the network model for high field transport
introduced by Chalker and Coddington \cite{Chalker1}.  In this case one
can also see immediately that the tilt angle is the Hall angle.

The original Chalker-Coddington model \cite{Chalker1,Fertig} describes
the quantum tunneling between the semi-classical orbits along the 
equipotential contours of the smooth random potential (see 
figure~\ref{network}). To derive the diffusive behavior of the probability 
density in this model we will
neglect interference effects and describe each node by the probability 
that a walker approaching it makes a step to the right ($R$) or
left ($T$); $T+R=1$. (This simplified model has been considered
by several earlier authors \cite{Kucera,Szafer2,Ruzin}; 
it is essentially classical and could serve
as a lattice realization of the classical behavior discussed in Sec.\ IA.) 
The links of the lattice can be divided into four sublattices 
$\alpha=A$, $B$, $C$, $D$ (see Fig.\ 2), and each unit cell of the 
lattice contains one of each of
the four classes of links. The nearest neighbor 
separation is $a$.
We use $\rho_{\alpha}(i,j,t)$ to denote the probability of being at 
link $\alpha$, site $i,j$ at time $t$. Assuming that it takes time $\tau$ 
for a particle to move {}from one link to the next, one can define a 
random walk problem on the network and write down a probability 
evolution equation:
\bea
\rho_B(i,j,t+\tau)&=&\rho_{D}(i,j,t)T+\rho_{A}(i,j,t) R,\non\\
\rho_C(i,j,t+\tau)&=&\rho_{A}(i,j,t) T+\rho_{D}(i,j,t) R,   \non\\
\rho_A(i,j,t+\tau)&=&\rho_{B}(i,j-1,t) T+\rho_{C}(i+1,j,t)R, \non\\
\rho_D(i,j,t+\tau)&=&\rho_{C}(i,j+1,t) T+\rho_{B}(i-1,j,t) R.
\label{evolution}            
\eea
This problem differs {}from the usual random walk in the respect that
on each link the walk is in only one direction, hence breaking
time-reversal symmetry. 

The above equation can be diagonalized in 
the Fourier space. One can show that the long-time large-distance modes 
have a diffusive spectrum $$-i\omega_{k}=Dk^2 $$ for $\omega \ll 1/\tau $ 
and $k\ll 1/a$, where $D$, the diffusion constant, is given by
$D=\frac{1}{4}a^2RT/(R^2+T^2)\tau$. 
The associated eigenmodes are of the following form in Fourier space:
$$ 
\left(\ba{c}\rho_A(k)\\\rho_B(k)\\\rho_C(k)\\\rho_D(k)
\ea\right)=\left(\ba{c}1+\sx^0 ik_xa +\sy^0ik_ya\\ 1 \\1+
(\sx^0-\sy^0-1)ik_xa +(\sx^0+\sy^0)ik_ya 
\\1+(-\sy^0-1)ik_xa+\sx^0 ik_ya
\ea\right),$$                        
where in anticipation of our discussion below we identify the two 
constants $\sx^0=RT/(R^2+T^2)$ and $\sy^0=-T^2/(T^2+R^2)$ as the  
the bare longitudinal and Hall conductivity of 
this model. Note that these conductivities satisfy the 
``semicircle relation'', 
\begin{equation}
\sigma_{xx}^2 + (\sigma_{xy}+N+1/2)^2=1/4, 
\end{equation}
with $N=0$ here in the case of the lowest Landau level,
which has been claimed to be a general, exact
result in the quantum Hall effect \cite{Ruzin}.
In real space, all four components of the 
probability distribution satisfy the same coarse-grained equation:
\be
-D\nabla^2 \ov{\rho}(\br,t)=-\partial_t \ov{\rho}(\br,t).
\ee
At the absorbing ends, the particle moves away {}from the tunneling region 
with constant velocity. The fact that the re-entry probability is zero 
gives the boundary condition at the leads: 
\be
\ov{\rho}=0 \;\;\mbox{in the lead}.
\ee 
Since the particles always move in the direction of the arrow on a link, the
density on a link can also be considered as its current. The differences among 
the four components define the diffusion current densities, which are suitable 
for coarse graining. For instance, we can define 
$j_y=[\rho_B(i,j)-\rho_D(i+1,j)]/a$ and $j_x=[\rho_D(i,j)-
\rho_C(i,j)]/a$.
For the low-frequency and long-wavelength modes, we can show 
that
$j_y=-(\sx^0\partial_{y}-\sy^0\partial_{x})\ov{\rho}(\br,t)$.
Along the reflecting walls, at zero frequency, the normal current is zero, e.g. 
$\rho_B(i,j)-\rho_D(i+1,j)=0$ at top edge, which gives the 
boundary condition:
\be
(\partial_n -\gamma \partial_t)\ov{\rho}=0,\label{TBC}
\ee
where $\hat{\bf n}$ is the outward normal, and $\hat{\bf t}=\hat{\bf n}
\times\hat{\bf z}$ 
is the tangential direction of the boundary,
and $\gamma=-T/R=\sy^0/\sx^0$ with our above identification 
of the bare conductivities in the model. These expressions for the current 
density are of the form used by KY and in Sec.\ IA\@.

It is interesting to note that, in the network model,
the definition for the local diffusion current density is not unique. 
One can also define, e.g., $j'_x=[\rho_D(i,j)-\rho_C(i,j+1)]/a$, 
$j'_y=[\rho_B(i,j)-\rho_D(i,j)]/a$. $j'_y$ and $j_y$ 
differ by a total derivative term $\partial_x\ov{\rho}$, and the corresponding
$x$ components differ by $-\partial_y\ov{\rho}$ and by a 
$\delta$-function boundary term 
$\ov{\rho}(x,W)\delta(y-W)-\ov{\rho}(x,0)\delta(y)$. 
In the presence of such an edge current, the boundary condition that 
ensures ``current conservation'', which in the
interior is the requirement $\nabla\cdot{\bf j}'=0$, becomes 
\begin{equation}
\partial_x\int_{W-0^+}^{W+0^+}dy'\, j'_x(x,y') - j'_y(x,W)=0
\end{equation}
at the top edge, 
which is still equivalent to (\ref{TBC}). These forms for the current density 
are similar to those of ML.
The edge contribution ensures that the total current 
through a cross-section transverse to the $x$-direction is the same, 
whichever definition of current is used. 
 
One can see that in the long-time, large-distance limit the lack 
of time-reversal symmetry affects the diffusion process only through the tilted 
boundary condition, which is present only because of the edges. 
As we will show explicitly later, these boundary conditions, 
although derived {}from a lattice here for convenience, are quite
general for conduction with broken time-reversal symmetry.  As one might
hope, in the long-time, large-distance limit 
the microscopic details of different models cease to matter.

As noted just above, the boundary condition derived in Eq. (1.7) applies for
$\gamma = -T/R$, the single-node transmission and reflection 
coefficients of the classical network model.
We now must further justify identifying this ratio as the bare Hall ratio.
There are two approaches to this. There is no applied 
electric field or electric potential in this problem so far, so one may 
simply define the chemical potential on each link as proportional to the 
current (or density) there (in analogy to the Landauer approach
for the entire sample).  This was done by Kucera and Str\v{e}da \cite{Kucera}
and leads to the formulas for $\sx^0$ and $\sy^0$ given above.  
In our view a somewhat more satisfactory method is to calculate the 
steady-state current for the network under periodic transverse
boundary conditions (i.e., a cylindrical system), when current is 
injected at only one end of the network, with unit current on each incoming
link at that end.
This is just the appropriate time-independent solution of Eq.\ 
(\ref{evolution});  in the absence of edges the solution is
the linear $k=0$ mode: 
$$\rho_B(x,y)=b_1 x+b_0 ,$$ 
where $b_1=1/(L/a+\sx^0)$, $b_0=1-[\sx^0/(L/a+\sx^0)]$, and     
$$ 
\left(\ba{c}\rho_A(x,y)-\rho_B(x,y)\\ \rho_C(k)-\rho_B(x,y)\\\rho_D(x,y)-
\rho_B(x,y)\ea\right)=\left(\ba{c}\sx^0 b_1  \\
(\sx^0-\sy^0-1)b_1  
\\(-\sy^0-1)b_1
\ea\right)$$
where the constants $\sx^0,\sy^0$ are as defined above Eq. (1.6).
The above solution has a uniform current distribution with a 
total longitudinal current, for $L\gg a $,  
$$I_x=g=\frac{W}{L}\frac{\sx^0}{1+ (\sx^0a/L)}\simeq \frac{W}{L} \sx^0 ,$$ 
and a total transverse current, circulating around the cylinder, 
$$I_y\simeq -\sy^0,$$ 
justifying the interpretation we have assumed.

To solve for the current in the presence of edges is much more
difficult, even in this classical model. We will see, however, that on large
scales the problem is equivalent to that solved by Rendell and Girvin
\cite{Girvin} and discussed in Sec.\ IA. The reason for the similarity is 
already clear; in the classical network model, we assumed that the chemical 
potential was proportional to the density, and, when the external electric 
field is zero, the coarse-grained current density was related to this 
potential in the same way in both cases.  

\subsection{Continuum action and propagator}

The diffusive behavior generated by the classical network model of the 
previous subsection can be reproduced in a simple continuum field theory. 
Consider the action 
\be
S_0=-\frac{\sx^0}{4} \int d^2 r\,\partial_{\mu} z \partial_{\mu} \ov{z} - 
\frac{\sy^0}{4} \int d^2 r\,\epsilon_{\mu\nu}\partial_{\mu} z \partial_{\nu}
\ov{z}. 
\label{S_0}   
\ee
Here $z(x)$ is a complex scalar field, $\ov{z}$ is its complex conjugate, 
and the geometry is the same as in 
Sec.\ IB. The second term is clearly a total derivative.
To obtain the equations of motion and boundary conditions in this model,
we first rewrite $S_0$ as 
\be
\frac{\sx^0}{4} \int d^2 r\,
 [z \nabla^2 \overline{z}] - 
\frac{\sx^0}{4} \int d^2 r\,
[\partial_{\mu}\{ z (\partial_{\mu}+\gamma \epsilon_{\mu\nu}
\partial_{\nu}) \overline{z} \}]. 
\label{S'_0}  
\ee
Then one can see that the second term is again a total derivative 
and it can be written as a boundary term. Taking the functional derivative with
respect to $z$, we
obtain the  equation of motion for $\overline{z}$
\bea
-\frac{\sx^0}{4}\nabla^2 \overline{z}(\br) =0 && \;\;\mbox{ in the bulk}\non\\
(\partial_{n}-\gamma \partial_{t}) \overline{z}(\br)=0  
&& \;\;\mbox{ at the reflecting walls}.
\label{eqforzd}
\eea     
Similarly for $z$, we get
\bea
-\frac{\sx^0}{4}\nabla^2 z(\br) =0 && \;\;\mbox{ in the bulk}\non\\
(\partial_{n}+\gamma \partial_{t}) z(\br)=0  && \;\;
\mbox{ at the reflecting walls}.
\label{eqforz}
\eea 
At the absorbing boundaries, we simply impose $z=\overline{z}=0$. 
These equations are equivalent to the zero frequency limit of those in Sec.\ 
IB\@. We observe that the boundary conditions on $z$ and on $\overline{z}$ are 
not consistent, which means we cannot find any nonzero configurations $z(\br)$ 
that satisfy both these conditions simultaneously. This is due to the fact 
that the differential operator, that appears in (\ref{S'_0}) between $z$ and 
$\overline{z}$, is not self-adjoint. As noted in KY and ML, 
equations similar to eqs.\ (\ref{eqforzd}) and (\ref{eqforz}) define the 
right and left eigenfunctions of this operator; these eigenfunctions are not 
complex conjugates of each other. The eigenfunctions are not simple to obtain 
in our geometry, because the $x$ and $y$ dependence does not separate, when 
$\gamma\neq 0$. The analysis of the eigenfunctions in KY and ML ignores the
boundary conditions at $x=0$, $L$, and is appropriate only for $L/W\rightarrow
\infty$. 

To obtain a (zero-frequency) diffusion propagator, we use $S_0$ as the action
in a functional integral, and define $d(\br,\brp)$ by
\begin{equation}
d(\br,\brp) \equiv \frac{\sx^0}{4}
\langle\langle \overline{z}(\br) z(\brp)
\rangle\rangle_0\equiv\frac{\sx^0}{4}\int D[z,\overline{z}]\,
\overline{z}(\br) z(\brp) e^{S_0(z,\overline{z})}/Z_0.
\label{ddef}
\end{equation}
(Here $Z_0$ is the same functional integral without the $z(\br)$, 
$\overline{z}(\br)$
inserted.) Then $d(\br,\brp)$ satisfies
\bea
-\nabla^2 d(\br,\brp)=-\nabla'^2 d(\br,\brp)&=&\dl(\br-\brp)\;\;
\mbox{in the bulk;}\non\\
(\partial_{n}-\gamma \partial_{t})d(\br, \brp)&=&0,\;\;\mbox{$\br$ at the 
reflecting walls;}\non\\
(\partial'_{n}+\gamma \partial'_{t})d(\br, \brp)&=&0,\;\;\mbox{$\brp$ at 
the reflecting walls.}  
\label{propagator}
\eea
The propagator exists and can be shown to satisfy the stated conditions. 
One notices that the propagator is not symmetric with respect 
to $\br$ and $\brp$ since the boundary conditions for $\br$ and $\brp$ 
at the edges differ by a sign. In principle, the propagator can be 
evaluated by expanding in the right and left eigenfunctions, as in KY and ML,
however as these are not readily available for our geometry, we will just
define it by eqns.\ (\ref{propagator}) (see also Sec.\ IIID below). 

The action (\ref{S_0}) has also appeared in the literature in connection with 
an open string with opposite electric charges attached to the ends, in a 
uniform magnetic field \cite{Callan}. The boundary conditions have also 
appeared there, along with explicit results for the diffusion propagator $d$ 
in some geometries simpler than ours. We will see later (in Sec.\ IE and in 
Sec.\ III) that (\ref{S_0}) also arises as the lowest-order part of the 
nonlinear sigma model action. In fact, the full Chalker-Coddington model 
with phase coherence \cite{Chalker1} 
is related to the nonlinear sigma model \cite{Read2} in a manner closely 
analogous to the relation between the models discussed in Sec.\ IB and here, 
which are just the linearized versions. In Sec.\ III, we will also discuss the 
expressions for the currents, like those in Sec.\ IB, {}from the point of view 
of the nonlinear sigma model action.

\subsection{Bilocal conductivity tensor and conductance}

In Sec.\ IA above we have used a classical formulation of the conductivity and
worked out some of the consequences for the two-probe conductance
in high magnetic field.  In this subsection we introduce a full quantum
formulation for the bilocal conductivity tensor and the conductance in a
disordered phase-coherent system in order
to treat both the average quantum conductance and its variance.

For a quantum conductor with phase coherence, that is, when $L_{in}$ is larger
than the sample size, the electron wave function is sensitive to the 
external field in the entire space. Equation~(\ref{Ohms}) cannot be used as
there is no definition of the chemical potential within the sample.
We apply standard linear response theory to a finite disordered region
(denoted by $A$) 
with Fermi energy $E_F$, connected to perfect leads held at fixed voltages, 
which induce a local electric field in the disordered region the detailed
form of which is not relevant.  
(For the two-probe case, the sample occupies the region $0\leq x\leq L$, 
$0\leq y\leq W$; see Figure~\ref{two probe geometry}.) 
Following the treatment of Baranger and Stone \cite{Baranger} one finds that
there is a non-local relation between the current response and the applied
electric field:
\be
j_{\mu}(\br)=\frac{e^2}{h}\int_A d^2 r'\,\sigma_{\mu\nu}(\br,\brp)E_{\nu}(\brp),
\label{linear1}
\ee  
where the bilocal conductivity tensor $\sigma(\br,\brp)$ (which has dimensions
of inverse length squared) at $T=0$ can be 
expressed in terms of a pair of Green's functions \cite{Baranger}:
\bea
\sigma_{\mu\nu}(\br,\brp)&=&\frac{\hbar^4}{4 m_e^2}\left 
[G^+(\br,\brp;E_F)\DD^*_\mu\DD'_\nu G^-(\brp,\br; E_F)\right]\non\\
& &\mbox{}-\frac{\hbar^4}{4 m_e^2}\int_{-\infty}^{E_F}
dE'\left[
\frac{d}{dE'}G^+(\br,\brp,E')\DD^*_\mu\DD'_\nu G^+(\brp,\br,E')\right.\non\\
& &\left.\mbox{}+  G^-
(\br,\brp,E')\DD^*_\mu\DD'_\nu \frac{d}{dE'}G^-(\brp,\br,E') \right],
\label{sigma}
\eea
where
$$G^{\pm}(E)=\frac{1}{E-H\pm i \eta} ,$$
and
\begin{eqnarray*}
G(\brp,\br)\stackrel{\leftrightarrow}{\bf D} G(\br,\brp) &=& G(\brp,\br)
[\nabla-i(e/\hbar){\bf A}_0(\br)]G(\br,\brp)\\
&&\mbox{}-G(\br,\brp)[\nabla+i(e/\hbar){\bf A}_0(\br)]G(\brp,\br).
\end{eqnarray*}
Here $H$ is the Hamiltonian, discussed further in Sec.\ II, ${\bf A}_0$ 
is the vector potential representing the background magnetic field, 
and we will also use $-i\hbar{\bf D}=-i\hbar\nabla-e{\bf A}_0$.
In the presence of the magnetic field the bilocal conductivity tensor is 
not entirely a Fermi-energy quantity. Even at $T=0$, the complete 
current response function, in the presence of magnetic
field, contains not only terms involving $G^+G^-$ at the Fermi energy, but also
terms involving $G^+G^+$ and $G^-G^-$ 
integrated over all energies $E$ up to the Fermi energy. 
We  denote the $G^+G^-$, $G^+G^+$ and $G^-G^-$ terms  as
$\sigma_{\mu\nu}^{+-}$, $\sigma_{\mu\nu}^{++}$ and 
$\sigma_{\mu\nu}^{--}$. In disordered systems, products of 
Green's functions that are both retarded or both advanced are generally 
short-ranged because of the amplitude cancellations among different wave 
fronts (they typically only extend over the range of the mean free path), 
so we can treat $\sigma^{++}$ and $\sigma^{--}$ as contact terms:
\be
\sigma^{aa}(\br,\brp)=\sigma^{aa}\dl(\br-\brp),
\ee
where $a=+,-$, $\sigma^{aa}=\int_A d^2 r\, \sigma^{aa}(\br,\brp)$.

In the presence of a magnetic field $B$, the current given by (\ref{linear1}) 
does not necessarily satisfy $\nabla\cdot{\bf j}=0$ even when $\bf E$ is 
time-independent, unless we also require $\nabla\times{\bf E}=0$
(i.e.\ that the component $B$ of the magnetic field perpendicular to the 2D
layer is also time-independent).
As shown by Baranger and Stone \cite{Baranger}, under this condition 
``current conservation'', $\nabla\cdot{\bf j}=0$, is satisfied, and from it we
can derive conditions on the bilocal conductivity.
Writing ${\bf E}(\brp)=-\nabla'\phi(\brp)$ (where $\phi$ is 
the electric potential), we have
\be
{\bf j}(\br)=\frac{e^2}{h}\int d^2 r'\, 
\sigma(\br,\brp)\cdot \stackrel{\leftarrow}{\nabla'}\phi(\brp)
-\frac{e^2}{h}\sum_{n}\phi_i \int_{C_i}\sigma(\br,\brp)\cdot d{\bf S}',
\ee
where $\phi_i$ is the (constant) potential in the $i$th lead, $C_i$ is a 
surface across the $i$th lead, and the boundary terms at the reflecting walls
vanish because the normal component $\sigma_{n \nu}(\br,\brp)$ vanishes for
$\br$ at the wall (and similarly for $\nu=n$, $\brp$ at the wall). Thus 
$\nabla\cdot{\bf j}=0$ implies that the 
following conditions are satisfied:
\bea
\stackrel{\rightarrow}{\nabla}\cdot\sigma(\br,\brp)\cdot
\stackrel{\leftarrow}{\nabla'}&=&0
\non\\
{\bf\nabla}\cdot\int_{C_i}\sigma(\br,\brp)\cdot d{\bf S}'&=&0\;\;\mbox{for all 
$i$}
\label{constraints}.
\eea 
The above identities have been verified in Ref.\ \cite{Baranger}
for the exact bilocal conductivity tensor. 

Using the second identity in (\ref{constraints}), one can transform the 
linear response equation (\ref{linear1}) into a different form. Assuming,
without loss of generality, that $\mu=E_F$ in all the leads (we always view
$E_F$ as a position-independent constant), the electric 
potentials there differ {}from the voltages (electrochemical potentials) only by
a constant, and the total current in the $i$th lead can be written as a 
function of only the voltages in the leads:
\be
I_i=\frac{e^2}{h}\sum_j g_{ij}V_j, 
\label{linear2}
\ee
where $g_{ij}$'s are  conductance coefficients. The $g_{ij}$'s are related 
to the bilocal conductivity tensor by
\be
g_{ij}=-\int \int d{\bf S}_i\cdot\sigma(\br,\brp)\cdot d{\bf S}'_j. 
\label{conductance}
\ee
where $S_i$ and $S_j$ are cross-sections in the $i$th and $j$th lead, and
$d{\bf S}_i$, $d{\bf S}_j$ are differentials of outward-pointing normals. 
(In eq.\ (\ref{linear2}), we used the relation 
$\sum_j g_{ij}=0$, which follows {}from eq.\ (\ref{conductance}) and the 
second of eqs.\ (\ref{constraints}), and implies that a constant $V$ 
produces no current in any lead.) For cross-sections $S_i$ not intersecting 
$S_j$, for $i\neq j$, the off-Fermi-energy terms 
$\sigma^{++}(\br,\brp)$ and $\sigma^{--}(\br,\brp)$  can be shown to give
zero contribution \cite{Baranger}, and therefore 
$g_{ij}$ can be expressed as a Fermi-energy quantity.    
It has been shown that $g_{ij}$ is proportional to the 
total transmission coefficient of the scattering states at the Fermi energy 
{}from the $i$th lead to the $j$th lead \cite{Buttiker,Landauer,Baranger}. 
In this article we will consider only the simple case of the 
two-probe conductance, in which certain further simplifications are possible.

Since the total currents at all cross-sections are the same for a 
two-probe set-up, we can average over all 
cross-sections to get a volume-integral form for the  
two-probe conductance:
\be
g=\frac{1}{L^2}\int_A d^2 r \int_A d^2 r'\,\sx(\br,\brp).
\label{two-probe}
\ee
Here we must be careful to include the 
off-Fermi-energy terms since they are needed to preserve the uniformity of 
the current across each section ({}from the technical point of view, since
$x$ and $x'$ can now coincide, we are no longer justified in dropping the
$\sigma^{aa}$ terms).
Despite this disadvantage the volume-integral form of $g$ is often more 
convenient to use in actual calculation than the surface-integral form
since volume-averaging still eliminates many diagrams which would be
non-zero in the surface-integral approach.  Thus
equations (\ref{sigma}), (\ref{conductance})  and (\ref{two-probe}) 
will serve as the starting points for the evaluation of quantum 
conductance and conductance fluctuations. 

We may gain further insight into the meaning of the current conservation
conditions, and the relation to the classical case, by use of the
self-consistent Born approximation (SCBA) results for the disorder average of
the bilocal conductivity tensor.
As we will show in Sec.\ II, in this approximation 
$\langle\sigma_{\mu\nu}(\br,\brp)\rangle$ (the single angle brackets will 
always denote the average over the disorder) is of the following form when 
$\br$ and $\brp$ are more than a mean-free path {}from the edges:
\bea
\langle\sigma_{\mu\nu}(\br,\brp)\rangle_{\rm SCBA}
&=&[\sx^0\dl_{\mu\nu}+(\sy^{I,0}+\sy^{II,0})\epsilon_{\mu\nu}]
\dl(\br-\brp)\non\\
&&\mbox{}-\frac{1}{\sx^0}
[\sx^0\partial_{\mu}+\sy^{I,0} \epsilon_{\mu\mu'}
\partial_{\mu'}][
\sx^0\partial'_{\nu}-\sy^{I,0} \epsilon_{\nu\nu'}\partial'_{\nu'}] 
d(\br,\brp),
\label{sigmaSCBA}
\eea
where $\sx^0$, $\sy^{I,0}$, $\sy^{II,0}$, $\sy^{I,0}+\sy^{II,0}=\sy^0$ are 
the SCBA conductivity parameters \cite{Pruisken1}, $\sy^{II,0}$ comes {}from 
the 
$\sigma^{++}$ and $\sigma^{--}$ part of equation (\ref{sigma}), and $d$ is the
diffusion propagator discussed in Sec.\ IC\@ (with $\sigma_{xy}^0$ appearing in
the boundary conditions).  
In the zero field limit, $\sy^{I,0}=\sy^{II,0}=0$, the above reduces to
$$\langle\sigma_{\mu\nu}(\br,\brp)\rangle_{\rm SCBA}=
\sx^0\left[\dl_{\mu\nu}\dl(\br-\brp)
-\partial_{\mu}\partial'_{\nu} d^0(\br,\brp)\right] 
$$ 
(in which $d^0$ is the diffusion propagator for $\sy^0=0$).  To our 
knowledge this basic result
first appeared in the mesoscopic physics literature in ref.\ \cite{Kane}, 
although it may well have been known earlier.  
It or (\ref{sigmaSCBA}) show that the current response to an electric field 
has a nonlocal part, the term containing $d^0$ or $d$ in the formulas, 
due to diffusion. For non-zero magnetic field, one finds {}from eq.\
(\ref{sigmaSCBA}) that
\be
\partial_{\mu}  \langle\sigma_{\mu\nu}(\br,\brp)\rangle_{\rm SCBA}=
\epsilon_{\mu\nu}\sy^{II,0}\partial_{\mu} \dl(\br-\brp) .
\label{nonzerodiv}
\ee
The divergence of the response current is therefore
\be
\nabla\cdot \langle{\bf j}(\br)\rangle = \frac{e^2}{h}\int d^2 r'\, \nabla\cdot 
\langle\sigma(\br,\brp)\rangle_{\rm SCBA}\cdot{\bf E}(\brp)
=\frac{e^2}{h}\sy^{II,0}\nabla\times {\bf E}(\br).
\label{divjnonzero}
\ee
In the presence of a magnetic field, the current is divergenceless only when 
$\nabla\times{\bf E}=-\frac{\textstyle \partial B}{\textstyle \partial 
t}=0$; otherwise there is a time dependence in the density,
$\partial\rho/\partial t=(e^2/h)\sy^{II,0}\partial B/\partial t$. 
To obtain a truly static response, we would have to impose 
$\nabla\times{\bf E}=0$. (There is also an edge-current contribution
involving $\sigma_{xy}^{II}$, which will be described in the SCBA case
in Sec.\ II.) This behavior is typical of the quantum Hall effect, in which 
$\sy^{II}\epsilon_{\mu\nu}\delta({\bf r}-{\bf r}')$ is the only part of 
$\langle\sigma(\br,\brp)\rangle$ that is nonzero in the interior of the system 
on scales larger than the localization length $\xi$, and $\sigma_{xy}^{II}$ 
is quantized to integer values. It is the local expression of the 
gauge-invariance argument \cite{Laughlin,Halperin}.

The measured experimental quantity is the two-probe conductance.
It is straightforward to show (see Sec.\ II), {}from an equation similar to 
(\ref{sigmaSCBA}), that the two-probe conductance within the SCBA, $g^0$, 
can be written in terms of the diffusion propagator as:
\be
g^0=-\sx^0\int_0^W dy'\, \int_0^W dy\,
(\partial_x+\gamma\partial_y)(\partial'_x-\gamma\partial'_y)d(\br,\brp)
\label{g-SCBA1}.
\ee
where $x\neq x'$ are arbitrary.
We note that although $\langle\sigma_{\mu\nu}(\br,\brp)\rangle$ in the bulk 
depends on $\sy^{I,0}$, the conductance depends only on the full $\sy^0$, 
due to additional edge current contributions to $\sigma$ which we omitted
in (\ref{sigmaSCBA}). These contributions are 
similar to those discussed in Sec.\ IB\@. 

It is in fact possible to show that the mean bilocal conductivity and 
conductance obtained in SCBA are identical to those obtained {}from the 
``classical'' formulas of Sec.\ IA, if $\sigma_{xy}^{II,0}$ is zero. To obtain
the response to an arbitrary electric field $\bf E$, we write the conditions of 
Sec.\ IA on the current density, using 
$\mbox{\boldmath$\cal E$}={\bf E}+\nabla\mu/e$, as 
\begin{equation}
\nabla^2\mu/e=-\partial_\mu(E_\mu+\gamma\epsilon_{\mu\nu}E_\nu)
\end{equation}
and the boundary conditions
\begin{equation}
(\partial_n-\gamma\partial_t)\mu/e=-(E_n-\gamma E_t)
\end{equation}
at the reflecting walls, $\mu/e=E_F/e$ in the leads.
These inhomogeneous equations for $\mu/e$ are equivalent for $\gamma=0$ 
to a 2D electrostatics problem with a mixed Dirichlet-Neumann boundary
condition \cite{Jackson}, and can be solved using a
Green's function technique. A slight generalization of the same technique works 
for $\gamma\neq0$. The required Green's function is precisely 
$d({\bf r},{\bf r}')$ as defined in Sec.\ IC, and one finds
\begin{equation}
\mu({\bf r})/e=E_F/e-\int
d^2r'\,[(\partial_\nu'-\gamma\epsilon_{\nu\nu'}\partial'_{\nu'})d({\bf r},
{\bf r}')]E_\nu({\bf r}').
\end{equation}
Using eqs.\ (\ref{Ohms}), the bilocal conductivity tensor that results is 
exactly of the form (\ref{sigmaSCBA}), with $\sigma_{xy}^{II,0}=0$, and no 
additional edge
contributions. Consequently, the two-probe conductances $g^0$ given by 
Eqs.\ (\ref{gclassical}) and (\ref{g-SCBA1}) are the same, for
the same values of $\sigma_{xx}^0$ and $\sigma_{xy}^0$, and 
since this involves only the total $\sigma_{xy}^0$, it remains true 
even if $\sigma_{xy}^{II,0}\neq 0$ is included as in the SCBA. This implies that
the bilocal conductivity in SCBA has just the form which follows from
a local relation between current and electromotive field, even though there
is no sensible definition of a local chemical potential in the phase-coherent
limit.  Hence all
the conclusions drawn in Sec.\ IA can also be applied within the SCBA.

In general, there is also an edge-current contribution, with coefficient 
$\sigma_{xy}^{II,0}$, to eq.~(\ref{sigmaSCBA}), which will be described in 
Sec.\ II\@. There is no reason why both the bulk and edge $\sigma_{xy}^{II,0}$ 
contributions should not also appear in the so-called classical formulation of
Sec.\ IA, even though they were not included in Ref.\ \cite{Girvin}, since 
terms of this form would presumably still be present even if there were
inelastic scattering. 

We may also connect the results of Secs.\ IA and ID with the field theory in
Sec.\ IC (again for $\sigma_{xy}^{II,0}=0$). In Secs.\ IB and IC, the external
electric field was zero. If we replace $\partial_\mu z$, 
$\partial_\mu \overline{z}$ by
$\partial_\mu z-2iA_\mu$, $\partial_\mu \overline{z}+2i\overline{A}_\mu$ in the
action $S_0$, eq.\ (\ref{S_0}), (this $\bf A$ should not be confused with ${\bf
A}_0$ or any other ``physical'' vector potential) and then define 
$j_\mu({\bf r})=\delta S_0[{\bf A}]/\delta A_\mu({\bf r})$ to be the current, 
we find $j_\mu=i\sigma_{\mu\nu}^0(\partial_\nu \overline{z}
+2i\overline{A}_\nu)/2$, and the conditions for an extremum of the action
(equations of motion) are $\partial_\mu j_\mu({\bf r})=0$ 
in the interior and $j_n=0$ at the reflecting walls. Thus, these equations have
the same form as the local classical conductivity equations, 
with $i\overline{z}/2$, $-\overline{A}_\mu$ in place of $(\mu-E_F)/e$, $E_\mu$, 
and since, for a
quadratic action like $S_0$, the linear response $\delta \langle\langle
j_\mu\rangle\rangle_0/\delta \overline{A}_\nu$ 
to the perturbation $\bf A$ can be equivalently obtained either {}from 
calculation of the correlation function
(similar to that in the definition of $d$, eq.\ (\ref{ddef})), or using the
equations of motion as above, we get the same bilocal conductivity in this
approach also. This calculation is done in detail in Secs.\ III and
IV, so we refrain {}from giving further details here. Sec.\ IIIC also includes
the corresponding ``classical'' equations for the case when $\sy^{II,0}\neq 0$.

The SCBA for the average conductance is the leading approximation 
in an expansion in powers of $1/\sx^0$, and it is not really surprising that
this leading approximation behaves identically to the classical case. 
When evaluating conductance fluctuations, or weak localization corrections, 
one must consider higher orders in $1/\sigma_{xx}^0$. In such calculations the 
tilted boundary condition is modified further. In the framework of diagrammatic
perturbation theory, this can be alternatively viewed as the appearance of 
additional boundary vertices describing interference effects. 
These vertices are more easily obtained and evaluated in the nonlinear 
$\sigma$-model approach to which we now turn.

\subsection{Nonlinear $\sigma$-model approach}

The approach of the previous subsection, in which the
self-consistent Born approximation is the leading contribution to conductivity
and conductance, can be developed as a diagrammatic expansion (see Sec.\ IIA).
However this approach becomes cumbersome in higher orders because 
all diagrams contain vertices which need to be 
evaluated in terms of the average single particle Green's functions, 
and are dressed with disorder lines in all possible ways.
However, when these vertices, which describe interference or ``interactions'' 
of diffusing modes, are calculated at small wavevectors, they are all 
found to be related to the same quantities $\sx^0$ and $\sy^0$. These complex
and often redundant calculations can be avoided
by using the nonlinear $\sigma$-model (NL$\sigma$M) representation of the 
problem.

The NL$\sigma$M approach starts by considering only Green's functions at the 
Fermi energy, which means that the non-Fermi-energy parts of the conductivity 
described in the previous subsection cannot necessarily be obtained, though the
conductance and its fluctuations can. After introducing replicas, the disorder 
is integrated out, followed by the variables representing the electrons 
propagating at the Fermi energy. After a Hubbard-Stratonovich decoupling, and 
neglecting modes that have no long-range effects, one is led to the action  
\cite{Pruisken1,Pruiskenrev} (more details will be given in Sec.\ III)
\be
S=-\frac{\sx^0}{8}\int_A d^2 r\, {\rm tr} [\partial_{\mu} Q \partial_{\mu} Q] 
-\frac{\sy^0}{8}\int_A d^2 r \, {\rm tr} [\epsilon_{\mu\nu}Q \partial_{\mu}
Q\partial_{\nu}Q],
\label{S0}
\ee
where $Q$ is a $2n\times 2n$ Hermitian matrix obeying $Q^2=I_{2n}$, ($I_{2n}$
is the identity matrix); $Q$ has $n$ eigenvalues equal to $+1$, $n$ equal to
$-1$. Using a parametrization given explicitly in Sec.\ III, it can be shown
that for small fluctuations about the maximum action configuration where $Q$ is
diagonal, $S$ reduces at quadratic order to $n^2$ copies of the earlier action 
$S_0$ in (\ref{S_0}). 

Before discussing perturbation theory for this action, we wish to mention some
general issues. The second term in $S$ is the so-called topological term. 
It is the only possible term that can be added to the first term in two
dimensions that is consistent with the symmetries of the problem and contains
only two gradients (higher derivatives would be irrelevant at long length
scales). On a compact, oriented manifold without boundary, such as a 
sphere or a torus (i.e. periodic boundary conditions), this term (with 
a factor of  $2\pi i\sy^0$ removed) is a topological invariant, which 
takes integral values. This follows {}from the fact that the term is a total 
derivative, and the absence of boundaries. Consequently, only the value of 
$\sy^0$ modulo $1$ is important. Moreover, because the term is a topological 
invariant, $\sy^0$ does not appear in perturbation theory at all, but only in 
non-perturbative effects involving configurations (``instantons'') for which 
the topological term is nonzero. However, in the integer quantum Hall effect, 
we expect to obtain plateaus at integral $\sy$ and transitions between them at 
$\sy$ half-odd-integral ($\sx$, $\sy$ denote the renormalized, large-scale,
parameters, as opposed to the bare values $\sx^0$ and $\sy^0$ at the cutoff
scale $l$). Because of the periodicity in $\sy^0$, the NL$\sigma$M 
predicts that all these plateaus and transitions will have identical universal 
properties. But by the same token, it is also unable to predict the integral 
part of $\sy$ that would be observed in a measurement; this information 
appears to be lost in going to the NL$\sigma$M. 

The apparent paradox is resolved on examining the action $S$ for a system with
reflecting boundaries. The ``topological'' term is a total derivative that can
be rewritten as a boundary term, just as for the action $S_0$ above. The
boundary term is not a topological invariant, so it can affect perturbation
theory, and since it takes arbitrary real values, the magnitude of $\sy^0$ is
important, and not just its value modulo integers. Thus when 
the boundaries of the system are correctly taken into account,
the value of $\sy$ can be obtained within the NL$\sigma$M formulation.
Some additional remarks about this point are made in the Conclusion.

Since the leading order part of the action is the same as $S_0$, the 
propagator for small fluctuations in $Q$ is the same as the propagator $d$ 
discussed earlier, and depends on $\sy^0$ through the boundary condition. 
In the work of KY and ML, this modified propagator was the only 
effect included, and was just inserted into the Xiong-Stone results for 
conductance fluctuations. However, the NL$\sigma$M shows that the boundary term 
also contributes at higher order, producing new vertices for
``interactions'' between diffusons, which are boundary interactions with
coefficient $\sy^0$, and which contribute to the fluctuations at leading order.
These terms must be present in order to maintain the full U($2n$) symmetry of 
the NL$\sigma$M, which essentially corresponds to preserving the continuity 
equation for the current. We have also obtained them
in the diagrammatic approach, but only with much additional effort.
(Pruisken \cite{Pruiskenrev} also 
discussed the relation of the topological term to edge states, but appears 
to infer an incorrect boundary condition. His boundary condition is very 
useful in instanton calculations \cite{LLP,Pruiskenrev,Pruisken2} but does not 
correctly represent the edge effects,
unlike the conditions to be discussed in this paper.) In this paper, we 
evaluate the effects of these terms to leading order in 1/$\sx^0$, and, 
to simplify the calculations, also to leading nontrivial order in 
$\gamma=\sy^0/\sx^0$. As well as calculating the mean and variance of the 
two-probe conductance, we show how the expression for the mean bilocal 
conductivity tensor in the SCBA can be recovered within the NL$\sigma$M,
including the non-Fermi-energy parts, by modifying the coupling of the
NL$\sigma$M to the external field, and we discuss the resulting form of current
conservation conditions to all orders in perturbation theory.
 
One may wonder if the boundary conditions for a system with boundary invalidate
the conclusions of the analysis of Pruisken and co-workers
\cite{LLP,Pruiskenrev,Pruisken2}, who studied effects
of instantons in a system without a boundary. Strictly
speaking, in a finite system with boundaries, there are no well-defined
topologically distinct sectors. However, small instantons which are well
localized inside the system boundary, so that $Q$ approaches a constant outside
the instanton core and satisfies the boundary conditions at the edges, while
probably not exactly local minima of the action, are still nearly so
when their size goes to zero compared with the system dimensions, and in this
limit their topological charge will still be an integer, and they will make the
same contribution to the action as they did for the other b.c.'s. Thus, the
effects on renormalization group flows for $\sx^0$ and $\sy^0$, obtained in the
interior of the system, should be unchanged. These effects will not be 
considered further in this paper, which emphasizes perturbation theory. 

In the one-dimensional case of the NL$\sigma$M, no other term can be added to
the basic gradient-squared term, and there are of course no sides on which the
boundary condition could be modified. Therefore, the conductance fluctuations 
in the unitary ensemble should be universal, and must be recovered in the
quasi-1D limit of the 2D system in a magnetic field which we are considering;
this constitutes a strong check on the 2D calculations. 
We will show that in this limit the only effect of $\sy^0$ is to modify the 1D
conductivity, which is known to cancel {}from the conductance fluctuations to
leave a universal number. KY and ML claimed that $\sy^0$ does affect the
quasi-1D limit; the present argument shows that their results must be 
incorrect. In the Conclusion, we will briefly mention the situation for
dimensions higher than two. 

\section{Diagrammatic Expansion for $\langle\sigma_{\mu\nu}(\br,\brp)\rangle$}

In this section, we evaluate the mean bilocal conductivity tensor for the 
short-ranged potential model using the diagrammatic impurity-averaging
technique \cite{Mahan}. We will first review the self-consistent Born
approximation (SCBA), which is the leading order of the perturbation 
expansion, and establish basic parameters such as the mean free path $l$, 
and the bare conductivities $\sx^0$ and $\sy^0$. A gradient expansion 
is used to treat the current vertex. Within the SCBA, 
$\langle\sigma_{\mu\nu}(\br,\brp)\rangle$ has a contact term as well as a 
long-ranged 
term which can be expressed in terms of the diffusion propagator. In the bulk, 
$\sy^{I,0}$ appears in the long-ranged term. Along the reflecting boundary, 
the edge currents give rise to $\dl$-function contributions proportional to 
$\sy^{II,0}$. As a result, it is $\sy^0=\sy^{I,0}+\sy^{II,0}$ that appears 
in the boundary condition and the two-probe conductance. We will also check 
that current conservation is respected within the SCBA.

\subsection{The model, edge current, and ideal leads}  

An electron in a system with edges, in a random potential and subject to
a perpendicular magnetic field is described by the following
Hamiltonian (we neglect spin throughout this paper):
\be
H=H_0+V(\br),\; \; H_0=\frac{1}{2m_e}(-i\hbar{\bf D})^2+U(y),
\ee
where $V(\br)$ is the random potential and it is confined in the region 
of $0\le x\le L$, and $U(y)$ is the confinement potential 
(see Figure~\ref{two probe geometry})\@. The uniform magnetic field is 
in the $z$-direction, and we choose the gauge ${\bf A}_0=-By\hat{\bf x}$.
For simplicity, let us assume the confinement $U(y)$ to be the hard wall
potential  with $U(y)=0$ for  $0\leq y \leq W$, and $U(y)=\infty$ for $y<0$ 
and $y>W$. The infinite potential barrier requires the wave function to 
vanish at $y=0$, $W$. The system is infinite in the $x$-direction, but the 
disorder is present only in the region $0\leq x\leq L$. 
For the random potential, we will take the simplest among
all short-ranged models, which has the statistics of the Gaussian white
noise with zero mean (angle brackets denote the disorder average):
\bea
\langle V(\br)\rangle &=&0\non\\
\langle V(\br)V(\brp)\rangle &=&u \dl(\br-\brp),\non\\
\langle V(\br_1)V(\br_2)\cdots V(\br_n)\rangle _{\rm connected}&= & 0,\;\;n>2,
\eea
where $u$ describes the degree of disorder.

In the absence of the random potential, the unperturbed wave
functions can be found by separating the variables \cite{Halperin}. 
The wave functions $\psi(\br)$ are labeled by the wavevector $k$ in the 
$x$-direction and by $N$ in the transverse direction; $N$ turns out to be the 
Landau level index. We have 
$$\psi_{Nk}(\br)=\frac{1}{\sqrt{L}}e^{ikx}\phi_{N,k}(y),$$ and
$\phi_{N,k} (y)$ satisfies:
\be
\left\{ \hbar^2[(-i \partial_y)^2+
(y - l_B^2 k)^2 l_B^{-4} ]/2m_e +U(y) \right\} \phi_{N,k}(y)=
E_{N,k}\phi_{N,k}(y),
\ee 
where $l_B^2=\hbar/eB$. Without the confinement potential, the 
Hamiltonian is simply that of a harmonic oscillator, with the harmonic 
potential centering at $y_k=kl_B^2$.
We have \cite{Halperin} $\phi_{N,k}(y)=\chi_N(y-l_B^2 k)$, 
$E_{N,k}=E_N=(N+1/2)\hbar\wc$ for $W\gg y_k\gg 0$, where $\chi_N$ is the
$N$-th wavefunction of the harmonic oscillator, and $N=0$, $1$, $2$, \ldots. 
The wave
functions spread an extent $R_c=\sqrt{2N+1}l_B$ around $y_k$. We can see 
that for wave functions which center at a distance more than $R_c$
away {}from the walls, the presence of the walls is hardly felt, but for those
which reside within a distance $R_c$ {}from the walls, their eigenenergies are
raised above $E_N$, because the wave functions are forced to zero at the
boundaries and thus made to oscillate more rapidly near the walls. Only 
the states within $R_c$ of the edges have non-zero group
velocity along the walls, i.e., the expectation value of the velocity operator 
$\langle N,k|v_x|N,k \rangle \neq 0$. {}From now on,  $R_c$ plays the role of 
the short length scale of the problem and we 
treat the edges as having zero width. We will later show that, 
in this sense, the inhomogeneity at the edges gives rise to $\dl$-function  
contributions to the current (see also \cite{niu}).

Although the above description of $H_0$ is very convenient for finding an
explicit solution for the energy eigenfunctions when the system is infinitely
extended in the $x$ direction, it does not provide a convenient description of
ideal leads in the presence of a non-zero magnetic field. For our purposes 
ideal leads should be perfect absorbers of all incident current,
i.e.\ they should behave as if they have essentially infinite 
conductivity compared to the sample.  The problem is that in the leads,
where $x<0$ or $x>L$, the states at the Fermi energy generically consist 
of a certain number of edge channels moving in each direction, and this number
is equal to the number of Landau levels below the Fermi energy in the bulk,
which is $N+1$ when the Fermi energy lies above the $N$th Landau level. 
Thus in the relevant sense, the leads have zero bulk conductivity and only
absorb and inject current at the corners.  There are several ways in
which we could modify our model to remedy this problem.  One way would be
to let the magnetic field drop to zero at the ends of the sample, so the leads
are like the usual 2D metallic leads in zero magnetic field. Another way,
which corresponds roughly to the ohmic metallic contacts used in real 
experimental systems, would be to ``thicken'' the system outside the sample 
so that it becomes three dimensional, thus increasing its conductance. There 
is however a third way, which most closely conforms to the perfect leads
used in the network model (see Fig.\ \ref{network}) and has some convenient 
properties. The links of the network can be viewed as edge channels, 
and outside the sample there are many of them, running
parallel, alternately right- and left-moving, without backscattering. A similar
setup can be produced in a 2D Hamiltonian model with a uniform magnetic
field, by replacing $U(y)$, {\em only} in the leads, by a potential $U_1(y)$,
which $=+\infty$ for $y>W$ or $y<0$, and has a sinusoidal form in $0<y<W$. If
the Fermi energy lies between the maxima and minima of $U_1$, there will 
be many ``internal'' edge channels at the Fermi energy, 
consisting alternately of $N+1$ channels moving in one direction and $N+1$
moving in the other. For an infinitely-long, translationally-invariant system
there will be no backscattering among these modes. In effect, we have many 
narrow leads in parallel, all connected to a single reservoir at $-\infty$ 
and to another at $+\infty$. Then the number of right-moving channels can be 
proportional to the width of the system, or arbitrarily large, and the current 
can be injected uniformly across the end of the sample. {}From now on, 
it is this model that we will implicitly use.  

\subsection{The self-consistent Born approximation (SCBA)}

In order to calculate $\langle\sigma_{\mu\nu}(\br,\brp)\rangle$, we first 
need to evaluate the disorder-averaged single-particle Green's 
function and  
two-particle Green's function, $\langle G\rangle$ and 
$\langle G G\rangle $, which can be expanded in power series in 
$1/(k_F l)$. 
The SCBA takes
into account all the non-crossing diagrams (Figure~\ref{diagrams1}) 
and it has been shown 
to be the leading contribution in $1/(k_F l)$ \cite{Chalker2}.

Within  the SCBA, the  single-particle Green's function 
(see Figure~\ref{diagrams1}(a)) satisfies 
\be
[E-H_0(\br)-\Sigma(\br)]\Gb(\br,\brp;E)=\delta(\br-\brp),      
\ee
where the self-energy in turn  depends on $\Gb$ (we use the over-bar 
to denote SCBA Green's function): $$\Sigma(\br,E)=u \Gb(\br,\br;E).$$   
The above equation can be solved analytically in the  
low field limit ($\wc\tau_0\ll 1$) and the high field 
limit ($\wc\tau_0\gg 1$) \cite{Ando1}.  In the intermediate field 
range, it can be solved numerically. For our purpose, we do not need the 
explicit solutions. We use the SCBA Green's function to define the 
effective scattering rate $1/\tau$ and the effective mean free path 
$l$ at the Fermi energy:
\be
\frac{1}{\tau}=\frac{\Sigma^+-\Sigma^-
}{i\hbar}=\left\{\ba{cc}1/\tau_0,&\wc\tau_0\ll 1\\
2\Gamma\sin\theta(E_F)/\hbar,&\wc\tau_0\gg 1\ea\right. ,
\ee
where $\Gamma=\sqrt{u/(2\pi l_B^2)}\sim \hbar \sqrt{\wc/\tau_0}$ is 
the width of broadened Landau levels, 
$\theta(E)=\cos^{-1}\left[\frac{E-E_N}{2\Gamma}\right]$, $0\leq\theta(E)
\leq\pi$ and 
\be
l^2=u \int d^2 r_1\, (x-x_1)^2
\Gbp(\br,\br_1,E_F)\Gbm(\br_1,\br;E_F)=\left\{\ba{cc} 
l_0^2,&\wc\tau_0\ll 1\\
R_c^2, &\wc\tau_0\gg 1\ea\right..   
\ee
Here $l_0$ is the mean free path in zero magnetic field, $l_0=\hbar k_F
\tau_0/m_e$.

Within the SCBA, the two-particle Green's function $S^{+-}(\br,\brp)=u
\ov{ G^+(\br,\brp)G^-(\brp,\br) } $ amounts to adding up all the
ladder diagrams (figure~\ref{diagrams1}(b)). The sum can be written in 
the form of an integral equation:
\bea
\lefteqn{S^{+-}(\br,\brp,E,E')=}   \non \\
&&u\Gbp(\br,\brp,E)\Gbm(\brp,\br,E')+ \int
d^2 r_1\, u \Gbp(\br,\br_1,E)\Gbm(\br_1,\br,E')
S^{+-}(\br_1,\brp,E,E').
\eea
We make use of the fact that $\Gbp(\br,\brp)\Gbm(\brp,\br)$ is short-ranged and
expand $S^{+-}(\br_1,\brp)$ in the vicinity of $\br_1=\br$. We get, for $E$,
$E'$ close to $E_F$,
\bea
\lefteqn{\int d^2 r_1\, u \Gbp(\br,\br_1)\Gbm(\br_1,\br)
S^{+-}(\br_1,\brp)=} \non\\
&&[C_0(\br;E,E')+{\bf C}_1(\br;E,E')\cdot\nabla+\frac{1}{2}C_2(\br;E,E')
\nabla^2+\cdots] S^{+-}(\br,\brp)
\eea
where 
\begin{eqnarray*}
C_0(\br;E,E')&=&u\int  d^2 r_1\,\Gbp(\br,\br_1,E)
\Gbm(\br_1,\br;E')\simeq 1+i(E-E')\tau/\hbar,\\
{\bf C}_1(\br;E,E')&=&u \int d^2 r_1\, (\br-\br_1)
\Gbp(\br,\br_1;E)\Gbm(\br_1,\br;E')\simeq 0, \\
C_2(\br;E,E')&=&u \int d^2 r_1\, (x-x_1)^2
\Gbp(\br,\br_1,E)\Gbm(\br_1,\br;E')=l^2.
\end{eqnarray*} 
It follows that $S^{+-}(\br,\brp,E-E')$
satisfies the diffusion equation:
\be
[- {\cal D}\tau \nabla^2 -i(E-E')\tau/\hbar ]
S^{+-}(\br,\brp;E-E') \simeq \delta(\br-\brp)
\ee
where  $\cal D$ is the diffusion constant: ${\cal D}(E)=l^2/(2\tau)$. One 
can see that $S^{+-}$ at $E=E'=E_F$, which is all that will be required in 
this paper, is proportional to the dimensionless diffusion 
propagator $d(\br,\brp)$ we defined in equation (\ref{propagator}):
$$
\frac{l^2}{2} S^{+-}(\br,\brp;0)=d(\br,\brp).
$$
We postpone derivation of the boundary conditions on $d$ until after we have
discussed the conductivity tensor.
The ladder sum 
for $S^{++}=u\ov{G^+ G^+}$  and $S^{--}=u
\ov{G^-G^-}$ can also be carried out in
similar fashion. It is easy to see  that $ S^{++}(\br,\brp)$ and 
$ S^{--}(\br,\brp)$ are generally short-ranged.  

Now we are ready to treat the mean bilocal conductivity tensor. Within 
the SCBA, $\langle\sigma_{\mu\nu}(\br,\brp)\rangle$ has two contributions: 
the simple bubble diagram and the sum of the ladder series 
(Figure~\ref{diagrams1}(c)):
\be
\langle\sigma_{\mu\nu}(\br,\brp)\rangle_{\rm SCBA}=\sigma_{\mu\nu}(\br,\brp)
_{\rm bubble}+\sigma_{\mu\nu}(\br,\brp)_{\rm ladder}.
\ee
The bubble diagram 
has the range of the mean free path $l$
and we treat it as  a $\dl$-function:
$$
\sigma_{\mu\nu}(\br,\brp)_{\rm bubble}=\sigma_{\mu\nu}^0\dl(\br-\brp).
$$
$\sigma_{\mu\nu}^0$ are the SCBA conductivity parameters, which are essentially
constant inside the sample:
\be
\sigma_{\mu\nu}^0=\int_A d^2r\, 
\sigma_{\mu\nu}(\br,\brp)_{\rm bubble}=\frac{1}{LW}\int_A d^2r \int_A d^2r'\, 
\sigma_{\mu\nu}(\br,\brp)_{\rm bubble}.
\ee
It follows {}from the definitions (c.f.\ (\ref{sigma})) \cite{Baranger,Streda} 
that
\bea
\sx^0&=& \frac{\hbar^2}{LW}{\rm Tr }\left[v_x\Gbp(E_F)
v_x \Gbm(E_F)\right]\non\\   
&& -\frac{\hbar^2}{LW} \int_{\infty}^{E_F} dE \left\{ 
{\rm Tr}\left[v_x\ov{\left(\frac{d}{dE}G^+\right)}
v_x\Gbp\right]+{\rm Tr}\left[v_x\Gbm v_x\ov{\left(\frac{d}{dE}G^-
\right)}\right]
\right\}\non\\
&=&-\frac{\hbar^2}{2LW}{\rm Tr}\left\{v_x\Delta \Gb(E_F) 
v_x \Delta \Gb(E_F) \right\},
\eea
where $\Delta \Gb=\Gbp-\Gbm$, ${\bf v}=-i\hbar{\bf D}/m_e$, and ${\rm Tr}$ 
denotes the trace of the matrix product, in which $\overline{G}
^\pm(\br,\brp)$ are viewed as $\br$, $\brp$ matrix elements. For 
$\sy^0$,
\bea
\sy^0&=&\sy^{I,0}+\sy^{II,0},                             \non\\
\sy^{I,0}&=&\frac{\hbar^2}{LW} {\rm Tr}\left[v_x\Gbp(E_F)
v_y \Gbm(E_F)\right],\non\\            
\sy^{II,0}&= & -\frac{\hbar^2}{LW} \int_{\infty}^{E_F} dE' \left\{ 
 {\rm Tr}\left[v_x\ov{\left(\frac{d}{dE'}G^+\right)}
v_y\Gbp\right]+  {\rm Tr}\left[v_x\Gbm v_y\ov{\left(\frac{d}{dE'}G^-
\right)}\right]
\right\}.\non\\ 
& &
\eea
The above expression for $\sy^0$ can be put in the same Fermi-energy 
form as in Ref.\ \cite{Streda}.
$\sx^0$ and $\sy^0$ have the following limiting behavior 
\cite{Ando1,Pruisken1,Pruiskenrev}:
\bea
\sx^0&=&h {\cal D}(E_F) \rho(E_F) =\left\{\ba{cc}h n_e\tau_0/m_e,&\wc\tau_0\ll 
1\\
(2N+1)\pi^{-1}\sin^2\theta(E_F),&\wc\tau_0\gg 1 \ea\right. 
\non\\
\sy^0 &=&\left\{\ba{cc}\sx^0\wc\tau_0,&\wc\tau_0\ll 1\\
                    N+\nu, &\wc\tau_0\gg 1, \ea\right.   
\eea
where $\rho(E)$ is the SCBA local density of states, $n_e$ is the electron 
density, $N$ is the highest Landau level index, $N=0$, $1$, \ldots, and
$\nu=1-\theta/\pi$ is the 
filling fraction of the highest Landau level, which is well-defined since there
is vanishing local density of states in the bulk in between the Landau levels
in the regime $\omega_c\tau_0\gg 1$, within SCBA\@. 
Often, only the peak value $(2N+1)/\pi$ of $\sigma_{xx}^0$ is quoted in the
literature; we emphasize that $\sigma_{xx}^0$ has oscillations and goes to 
zero when the Fermi energy lies in one of the gaps in the bulk density of 
states.

Now we treat the ladder diagrams.
Since $S^{++},S^{--}$ are short-ranged, $\sigma^{++}$ and $\sigma^{--}$ 
are also short-ranged. One can 
further show that the ladder series for $\sx^{++}$ and $\sx^{--}$ 
do not give additional contributions to the $\dl$-function term.
The long-ranged term of $\sigma_{\mu\nu}(\br,\brp)$ comes {}from
$$\sigma_{\mu\nu}^{+-}(\br,\brp)_{\rm ladder} =\hbar^2u
\int d^2 r_1\int d^2 r_2\,J_{\mu}^{+-}(\br,\br_1)S^{+-}(\br_1,\br_2) 
J_{\nu}^{-+}(\brp,\br_2).$$ 
The current vertex
${\bf J(\br;\br_1)}$ is short-ranged:
\be
{\bf J}^{ab}(\br, \br_1)=\frac{-
i\hbar}{2m_e}
\Gb^{b}(\br_1,\br)
{\bf  \DD} \Gb^{a}(\br,\br_1),
\ee                              
where $a$, $b=+$, $-$, and ${\bf \DD}$ was defined in Sec.\ IC\@. 
We can carry out the integral by expanding $S^{+-}(\br_1,\brp)$ in 
the neighborhood of $\br$:
\be
\int d^2 r_1\, {\bf J}(\br,\br_1) S^{+-}(\br_1,
\br_2)=
\left [{\bf J_0(\br)} + J_1(\br)\cdot
{\bf \nabla
}+\cdots
\right ] S^{+-}(\br,\br_2).
\ee
The first term is
\bea
{\bf J}^{+-}_0(\br)&=&
\left(\frac{-i\hbar}{2m_e}\right)\int d^2 r_1\,
\Gbm(\br_1,\br){\bf \DD}\Gbp(\br,\br_1)
\non\\
&\simeq &\frac{1}{\Sigma^+ -\Sigma^-}\langle\br|{\bf v}(\Gbp-
\Gbm)|\br\rangle.
\eea
In the last step, we have used the identity 
$\Gb^+\Gb^-=(\Gbp-\Gbm)/(\Sigma^+-\Sigma^-)$. Rewriting 
$$\langle\br|{\bf v}(\Gbp-\Gbm)|\br\rangle\;\;\;\mbox{ as }\;\;\;
2\pi i \ov{\sum_{\al}\dl (E_F-E_{\al})
\psi_{\al}(\br){\bf v} \psi_{\al}^*(\br)},$$ it is easy to recognize that
${\bf J}^{+-}_0(\br) $ is proportional to the impurity-averaged 
equilibrium current
density at position $\br$ and at the Fermi energy $E_F$. 
In the bulk, this should be zero by isotropy, however, at the edges 
isotropy is
broken, ${\bf J}^{+-}_0$ 
parallel to the boundary is not zero. Hence we can write    
\be
 J^{+,-}_{0,x}(\br)=\frac{2\pi i}{e(\Sigma^+ -\Sigma^-)}\left[
\left.\frac {\partial
I_e(E)}{\partial E}\right|_{E_F}\dl(y-W)-
\left.\frac{\partial I_e (E)}{\partial E}\right|_{E_F}\dl(y)\right],
\ee
where $ I_e(E_F)$ is the total edge current. 
It has been shown in Ref.\ \cite{Pruiskenrev} that:
\be
\left.\frac{\partial I_e}{\partial E}\right|_{E_F}
=\left.\frac{\partial M}{\partial E}\right|_{E_F}
=-\frac{e\sy^{II,0}}{h},
\ee
where $M(E)$ is the total magnetization.

The coefficients for the first-derivative terms are:
\begin{eqnarray*}
J_{1,\mu\nu}^{+-}&=&\frac{1}{LW}\left
\{  {\rm Tr}[v_{\mu}\Gbp r_{\nu}\Gbm]-
 {\rm Tr}[r^{\nu}v^{\mu}\Gbp\Gbm]+\frac{i\hbar}{2m}{\rm Tr}[\Gbp\Gbm]\dl_{
\mu\nu}\right\},\\ 
J_{1,\mu\nu}^{-+}&=&\frac{1}{LW}\left\{{\rm Tr}[v_{\mu}\Gbm r_{\nu}\Gbp]-
 {\rm Tr}[r^{\nu}v^{\mu}\Gbm\Gbp]+\frac{i\hbar}{2m} {\rm Tr}[\Gbm\Gbp]\dl_{
\mu\nu}\right\} .
\end{eqnarray*}
Expressing the coefficients in terms of the SCBA conductivities and 
the mean
free path, we have:
\begin{eqnarray*}
u \hbar^2J^{+-}_{1,xx}J^{+-}_{1,xx}&=&\sx^{0}l^2/2, \\
u\hbar^2 J^{+-}_{1,xx}J^{+-}_{1,xy}&=&\sy^{I,0}l^2/2.
\end{eqnarray*}

Putting all the pieces together, we obtain for the full bilocal large-scale 
conductivity tensor within the SCBA, including the edge effects:
\bea
\langle\sigma_{\mu\nu}(\br,\brp)\rangle_{\rm SCBA}&=& 
\left[\sx^0\dl_{\mu\nu}+(\sy^{I,0}+\sy^{II,0})\epsilon_{\mu\nu}\right]
\dl(\br-\brp)\non\\
&&\mbox{}-\frac{1}{\sx^0}\left[\sx^0\partial_\mu
+\sy^{I,0}\epsilon_{\mu\mu'}\partial_{\mu'}+\sy^{II,0}\dl_{\mu x}(\dl(y-W)-
\dl(y))\right]\non\\
&&\mbox{}\times\left[
\sx^0\partial'_{\nu}
-\sy^{I,0}\epsilon_{\nu\nu'}\partial'_{\nu'}-\sy^{II,0}\dl_{\nu x}(\dl(y'-
W)-\dl(y'))\right] d(\br,\brp).      
\label{smunuSCBA}
\eea
Within the SCBA, the conductivity tensor obeys 
$\langle\sigma_{y\nu}(\br,\brp)\rangle=0$, for $\br$ at the reflecting 
edges and $\br$ sufficiently far {}from $\brp$, as an exact relation 
{\em before} the long-wavelength approximation is made. 
Together with (\ref{smunuSCBA}), this implies the large-scale boundary 
conditions ({\ref{propagator}) on the diffusion propagator $d$; see 
Appendix A\@. Because of the edge current, this is very similar to the 
discussion of the conservation of the ${\bf j}'$ current at the edge in 
Sec.\ IA (where $\sy^{II,0}$ was $-1$).

\subsection{The SCBA for the two-probe conductance}

We are finally ready to express the two-probe conductance
in terms of the surface integral of the diffusion propagator. 
For the integrated currents at any cross-sections, the two
opposite edge currents at the boundaries can be transformed to an
additional bulk derivative in the transverse direction:
$$ \int_0^W dy\, [\dl(y-W)-\dl(y)] d(\br,\brp)=\int_0^W dy\, 
\partial_y d(\br,\brp);$$
therefore, the second and third terms in the square brackets of 
(\ref{smunuSCBA}) combine in this case to give $\sy^0$. We get the 
form for the SCBA conductance $g^0$ (\ref{g-SCBA1}) expressed in terms
of an integral over two cross-sections.  This can be further transformed
into the following volume-integral form:
\be
g^0=\frac{1}{L^2}\int_A d^2 r \int_A d^2 r'\,\sx^0 [\dl(\br-\brp)- 
(\partial_x+\gamma\partial_y)(\partial'_x-\gamma\partial'_y)d(\br,\brp)].
\label{g-SCBA2}
\ee     
In the next section, we will see that the above expression for 
conductance can also be obtained using the NL$\sigma$M formalism 
which we develop below.  

\section{Field-theoretical Approach} 

In this section, we first set up (in Sec.\ IIIA) a generating function 
{}from which we can obtain any expressions involving  
the Green's functions at the Fermi energy. As we have seen in the 
introduction \cite{Baranger}, this allows calculation of the mean and the 
variance of the two-probe conductance. In Sec.\ IIIB, we discuss the coupling 
to a U($2n$) vector potential, which can be used to generate the NL$\sigma$M 
conductivity. For a certain form of coupling, we can in fact recover, 
at lowest order in $1/\sx^0$, the SCBA form of the bilocal conductivity 
tensor. We discuss the physical meaning 
of the various terms in relation to current conservation. The results show 
that the variance of the two-probe conductance can be calculated within the 
NL$\sigma$M\@. We then turn to the perturbation expansion itself.  

\subsection{The nonlinear $\sigma$ model}

In setting up the partition function (or generating function for average
Green's functions), we will use the replica method to perform the average 
over disorder. We define
\be 
Z=\int D[V]P[V] \int 
D[\varphi,\fba ]\exp \int d^2 r\,\fba(\br)[E-H+i\eta\Lambda]\varphi(\br),  
\label{genfunct}
\ee        
where, as in Sec.\ II, $H=H_0+V(\br)$, $\varphi=\ldots \varphi^a_i \ldots $ 
($i=1$, $2$, \ldots, $n$, $a=+$, $-$) is a  $2n$-component vector of 
complex Grassmann numbers, $\eta=0^+$, and
$$
\Lambda=\left(\ba{cc} I_n &0\\0&-I_n \ea\right)$$
($I_n$ is the $n\times n$ identity matrix).  
Here we remark that the choice of Grassmann number (anticommuting) fields, 
which leads to the symmetry group being U($2n$), as will be discussed below, 
is not essential. If one uses bosonic (commuting) fields, the symmetry is the
noncompact group U($n$,$n$). This is usually not used in the Hall effect
because there is no topological term in this case for $n>0$ integer, and so the
$\sigma_{xy}$ dependence found {}from non-perturbative instanton effects 
is not seen \cite{Pruisken2}. However, there is nonetheless a boundary term 
of the same structure as in the U($2n$) case, and for a system with boundaries 
this has effects even in perturbation theory, and these will be the same in the 
$n\rightarrow0$ limit for either choice of symmetry. We will continue to work
with the choice that leads to the compact symmetry. 
The random potential  $V(\br)$ has the Gaussian distribution 
$$ P[V]\propto e^{-\int d^2r\,V^2(\br)/2u},$$ as in Sec.\ II.
For $\eta=0$, the action has global ($\br$-independent) U($2n$) symmetry, 
which acts on $\varphi$ as $\varphi(\br)\rightarrow U\varphi(\br)$, where $U$ 
is an element of U($2n$). For $\eta>0$, the symmetry is broken
to U($n$)$\times$U($n$). 
In discussing the conducting properties, it is useful to introduce a source
term that will generate current correlations. This is done by introducing the
vector potential ${\bf A}(\br)$, where ${\bf A}$ is a $2n\times 2n$ 
hermitian-matrix-valued vector field (not to be confused with the vector 
field ${\bf A}_0$ associated with the constant magnetic field 
${\bf B}=\nabla\times{\bf A}_0$). It is introduced into $Z$ by replacing the 
covariant derivative $\bf D$ (viewed as multiplied by $I_{2n}$) by 
${\bf D}-i{\bf A} $. 
The generating functional $Z[{\bf A}]$ then has gauge invariance, since 
under a local U($2n$) gauge transformation 
$\varphi(\br)\rightarrow U\varphi(\br)$,
$A_\mu\rightarrow U A_\mu U^{-1}+U\partial_\mu U^{-1}$, the action $S[{\bf A}]$
is invariant, and so is the integration measure $D[\varphi,\fba]$.
Performing the functional integral over $\varphi$ and 
$\fba$, we get
\be
Z[{\bf A}]=\int D[V]P[V]\exp {\rm tr}\,{\rm Tr}\log [E-H-i\dl \Lambda-
\hbar^2({\bf A}^2 +i{\bf D}\cdot{\bf A}+i{\bf A}\cdot{\bf D})/2m_e],
\label{genfunc}
\ee
where ${\rm Tr}$ denotes the trace over functions in real space as before, and 
$\rm tr$ is the trace over the $2n$ replicas.

{}From the above partition function, we cannot get 
the exact bilocal conductivity tensor $\sigma_{\mu\nu}(\br,\brp)$ as given in
(\ref{sigma}), since we cannot generate the below-Fermi-energy contributions. 
However, on length scales greater than the mean free path, the latter simplify
and can be re-expressed as Fermi-energy terms, as will be shown below, and 
these can be reproduced {}from our partition function. An exception to this is 
the $\sy^{II,0}$ term in the bulk, which therefore describes true 
non-Fermi-energy physics. We can however get the expression for the 
two-probe conductance, which can be written in terms of Fermi-energy 
quantities alone \cite{Baranger}. 
Let us assume that the source field $\bf A$ is independent of $y$, then 
using $\delta/\delta A(x)$ to denote a functional derivative for a
$y$-independent variation, we can show that
\bea
\lefteqn{-\lim_{{\bf A}\rightarrow 0}
\frac{\dl ^2 Z[{\bf A}]}{\dl A^{+-}_{x,11}(x) \dl A^{-+}_{x,11}(x') }
= \int D[V]P[V]
 \int dy \int dy'\, \biggl\{ e^{{\rm trTr}\log (E-H-i\eta \Lambda)} }
\non\\
& & \biggl[\hbar^2 \frac{
G^+(\br,\brp)+G^-(\br,\brp)}{2m_e}\dl(\br-\brp)-\hbar^2 G^-(\brp,\br) 
\left(\frac{-i\hbar}{2m_e} \DD_{x'}^*\right) 
\left(\frac{-i\hbar}{2m_e}\DD_x\right) G^+(\br,\brp)   \biggr] \biggr\}. \non\\
                               \label{dAdA}\eea                          
The $\dl$-function term is related to $\sigma_{xx}^{++}(\br,\brp)$ and 
$\sigma_{xx}^{--}(\br,\brp)$ (see Sec.\ ID). We have argued in Sec.\ ID
that $\sigma_{xx}^{aa}(\br,\brp)$ ($a=+,-$) in disordered systems are 
short-ranged and can be treated as contact terms  
$\sigma^{aa}_{xx}\dl(\br-\brp)$.  One can show  using the commutation relations 
$v_x=i[\hat{H},x]/\hbar$ and $ [v_x,x]=\hbar/im_e$ \cite{Baranger,Streda,Xiong}
that   
$$
\sx^{aa}=-\frac{\hbar^2}{2LW}{\rm Tr}\left[G^a(E_F) v_xG^a(E_F) v_x\right]=
\frac{\hbar^2}{2m_e} G^a(\br,\br).
$$
Therefore, the expression in the square bracket in (\ref{dAdA}) gives an
approximate version of the unaveraged $\sigma_{xx}(\br,\brp)$, eq.\
(\ref{sigma}), valid on 
scales greater than the mean free path $l$. The $\dl$-function term drops out 
for two cross-sections far apart. Taking the limit  $n\rightarrow 0$, 
in which case $e^{{\rm tr\,Tr} \log (E-H-i\eta \Lambda)}\rightarrow
1$, equation~(\ref{dAdA}) gives the disorder-averaged two-probe 
conductance:  
\be
\langle g(x,x') \rangle=- 
\lim_{n\rightarrow 0}\lim_{{\bf A}\rightarrow 0}
 \frac{\dl ^2 Z[{\bf A}]}{\dl A^{+-}_{x,11}(x) \dl A^{-+}_{x,11}(x')}.
\label{g-sigma}     
\ee
Similarly, the second moment of the conductance can be obtained by applying 
four derivatives to the partition function:
\be
\langle g(x_1,x'_1)g(x_2,x'_2) \rangle  =  
\lim_{n\rightarrow 0}\lim_{{\bf A}\rightarrow 0}
     \frac{\dl ^4 Z[{\bf A}]}{\dl A^{+-}_{x,11}(x_1) \dl A^{-+}_{x,11}(x_1')
\dl A^{+-}_{x,22}(x_2) \dl A^{-+}_{x,22}(x_2') }.
\label{gg-sigma} 
\ee
These expressions should be independent of $x_1$, \ldots, $x_2'$. This will be
discussed in the next subsection.

An effective NL$\sigma$M action can be derived using the Hubbard-Stratonovich
transformation \cite{Pruisken1,Pruiskenrev}; a mean-field approximation then 
corresponds 
to the SCBA. The fluctuations about this mean-field theory produce all of 
the higher-order effects. An effective action for the long-range effects 
can then be derived; we get, retaining only terms with no more than two 
derivatives, and omitting the gauge field $A$ until the next subsection, 
\bea
S&=&\int d^2 r \,\left\{-\frac{1}{8}\sx^0 {\rm tr} [\partial_{\mu}
Q\partial_{\mu} Q]
-\frac{1}{8}\sy^0 {\rm tr } [\epsilon_{\mu\nu} Q
\partial_{\mu}Q\partial_{\nu} Q]+ \eta  {\rm tr } [Q\Lambda] \right\},
\label{S[0]}\\
Z&= &\int D[Q]\,e^{S}.
\label{Z[0]}
\eea
The field $Q$ is a $2n\times 2n$ Hermitian matrix obeying
(at each $\br$) $Q^2=I_{2n}$, of which $n$ eigenvalues are equal to $+1$, 
$ n$ to $-1$. This means that, at each $\br$, $Q$ takes values in the coset 
manifold U($2n$)/U($n$)$\times$U($n$). 
The action $S$ has the same symmetry as the
original action, $Q(\br)\rightarrow UQ(\br)U^{-1}$,
but $Q$ is invariant under the diagonal U(1) subgroup of
U($2n$). The remaining SU($2n$) symmetry is again broken to
S(U($n$)$\times$U($n$)) for non-zero $\eta$. The parameters $\sx^0$ and $\sy^0$
are bare conductivity parameters, like those resulting {}from the SCBA, 
(but may differ by finite renormalizations, corresponding to short-range 
effects that are not included again by the NL$\sigma$M)
and describe the response of the system at the scale of the short distance 
cutoff, which is of order the mean free path $l$. The measure $D[Q]=\prod_\br 
dQ$ is the product over points $\br$ inside the sample of the unique 
SU($2n$)-invariant measures $dQ$ on the space U($2n$)/U($n$)$\times$U($n$) 
for each point $\br$. At the ends $x=0$, $L$, we insist that $Q=\Lambda$ 
to represent the absorbing boundary condition.
   
\subsection{Gauge invariance, current conservation, and boundary condition}

In this subsection, we discuss the way in which the gauge potential $\bf A$
enters the NL$\sigma$M action, and the related questions of current
conservation, the equation of motion, and the tilted boundary condition. 
We begin by requiring that the action be gauge invariant. In Sec.\ IIIC, 
we will modify it to a non-gauge-invariant form to bring the conductivities 
into line with those discussed in the previous sections. 

In view of the gauge invariance of
the generating functional $Z[{\bf A}]$, the action $S[{\bf A}]$, including $\bf
A$, should also be invariant (when $\eta=0$) under the local gauge 
transformation $Q\rightarrow U(\br)Q(\br)U^{-1}(\br)$, 
$A_\mu\rightarrow U A_\mu U^{-1}-iU\partial_\mu U^{-1}$ (it is assumed that $U$
respects $Q=\Lambda$ at the ends; the invariance of the functional integration 
measure implies that invariance of the action ensures invariance of 
$Z[{\bf A}]$).   
The simplest way to introduce the external source field ${\bf A}$ 
into the NL$\sigma$M is to replace the partial derivative $\partial_{\mu}Q $ 
by the covariant derivative $D_{\mu}Q=\partial_{\mu}Q +i[A_\mu, Q]$
everywhere in $S$. This leads to a manifestly gauge-invariant action, which is
given below as the $\sigma_{xx}^0$ and $\sigma_{xy}^{I,0}$ terms in 
eq.~(\ref{S[A]}). However, this is not the only way. The second
way is less obvious and will follow a brief digression.

To obtain the other gauge-invariant coupling to $\bf A$, let us first point out
that the topological term, without any $\bf A$-dependence, is a total 
derivative, which is a function of the values of $Q$ on the boundary and  
the homotopy class of $Q$ in the interior for given boundary values, but not 
of the detailed form of $Q$ in the interior. That is, it is possible to change 
$Q$ in the interior, leaving its boundary values fixed, in such a way that the 
topological term {\em changes}. Now any change in $Q$ can be viewed as the 
result of a gauge transformation, and a transformation $U$ that leaves the 
boundary values of $Q$ unchanged must reduce at each point on the boundary 
to some element in a certain S(U($n$)$\times$U($n$)) subgroup, determined by 
$Q$ at that point. Such a gauge transformation is characterized by an 
integer-valued winding number $q$, say, which describes the winding of the 
gauge transformation $U$ at the edge, and it changes the topological term by 
$2\pi i\sigma_{xy}^{0}q$ (for $\sigma_{xy}^{0}$ equal to an integer, this has 
no effect on the exponential of the action). In particular, there is a 
continuously-connected class of transformations for which $q=0$, so that 
there are continuously-connected classes of configurations $Q$ with given 
boundary values throughout which the topological term takes the same value.
Further discussion of the topological issues, related to edge states and
quantization, is contained in Appendix C.

Now, because the topological term is (apart {}from the topological effects just
discussed) a function of only the boundary values of $Q$, this suggests that 
we can attempt to compensate for a gauge transformation by including a 
coupling to $\bf A$ on only the reflecting (hard) walls. The form of this 
coupling can be easily obtained, and on including both forms of coupling with 
different coefficients $\sigma_{xy}^{I,0}$, $\sigma_{xy}^{II,0}$, whose sum is
$\sigma_{xy}^{0}=\sigma_{xy}^{I,0}+\sigma_{xy}^{II,0}$, in the respective
forms of the topological term, we obtain the action which is gauge-invariant
when $\eta=0$,
\bea 
S[{\bf A}]&=&\int d^2 r \,\left[-\frac{1}{8}\sx^0 {\rm tr} (D_{\mu}
QD_{\mu} Q)-\frac{1}{8}\sy^{I,0} {\rm tr } (\epsilon_{\mu\nu} Q
D_{\mu}QD_{\nu} Q) \right.\non\\
&&\left.\mbox{}-\frac{1}{8}\sy^{II,0} {\rm tr} (\epsilon_{\mu\nu} Q
\partial_{\mu}Q\partial_{\nu} Q) 
+ \eta {\rm tr } (Q\Lambda) \right]\non\\
&&\mbox{}+\frac{i}{2}\sy^{II,0}\int dx\, \left[{\rm tr} (A_x(x,W)Q(x,W))-{\rm
tr}(A_x(x,0)Q(x,0))\right].
\label{S[A]}
\eea
The justification for identifying this split of $\sigma_{xy}^0$ into two
pieces $\sigma_{xy}^{I,0}$, $\sigma_{xy}^{II,0}$, with that found in the
previous sections, as implied by the choice of notation, will be given below.
We note that the action is independent of the trace of $\bf A$.
To see this, we may split $\bf A$ into the sum of a traceless part and the
trace multiplied by $I_{2n}/2n$. The latter part is the gauge potential 
corresponding to the diagonal U($1$) subgroup generated by $I_{2n}$, and it
does not contribute to $D_\mu Q$ (because $[I_{2n},Q]=0$), or to the
$\sigma_{xy}^{II,0}$ edge coupling (because ${\rm tr}\,Q=0$). Hence, $S[{\bf
A}]$ is independent of it, but it may be left in for convenience. The action 
$S[{\bf A}]$ is easily verified to be invariant under any gauge transformation,
including the topologically-nontrivial ones that leave $Q$ on the boundary 
unchanged. 

Although we have not given a derivation of our gauge-invariant action 
$S[{\bf A}]$, eq.\ (\ref{S[A]}), {}from the gauge-invariant generating 
functional (\ref{genfunc}) for averages of products of Green's functions at the 
Fermi energy, it is not very difficult to extend the existing 
derivations (see, e.g., \cite{Pruiskenrev}) to include $\bf A$, and obtain 
eq.\ (\ref{S[A]}). It is almost self-evident that this will be obtained, 
by comparing (i) the diagrams for the response to $\bf A$ {\em at the Fermi 
energy} that are obtained {}from eq.\ (\ref{genfunc}) with (ii) those studied 
in detail for the mean bilocal conductivity tensor within the SCBA in Sec.\ II,
and (iii) those obtained in the perturbation theory for the NL$\sigma$M 
constructed below.

We now use the gauge invariance of $S$ in the $\eta\rightarrow 0$ limit
to derive some current-conservation relations for the NL$\sigma$M; only
infinitesimal, topologically trivial gauge transformations are needed for this.
The tilted boundary condition is one consequence of current conservation. 
We will recover the expression for the bilocal conductivity in the SCBA 
as the leading-order term in an expansion in $1/\sigma_{xx}^0$.
We will also show that the conductance we obtained using
equation~(\ref{g-sigma}) is independent of the positions of the cross 
sections so that the volume-integral form for the conductance can be used. 

We begin by considering the equations of motion that follow {}from the action
$S[{\bf A}]$. As their name implies, these are the equations of motion that are
obtained if $S[{\bf A}]$ is used as the action of a classical nonlinear field
theory (in one space and one imaginary time dimension). The canonical way to
obtain these equations is to seek an extremum of $S$, such that 
$\delta S/\delta Q=0$, where the variation, which is the usual one that varies
both the $x$ and $y$ dependence of $Q$, respects the restrictions $Q^2=I$,
$Q=Q^\dagger$, and the boundary condition
$Q=\Lambda$ at the open ends (we note that this is imposed on all
configurations in the functional integral). The resulting equations also serve
as operator relations in the quantum field theory that we take to be defined by
the functional integral 
$$Z[{\bf A}]=\int D[Q]\,e^{S[{\bf A}]}.$$
In the functional integral language, the equations of motion become 
identities among correlation functions. They are obtained in general by the
following argument: Consider a small change in $Q$, $Q\rightarrow Q' = 
Q+\delta Q$, as a change of
variable in the functional integral. Since $Q$ is integrated over, such a
change can have no effect on $Z[{\bf A}]$. On the other hand, it changes $S$,
and provided the change is such that the Jacobian resulting {}from the change in
measure is $1$, we obtain the identity
$$
0=\int D[Q]\,\frac{\delta S}{\delta Q}\,e^{S[{\bf A}]},
$$
which is the equation of motion.

A variation of $Q$ that respects its form can be parametrized as
$Q'=UQ U^{-1}$, where $U=\exp iR$ and $R(\br)$ is a $2n\times 2n$ hermitian 
matrix function of $\br$, and thus is a gauge transformation, which leaves the
integration measure unchanged; however, we vary
$Q$ while leaving $\bf A$ fixed. If we view $S[{\bf A}]$ as a functional of $Q$
as well as of $\bf A$, thus writing $S[{\bf A}]\equiv S\{Q,{\bf A}\}$ (the
curly brackets are used to avoid confusion of the two arguments of the 
functional with a commutator), 
then the equations of motion are equivalent to $\delta
S\{Q',{\bf A}\}/\delta R=0$, evaluated at $R=0$. This may be re-expressed by
making use of the gauge invariance of $S\{Q,{\bf A}\}$. Gauge invariance tells 
us that if $Q'=UQU^{-1}$, ${\bf A}'=U{\bf A}U^{-1}-iU\nabla U^{-1}$, then 
\begin{eqnarray}
S\{Q,{\bf A}\}&=&S\{Q',{\bf A}'\}\non\\
           &=&S\{Q+i[R,Q],{\bf A}\}+S\{Q,{\bf A}+i[R,{\bf A}]-\nabla R\}
                  -S\{Q,{\bf A}\},
\label{gaugeinvS}
\end{eqnarray}
to first order in $R$. It follows that, for ${\bf r}$ in the interior of 
the system, we obtain
\begin{eqnarray}
\frac{\delta S}{\delta R_{\beta\alpha}}&=&-\partial_\mu\frac{\delta S}
{\delta A_{\mu,\beta\alpha}} -iA_{\mu,\alpha\gamma}\frac{\delta S}
{\delta A_{\mu,\beta\gamma}}+ iA_{\mu,\gamma\beta}\frac{\delta S}
{\delta A_{\mu,\gamma\alpha}}\non\\
 &=&-(D_\mu j^\sigma_\mu)_{\alpha\beta}.
\end{eqnarray}
Here we have used $\alpha$, $\beta$, $\gamma$, \ldots, for indices running 
{}from $1$ to $2n$ (the first $n$ values being $i$, $+$, $i=1$, \ldots, $n$, 
the remainder $i$, $-$, $i=1$, \ldots, $n$), with a summation convention, and 
introduced the definition of the current in the NL$\sigma$M, 
\begin{equation}
j^\sigma_{\mu,\alpha\beta}({\bf r})=\frac{\delta S[{\bf A}]}{\delta
A_{\mu,\beta\alpha}({\bf r})},
\label{jsigdef}
\end{equation} 
which is implicitly a local function of $Q({\bf r})$ and ${\bf A}({\bf r})$.
The covariant derivative of ${\bf j}^\sigma$ is defined in the same way as that
of $Q$. Therefore, using the notation
$$ 
\langle\langle\cdots\rangle\rangle = \int D[Q]\, 
\cdots\, e^{S[{\bf A}]}/Z[{\bf A}],
$$ 
where the $\cdots$ represent any functional of $Q$, and using the fact that the
integration measure is invariant under U($2n$) gauge transformations, 
the equation of motion becomes the matrix equation, 
\begin{equation}
{\bf D}\cdot \langle\langle {\bf j}^\sigma\rangle\rangle=0. 
\label{wardid}
\end{equation}
This can be read as the statement that the covariant divergence of the current
is zero in the interior of the system. As such it is known as a Ward identity, 
and is a consequence of the U($2n$) symmetry of the original action $S$; 
the existence of a covariantly-conserved current as a consequence of a 
gauge-invariant coupling to a gauge potential is the essential content of 
Noether's theorem. A feature of the NL$\sigma$M is that the Ward identities 
are not just consequences of, but are equivalent to, the equations of motion.
There are also many similar Ward identities when the functional average 
contains ${\bf D}\cdot {\bf j}^\sigma$ times other functionals of $Q$. 
Particular cases of these, including all those that will be of interest in this 
paper, are those that contain other currents ${\bf j}^\sigma$. These may be 
obtained by taking functional derivatives of the basic Ward identity
(\ref{wardid}) with respect to $\bf A$, since the left-hand side is still a 
functional of $\bf A$. Functional derivatives of the action yield currents, 
however, ${\bf j}^\sigma$ contains $\bf A$, and so does the covariant 
derivative $\bf D$, so there are additional $\delta$-function terms. 
The delta-function terms in the response functions that result {}from the 
$\bf A$ in ${\bf j}^\sigma$ will be referred to as contact terms, and 
correspond to those in earlier sections.

The above Ward identity, or equation of motion, (\ref{wardid}), is only 
the bulk part of the system of equations. There are also boundary equations on 
the reflecting (hard) wall or edge. These come {}from two sources: (i) the 
boundary terms in the action $S[{\bf A}]$, (ii) the boundary term that appears 
when integrating by parts to transfer the derivative {}from $\nabla R$ to 
$\delta S/\delta A$ in eq.\ (\ref{gaugeinvS}) when taking the functional 
derivative with respect to $R$. The boundary part of the equation of motion 
can be obtained in either of two equivalent ways. One way is to take the 
functional derivative with respect to the full dependence of $R$ on the 
coordinates $x$ and $y$, and obtain a single equation like (\ref{wardid}) 
but containing delta-function terms at the edge. Since an equation that sets 
a sum of a finite function and a delta function to zero implies that each piece 
separately vanishes, we obtain the bulk equation of motion (or Ward identity) 
as in (\ref{wardid}), together with a boundary condition that states that the 
function multiplying the delta-function at the edge is zero. The other method 
is to separate the change in the action due to $R$, after integrating by parts 
to remove derivatives {}from $R$, into a bulk part that yields the bulk 
equation above, and a boundary part, then take a functional derivative with 
respect to the single coordinate $x$ that descibes position on the boundary, 
to yield the boundary conditions. This was the method followed in Sec.\ IC for 
the quadratic action there, eq.\ (\ref{S_0}). With either method, it is 
straightforward to obtain the results below.

First, we should record the actual expression for the current density, 
obtained {}from the definition (\ref{jsigdef}). It is 
\begin{equation}
j_\mu^\sigma({\bf r})=-i[\sx^0 Q({\bf r}) D_\mu Q({\bf r})
-\sy^{I,0} \epsilon_{\mu\nu} D_\nu Q({\bf r})]/2 
+i\sigma_{xy}^{II,0}\delta_{\mu x}Q({\bf r})[\delta(y-W)-\delta(y)]/2
\label{jsigexpr}
\end{equation} 
(we used the identity $Q\partial_\mu Q=-\partial_\mu Q.Q$ which follows 
{}from $Q^2=I_{2n}$). Here we see explicitly the edge contribution with 
coefficient $\sigma_{xy}^{II,0}$. It is simply proportional to 
$Q$ at the edge. 

The boundary condition at the reflecting walls (assumed parallel to the
$x$-axis as always in this paper), that is obtained by either of the methods
described above, can be written
\begin{equation}
\sigma_{xx}^0 QD_yQ+\sigma_{xy}^{I,0}D_xQ+\sigma_{xy}^{II,0}D_xQ=0.
\label{wardidbc1}
\end{equation}
As in the case of the bulk equation of motion, in the quantum field theory of
the NL$\sigma$M (defined by the functional integral), this is valid only 
when inserted in the average $\langle\langle\ldots\rangle\rangle$. This 
equation can be interpreted as
stating the covariant conservation of the current at the edge, in a very
similar way to that discussed in Appendix A within SCBA. For
$\sigma_{xy}^{II,0}=0$, we would have only the first two terms, and 
it would state simply that the normal component of the
(bulk) current in eq.\ (\ref{jsigexpr}) tends to zero at the edge. These terms
originate {}from the edge term left after integrating the bulk part of the 
action by parts in order to take $\delta/\delta R$. In the presence of a 
non-zero $\sigma_{xy}^{II,0}$, this boundary condition is modified to include 
the last term, which is the covariant derivative (along the edge) of the edge 
part of the current in eq.\ (\ref{jsigexpr}). Thus eq.\ (\ref{wardidbc1}) is
equivalent to
\be 
j^\sigma_{y,{\rm bulk}}-D_xj^\sigma_{x,{\rm edge}}=0,
\label{wardidbccurr}
\ee
where the edge contribution is obtained as $j^\sigma_{x,{\rm
edge}}(x,W)=\int_{W-0^+}^{W+0^+}dy\, j^\sigma_x(x,y)$, and similarly for 
the edge at $y=0$, as in Appendix A. The edge term originates, of course, 
{}from taking $\delta/\delta R$ on the edge term in the action itself, 
eq.\ (\ref{S[A]}). In the boundary condition, the last two terms can be 
combined to leave
\begin{equation}
\sigma_{xx}^0 QD_yQ+\sigma_{xy}^0D_xQ=0.
\label{wardidbc2}
\end{equation}
This is the analogue of the tilted boundary condition discussed in Secs.\ I and
II, generalized to the full {\em nonlinear} field theory, and including 
$\bf A$; as in those discussions, only the total $\sigma_{xy}^0$ enters the
boundary condition. We will show, in Sec.\ IIIC below, that, in leading order 
in perturbation theory, this boundary condition reduces to exactly the one 
used in earlier sections.

To close this subsection, we obtain Ward identities that apply to moments of
the two-probe conductance, in which the currents are integrated across sections
parallel to the $y$ axis. We have already seen that these should be calculable
using only Green's functions at the Fermi energy, which can be obtained using
our generating function or the NL$\sigma$M. As in Sec.\ IIIA, we will therefore
here specialise $\bf A$ to be in the $x$ direction and independent of $y$.
Functional derivatives of $Z[{\bf A}]$ of the form $\delta/\delta A_x(x)$ then
produce the desired mean and variance of the two-probe conductance. If we
specialise to such $\bf A$ in the NL$\sigma$M action $S[{\bf A}]$ (noting that
$A_x(x,W)=A_x(x,0)=A_x(x,y)$ for all $x$, $y$), then we see that the $\sy^0$
terms can be combined, using an integration by parts, to leave
\be 
S[A_x]=\int d^2 r \,\left[-\frac{1}{8}\sx^0{\rm tr}\,(D_{\mu}QD_{\mu}Q)
-\frac{1}{8}\sy^0{\rm tr}\,(\epsilon_{\mu\nu}QD_{\mu}QD_{\nu} Q) 
+ \eta {\rm tr}\,(Q\Lambda)\right],
\label{S[A_x]}
\ee
in which, we emphasize, $\bf A$ takes the specialised form ${\bf A}({\bf r})
=(A_x(x),0)$, independent of $y$. This action is gauge invariant, and $A_x$ 
remains
$y$-independent, for gauge transformations $U$ that are independent of $y$.
Using such a transformation, we can obtain, similarly to the above derivation,
the identity
\be
D_x\int dy\,\langle\langle j_x^\sigma(x,y)\rangle\rangle=0,
\label{wardidsec}
\ee
in which $j_x^\sigma$ and the action $S[A_x]$ to be used in calculating the 
average still contain $A_x$. We note that, in $\int dy\, j_x^\sigma$, using
(\ref{jsigexpr}), $\sy^{I,0}$ and $\sy^{II,0}$ terms can be combined into a
single term, as implied by the action $S[A_x]$, in which the same is true. Thus
{}from this point on, only the total $\sy^0$ enters the calculations for the
conductance and its moments. 

On taking further functional derivatives $\delta/\delta A_x(x)$ of
(\ref{wardidsec}), we can obtain identities involving the mean and variance (or
alternatively, second moment) of the conductance. {}From the discussion in
earlier sections, we should have 
\be
\partial_x\langle g(x,x')\rangle =0,
\ee
for all $x$, $x'$ inside the sample, including $x=x'$, which expresses the 
independence of the conductance on the location of the cross-sections,  
and similar statements should hold for the higher moments and for the 
dependence on the other variables $x'$, \ldots. One may be concerned that $D_x$
appears in (\ref{wardidsec}), not $\partial_x$, and that this might lead to
additional terms containing e.g.\ $\delta(x-x')$ when further functional
derivatives are taken. However, we have verified that such terms vanish, 
to all orders in perturbation theory,
for the replica components we require to produce the mean and variance of 
the conductance, as in (\ref{g-sigma}) and (\ref{gg-sigma}). Thus, 
we have performed all the steps in a derivation
showing that the calculation of the two-probe conductance and its variance, 
in the rectangular geometry, can be carried out within the NL$\sigma$M, 
including the independence of the locations of the cross-sections, which 
allows the use of an average over these locations. The explicit expressions 
for the mean and variance are given in Secs.\ IV and V, and evaluated to 
leading order in the perturbation expansion.

\subsection{Perturbation expansion, current conservation, and bilocal
conductivity}

The approach given in the previous subsection is sufficient for the two-probe 
conductance, provided the cross-sections used are parallel to the $y$ axis. 
For more general cross-sections, and to consider the non-Fermi-energy effects 
and the bilocal conductivity, a deeper analysis is required, which is 
contained in the present subsection, but which may be skipped (apart {}from 
the first part introducing the perturbation expansion) by readers interested 
only in the two-probe conductance.

Our goal here is to compare the consequences of the definition of the current
and the Ward identities with the properties of the currents in our model of 
the original electron system. To provide motivation, we will compare results in
the NL$\sigma$M formulation with those in the SCBA in the previous section, and
for this purpose we will now introduce the perturbation expansion of the model.
We then show how to modify the coupling of $\bf A$ to the NL$\sigma$M so as to
reproduce the properties found in Sec.\ II. 

The matrix $Q$, which obeys $Q^{\dagger}=Q$, $Q^2=I_{2n}$, can be parametrized 
in the following way:
\be
Q=\left(\ba{cc} \sqrt{1-z\zd}& z\\
                   \zd&-\sqrt{1-\zd z}\ea \right),
\ee
where $z$ is an $n\times n$ complex matrix.
Expanding the NL$\sigma$M action in terms of the $z$-matrix, we get
\be
S[{\bf A}]=S_0[{\bf A}] +S_1[{\bf A}].
\ee
where $S_0[{\bf A}]$ is the part quadratic in $z$, $\zd$ and $\bf A$, which is 
the same as in eq.~(\ref{S_0}), except that there are now $n^2$ copies of $z$, 
and we include the gauge potential $\bf A$. For $\bf A$ of the restricted form 
\begin{equation}
{\bf A}=\left(\begin{array}{cc}0&{\bf A}^{+-}\\
                               {\bf A}^{-+}&0\end{array}\right),
\end{equation}
where ${\bf A}^{+-}$ is a complex $n\times n$-matrix--valued vector
field, and ${\bf A}^{-+}$ is its adjoint (these are the only components of
$\bf A$ that will be used below), we have
\begin{eqnarray}
S_0[{\bf A}]&=&-\frac{\sx^0}{4} \int d^2 r\,{\rm tr}\,\left((\partial_{\mu}
z-2iA_\mu^{+-}) (\partial_{\mu} \zd +2iA_\mu^{-+})\right) \non\\
&&\mbox{}-\frac{\sy^{I,0}}{4} \int d^2 r\,\epsilon_{\mu\nu}
{\rm tr}\,\left((\partial_{\mu} z-2iA_\mu^{+-}) 
(\partial_{\nu}\zd+2iA_\nu^{-+})\right) \non\\
&&\mbox{}-\frac{\sy^{II,0}}{4} \int d^2 r\,\epsilon_{\mu\nu}
{\rm tr}\,(\partial_{\mu} z \partial_{\nu}\zd) 
-\frac{i\sy^{II,0}}{2}\oint dl_\mu\,{\rm tr}\,(A_\mu^{+-}\zd-zA_\mu^{-+}). 
\label{S_0[A]}
\end{eqnarray}
The line integral $\oint d\l_\mu$ is taken in the counterclockwise direction
around the edge of the sample.
Here and below, we use the symbol $\rm tr$ for a trace on the $n$-dimensional 
space, as well as for that on the $2n$-dimensional one; it should be clear 
{}from the context which is meant. $S_1$ describes the interaction between the 
diffusion modes caused by quantum interference effects. We give it here only 
for ${\bf A}=0$, and to the order $O[(z\zd)^3]$ required for our later 
calculations, 
\bea
S_1[0]&=&\int d^2 r\,\left\{
 -\frac{\sx^0}{32} {\rm tr} [\partial_{\mu} (z\zd)\partial_{\mu} (z\zd)+
\partial_{\mu} (\zd z)\partial_{\mu} (\zd z)]\right. \non\\
& &-\frac{\sy^0}{32} {\rm tr} [\epsilon_{\mu\nu}\partial_{\mu} (z\zd z 
\partial_{\nu}\zd )-\epsilon_{\mu\nu}\partial_{\mu}(\zd z 
\zd \partial_{\nu} z) ]\non\\ 
& &-\frac{\sx^0}{64}  {\rm tr} [\partial_{\mu} (z\zd)\partial_{\mu} (z\zd 
z\zd)+
\partial_{\mu} (\zd z)\partial_{\mu} (\zd z\zd z)]\non\\
& &\left.-\frac{\sy^0}{64} {\rm tr} [\epsilon_{\mu\nu}\partial_{\mu} (z\zd   
z\zd  z
\partial_{\nu}\zd )-\epsilon_{\mu\nu}
\partial_{\mu}(\zd z\zd z\zd \partial_{\nu}z)]+ 
O[(z\zd)^4] \right\}.
\eea
Notice that terms proportional to $\sy^0$ can all be written as total
derivatives, therefore they can be expressed as boundary terms.
To calculate the ensemble average of any quantity $X[z,z^\dagger]$, 
we perform the following expansion:
\be
\langle\langle X  \rangle\rangle  =\lim_{n\rightarrow 0}
  \int D[z,\zd]\, I[z,\zd] X[z,\zd]
e^{S_0\{z,\zd,0\}}\sum_{m=0}^{\infty}\frac{1}{m!}S_1^m[z,\zd].
\ee
Here $I[z,\zd]$ is the Jacobian needed to make the measure in
the $z$, $\zd$ space invariant under a $U(2n)$ rotation at each $\br$. 
The explicit form of this Jacobian will not be needed. Its only role is to
cancel quadratically-divergent diagrams that arise in perturbation theory, in a
manner that is standard for all NL$\sigma$M's (see, e.g., Ref.\ \cite{Amit}).
The terms in the expansion can be written in terms of averages calculated using
the quadratic action with ${\bf A}=0$, defined by 
\begin{equation} 
\langle\langle\cdots\rangle\rangle_0 = \int D[z,\zd]\,\cdots\, 
e^{S_0[{\bf A}=0]}/Z_0[0],
\end{equation}
in which the functional integral in the denominator is the same as the
numerator but with the insertion $\cdots$ omitted.
The expressions can be evaluated by contracting pairs of $z$ and 
$\zd$, which gives the diffusion propagator, 
\begin{equation}
\frac{\sx^0}{4}\langle\langle \zd_{ij}(\br) z_{kl}(\brp)\rangle\rangle_0
=\delta_{il}\delta_{jk}d(\br,\brp),
\label{diffprop}
\end{equation}
which obeys the same conditions (\ref{propagator}) as in earlier sections.
The basic perturbation expansion is now a series in powers of
$1/\sigma_{xx}^0$, though it will also be convenient to expand in powers of
$\gamma=\sigma_{xy}^0/\sigma_{xx}^0$, to obtain a double expansion.

We now return to the physical meaning of the Ward identities (\ref{wardid}), 
(\ref{wardidbc2}), that resulted {}from the gauge invariance of the action
$S[{\bf A}]$. We wish to compare these with our physical expectation that the
current is divergenceless inside the sample, and that no current flows in or
out of the sample at the reflecting walls (as we have shown in Sec.\ II, and
discussed in Sec.\ I, the current response obtained in the SCBA is not
divergenceless, because of the bulk $\sy^{II,0}$ term, but that is a
non-Fermi-energy effect that will not be considered in the present formalism
until later in this subsection). The first difficulty that seems to arise 
(as with eq.\ (\ref{wardidsec})) is
that the Ward identity states ${\bf D}\cdot{\bf j}^\sigma=0$, not
$\nabla\cdot{\bf j}=0$, as we might have expected. The vector potential $\bf A$
present in $\bf D$ will generate $\delta$-function terms when further
functional derivatives are taken to obtain Ward identities, as must be done 
for the bilocal conductivity tensor and analogous correlators of more than two
currents. However, as we mentioned in connection with the conductance in the
previous subsection, in practise, for the particular components of $\bf A$ that
yield the physically-relevant conductivities, this does not seem to occur. For
example, in addition to the results cited in the previous subsection, we can
show that the Ward identity implies
\be
\partial_\mu\langle\sigma_{\mu\nu}(\br,\brp)\rangle=0
\ee
for the mean bilocal conductivity tensor calculated in the NL$\sigma$M using
the action $S[{\bf A}]$, and this is valid for all $\br$, $\brp$ inside the
sample, including $\br=\brp$, to all orders in perturbation theory.

The boundary condition (\ref{wardidbc1}) also involves the tangential 
covariant derivative of the edge current, not the usual partial derivative as
one would want in the electronic system. In this case, we do find a clash
between the theory as formulated and our intuition. Eq.\ (\ref{wardidbccurr}) 
is more explicitly 
\be 
j^\sigma_{y,{\rm bulk}}-\partial_xj^\sigma_{x,{\rm edge}}
    =-\frac{1}{2}\sy^{II,0}[A_x,Q].
\label{wardidbccurr2}
\ee
The left-hand side is the combination one might have expected to be zero.
However, it is nonzero when $A_x$ is nonzero, implying that conservation is
violated by $\delta$-function terms on the edge in the bilocal conductivity 
and its moments. Note that similar commutators $[A_x,Q]$ and $[A_y,Q]$ appear in
$j^\sigma_{y,{\rm bulk}}$ but are not a problem. 

To illuminate the point further, we can use the perturbation expansion and
compute the mean bilocal conductivity within the NL$\sigma$M as formulated so
far. {}From $S_0[{\bf A}]$, the equation of motion for $\zd$ (formulas for $z$
are similar) is, in the bulk
\be
\sx^0 \nabla^2\zd=-2i\left(\sx^0\partial_\mu
A_\mu^{-+}+\sy^{I,0}\epsilon_{\mu\nu}\partial_\mu A_\nu^{-+}\right),
\ee
(as in Sec.\ IA, where, however, $\sy^0=\sy^{I,0}$ and $\sy^{II,0}=0$), 
and at the edge is the tilted boundary condition 
\be
\sx^0 (\partial_y \zd +2iA_y^{-+})-\sy^0(\partial_x \zd+2iA_x^{-+})=0.
\ee
These equations give the generalization of the ``classical'' theory of Sec.\ IA
to include the edge currents with coefficient $\sy^{II,0}$, on using the
identifications given in Sec.\ ID. The currents are given, in the
present approximation, by
\bea
\delta S_0/\delta A_\mu^{+-}\equiv j_\mu^{-+}&=&\frac{1}{2}i[\sx^0
(\partial_\mu\zd+2iA_\mu^{-+})+\sy^{I,0}\epsilon_{\mu\nu}(\partial_\nu\zd
+2iA_\nu^{-+}) \non\\
&&\mbox{}+\sy^{II,0}(\delta(y-W)-\delta(y))\delta_{\mu x}\zd].
\label{j-+}
\eea
The bulk equation of motion can therefore be written $\partial_\mu
j_\mu^{-+}=0$, while at the edge we have 
\be
j_{y,{\rm bulk}}^{-+}-\partial_x 
j_{x,{\rm edge}}^{-+}=-\sy^{II,0}A_x^{-+},
\label{wardidbccurr3}
\ee
which can also be written
\be
j_{y,{\rm\,bulk}}^{-+}-D_x j_{x,{\rm\,edge}}^{-+}=0,
\ee
where the covariant derivative is the linearized version of that in the full
NL$\sigma$M, namely $D_\mu \zd\equiv\partial_\mu \zd+2iA_\mu^{-+}$, 
while $D_\mu D_\nu \zd\equiv \partial_\mu D_\nu \zd$. Thus, for current 
$j^{-+}$ as defined here, current conservation in the naive form is violated 
by $\delta$-function terms at the edge. This can be rectified, but before 
doing so, we calculate the bilocal conductivity of the present model in the 
same approximation.

The mean bilocal conductivity in the present approximation is obtained as 
(compare eq.\ (\ref{g-sigma}); 
we leave implicit the choice of all replica components equal to 1, and the
$n\rightarrow0$ limit)
\bea
\langle\sigma_{\mu\nu}(\br,\brp)\rangle_0&\equiv&\mbox{}-\lim_{{\bf A}
\rightarrow0}\frac{\delta}{\delta A_\nu^{-+}(\brp)}\langle\langle
j_\mu^{-+}(\br)\rangle\rangle_0\non\\
&=&(\sx^0\delta_{\mu\nu}+\sy^{I,0}\epsilon_{\mu\nu})\delta(\br-\brp)
-\langle\langle j_\mu^{-+}(\br)j_\nu^{+-}(\brp)\rangle\rangle_0.
\eea
On evaluating this using eq.\ (\ref{j-+}) (with ${\bf A}=0$), the similar 
formula for
$j^{+-}$, and the definition (\ref{diffprop}) of $d$, we obtain the same result
as in eq.\ (\ref{smunuSCBA}), except that the bulk $\sy^{II,0}$ term is not
present. This result obeys $\partial_\mu\langle\sigma_{\mu\nu}(\br,\brp)
\rangle_0=0$ for all $\br$, $\brp$ in the bulk, and eq.\ (\ref{wardidbccurr3})
implies that 
\be
\langle\sigma_{y\nu}(\br,\brp)\rangle_0-\partial_x\int_{W-0^+}^{W+0^+}dy\,  
\langle\sigma_{x\nu}(\br,\brp)\rangle_0=\sy^{II,0}\delta_{\nu x}\delta(\brp-\br)
\ee
for $\br$ at the upper edge $y=W$ (and similarly for the lower).
In effect, in the edge channel,
$A_x^{-+}$ simply creates current, so the naive conservation law is violated.

In the SCBA, we did not directly address this issue, but derived
$\partial_xj_{x,{\rm edge}}^{\rm SCBA}-j_{y,{\rm bulk}}^{\rm SCBA}=0$ 
for $\br\neq \brp$ only. 
On the other hand, in the SCBA we also found a non-Fermi-energy contribution
$\sy^{II,0}\epsilon_{\mu\nu}\delta(\br-\brp)$ in the bulk, which means that in
the presence of $\bf E$, there is an additional part
$(e^2/h)\sy^{II,0}\epsilon_{\mu\nu}E_\nu$ in the bulk current
\be
j^{\rm SCBA}_\mu=j^{\rm SCBA}_{\mu,E=E_F} +\frac{e^2}{h}\sy^{II,0}
\epsilon_{\mu\nu}E_\nu,
\ee
within the SCBA. If we introduce a corresponding change in the bulk current
here, so that
\be 
j_{\mu,{\rm mod}}^{-+}=j_\mu^{-+}-\sy^{II,0}\epsilon_{\mu\nu}A_\nu^{-+},
\label{jmumod}
\ee
then the tilted boundary condition (\ref{wardidbccurr3}) becomes
\be
j_{y,{\rm bulk,mod}}^{-+}-\partial_x j_{x,{\rm edge,mod}}^{-+}=0
\ee
in the presence of $\bf A$. Thus the modified current is conserved (not
covariantly) at the edge (and so is $j^{\rm SCBA}_\mu$); 
the current in the edge channel comes {}from the
bulk. Physically, there are two modes of conduction response to an electric
field in the system. One is the ``sliding'' of the total charge density, which
gives the bulk Hall conductivity $\sy^{II,0}$. This is a non-Fermi-energy
effect, and is a local ($\delta$-function) response to an electric field. The
other is the Fermi-energy response, which is diffusive in the bulk (including
the Hall effect with coefficient $\sy^{I,0}$) and is chiral along
the edge. As discussed in Sec.\ I, the $\sy^{II,0}$ bulk effect implies
$\nabla\cdot{\bf j}\neq 0$, meaning that $\partial\rho/\partial t \neq 0$. At 
the edge, there is no charge accumulated. A tangential electric field at the 
edge can produce a bulk current normal to the edge, and also a Fermi-energy 
edge current that increases along the edge. These effects involve the same 
coefficent $\sy^{II,0}$, and the result
is that no current is created, so no charge accumulates at the edge. This occurs
because of a version of the Laughlin-Halperin gauge-invariance argument
\cite{Laughlin,Halperin}. A
change in the potential (which is essentially what $z$ is) would accumulate a
charge density of order the inverse velocity of the edge states, but the same
velocity also appears in the edge current, which carries away the charge.

We now propose a modification of the NL$\sigma$M action which incorporates
this non-Fermi-energy effect so as to recover the SCBA bilocal conductivity
tensor in full, and maintain current conservation at the edge (though not in
the bulk), to all orders in perturbation theory. Our proposed action (in which
we reinstate the full $2n\times 2n$ matrix $\bf A$) is
\bea 
S_{\rm mod}[{\bf A}]&=&\int d^2 r \,\left\{-\frac{1}{8}\sx^0 {\rm tr} (D_{\mu}
QD_{\mu} Q)
-\frac{1}{8}\sy^{I,0} {\rm tr} (\epsilon_{\mu\nu} Q
D_{\mu}QD_{\nu} Q) \right.\non\\
&&\left.\mbox{}-\frac{1}{8}\sy^{II,0} {\rm tr} (\epsilon_{\mu\nu} Q
\partial_{\mu}Q\partial_{\nu} Q) 
+\frac{1}{8}\sy^{II,0} {\rm tr} (\epsilon_{\mu\nu} Q
[A_{\mu},Q][A_{\nu},Q]) 
+ \eta {\rm tr } (Q\Lambda) \right\}\non\\
&&\mbox{}+\frac{i}{2}\sy^{II,0}\int dx\, \left\{{\rm tr}(A_x(x,W)Q(x,W))-{\rm
tr}
(A_x(x,0)Q(x,0))\right\}.
\label{Smod}
\eea
The added $\sy^{II,0}$ term maintains SU($2n$) global, but not local gauge, 
symmetry,
corresponding to the nonconservation of the corresponding modified current
which contains an additional term in the bulk:
\be
j^\sigma_{\mu,{\rm mod}}=\frac{\delta S_{\rm mod}}{\delta A_\mu}
=j^\sigma_\mu-\frac{1}{2}\sy^{II,0}\epsilon_{\mu\nu}[A_\nu,Q].
\ee
The modified equation of motion must be obtained {}from 
$\delta S_{\rm mod}\{Q',{\bf A}\}/\delta R=0$ (as in Sec.\ IIIB) without using
gauge invariance. It can be written in terms of the modified current, 
to give the modified Ward identity in the bulk,
\be
D_\mu j^\sigma_{\mu,{\rm mod}}=-\frac{1}{2}\sy^{II,0}\epsilon_{\mu\nu}
\left\{\partial_\mu\left([A_\nu,Q]\right)+i[A_\mu,[A_\nu,Q]]\right\}.
\ee
The modified current is not covariantly conserved, because the modified action 
is not gauge invariant. However, the boundary condition eq.\ (\ref{wardidbc2}) 
is unchanged, because the added term in $S_{\rm mod}$ contains no derivatives, 
so does not give rise to any boundary terms. Nonetheless, the 
{\em interpretation} of the boundary condition changes, because the current 
has been modified. In terms of the modified current, the boundary condition 
states that at the edge
\be
j^\sigma_{y,{\rm mod}}-\partial_x j^\sigma_{x,{\rm edge,mod}}=0,
\ee
which is ``current conservation''. Strictly, our arguments imply that this
modification of the action applies only for the $+-$ and $-+$ components of 
${\bf A}$; for the other components, the correct form may depend on what is 
assumed in the underlying model the NL$\sigma$M is supposed to represent. 

For the quadratic part $S_{0,{\rm mod}}$ of the modified action $S_{\rm mod}$,
the corresponding formulas are: for the action, 
\be
S_{0,{\rm mod}}=S_0-\sy^{II,0}\int d^2r\,\epsilon_{\mu\nu}{\rm
tr}\,(A^{+-}_\mu A^{-+}_\nu);
\ee
for the current,
\be
j_{\mu,{\rm mod}}^{-+}=j_\mu^{-+}-\sy^{II,0}\epsilon_{\mu\nu}A_\nu^{-+},
\ee
as in eq.\ (\ref{jmumod}) above; for the modified equation of motion,
\be
\partial_\mu j_{\mu,{\rm mod}}^{-+}=-\sy^{II,0}\epsilon_{\mu\nu}\partial_\mu 
A^{-+}_\nu,
\ee
which corresponds to the earlier eq.\ (\ref{divjnonzero}); and in the bilocal
conductivity tensor, the bulk contact term
$\sy^{II,0}\epsilon_{\mu\nu}\delta(\br-\brp)$ appears as in the full SCBA
result, eq.\ (\ref{smunuSCBA}). The boundary conditions on $z$, $\zd$  are, 
however, unmodified.

When using the modified action for calculations of conductance, there is no 
change to the results, as long as one uses cross-sections that are parallel 
to the $y$-axis, and therefore the expressions contain the integrals of the 
$x$-components of the currents, as we were doing earlier. For those
calculations, we already showed at the end of Sec.\ IIIB that the conductance is
independent of the positions of the sections. Since the vector potentials used
there have $x$-components only, the extra term in $S_{\rm mod}$ is zero. In
addition to the usefulness of the general $S_{\rm mod}$ for reproducing the
bilocal conductivity tensor and maintaining the current-conservation properties
at the edge to all orders in perturbation theory, it is also crucial for the
conductance if one calculates the flux of current through more general
cross-sections than those specified above. In general, a cross-section could
be any curve that intersects the edges of the sample just twice, once on each
of the reflecting walls. Two such sections may intersect at
isolated points (instead of along their whole length), and the intersections 
are then said to be {\em transversal}, that is, the normals to the curves 
at the point of 
intersection are not parallel (nor antiparallel). The question arises whether 
the conductance and its moments calculated using such sections is independent 
of the position and shape of the sections, including the case in which they
intersect. When they intersect transversally, the bulk $\sy^{II,0}$ contact
term will contribute at the intersection point, even within the SCBA. Our 
preliminary investigations of this, which will not be included here, show that 
these contributions are needed to cancel effects of the $\sy^{II,0}$ terms at 
the edge, so as to maintain conservation of the total current, and that the
conductance obtained is the same as for the straight sections, for any shape
and position. We believe this to be true in general, to all orders in
perturbation theory. This shows that the use of the modified action is 
obligatory for such general calculations. Note that, for more general 
geometries, such as four probes, such intersecting sections will be common.

\subsection{Further details of perturbative calculations}

\subsubsection{Boundary perturbation expansion for the diffusion 
propagator}  

It is difficult to calculate the diffusion propagator 
as defined by eq.~(\ref{propagator})
explicitly (however, see reference \cite{Girvin}), although 
propagators for simpler geometry such as a half-plane \cite{Callan}, 
an infinite strip \cite{Callan,Yosef} or an annulus \cite{Callan} can be found 
(the results for the infinite strip can be
obtained by conformal mapping {}from the half-plane \cite{Callan}). 
We can  perform a boundary perturbation expansion \cite{Math} in powers of 
$\gamma$ using the propagator at $\gamma=0$, which we define as $d^0(\br,\brp)$.
$d^0(\br,\brp)$ can be constructed out of the solutions for the following 
eigenvalue problem:
\be
- \nabla^2 \phi^0=\Lambda \phi^0
\ee
with the  boundary conditions
$
\partial_y \phi^0(x,W)=\partial_y \phi^0(x,0)=0
\;\mbox{and}\;\phi^0(0,y)=\phi^0(L,y)=0.
$
The eigenfunctions are
$
\phi_{nm}^0(x,y)=\frac{2}{\sqrt{LW}}\cos(n\pi y/W)\sin(m\pi x/L)
$
with corresponding eigenvalues
$\Lambda_{nm}^0=(n\pi/W)^2+(m\pi/L)^2$, where
$n=0$, $1$, $2$, ...,
$m=1$, $2$, \ldots . We have
\be
d^0(\br,\br')=\sum_{n=0}^{\infty}\sum_{m=1}^{\infty}\frac{\phi_{
nm}^0(x,
   y)\phi^
0_{nm}(x',y')}{\Lambda^0_{nm}}.
\label{d^0}
\ee 
Using the bulk diffusion equations for $d^0(\br,\brp)$ and 
$d(\br,\brp)$, 
we get
\be
d(\br_2,\br_1)=d^0(\br_1,\br_2)+ \int_C d{\bf S}\cdot
\left[d^0(\br,\br_2){\bf
\nabla}d(\br,\br_1)-d(\br,\br_1){\bf
\nabla}d^0(\br,\br_2)\right],
\ee
where $C$ is a closed surface enclosing the disordered region 
including the edges (see figure ~\ref{two probe geometry}). 
Let us divide the surface $C$ into four parts $C_1$,
$C_2$, $C_3$ and $C_4$. Applying the
boundary conditions for the propagators for different sections, we 
have the following:
$$
d(\br_2,\br_1)=d^0(\br_1,\br_2)+\int_{C_1+C_3}  dS_n\,
d^0(\br,\br_2)
\partial_n d(\br,\br_1).
$$
Plugging in the boundary condition $\partial_n d=\gamma 
\partial_t d$, we have
\be
d(\br_2,\br_1)=d^0(\br_1,\br_2)+\gamma \int_{C_1+C_3}  dS_n\,
d^0(\br,\br_2)
\partial_t d(\br,\br_1).
\ee
The above equation generates an expansion in terms of $d^0$ and in 
powers of $\gamma$. Using ${\cal B}$ to denote the operation 
$\int_{C_1+C_3}dS_n\,\gamma \partial_t$, we can write schematically:
\be
 d=d^0+d^0 {\cal B }d^0 +d^0{\cal B} d^0 {\cal B} d^0 +\ldots. 
\label{PGT EXP}
\ee

\subsubsection{The 1D propagator}
 
The above expansion is not valid for extremely narrow samples with $W\ll L$, 
where equation~(\ref{PGT EXP}) needs to be summed exactly. 
However, the propagator itself approaches the 1D diffusion propagator with 
conductivity $\sx^0(1+\gamma^2)W$. Such limiting behavior can be 
demonstrated by rewriting the quadratic action in the following way:
\be
S_0(z,\zd)=-\frac{\sx^0}{4}\int d^2 r\, 
\left\{
(1+\gamma^2){\rm tr}[\partial_x z\partial_x \zd]+ 
{\rm tr}[(\partial_y z+\gamma\partial_x z) (\partial_y \zd 
-\gamma\partial_x \zd)] 
\right\}.   
\ee
In the limit $W\ll L$, the main contribution to $\langle\langle\zd z 
\rangle\rangle_0$ comes {}from the low-lying eigenmodes of $-\nabla^2$ 
which satisfy 
$(\partial_y +\gamma\partial_x)\phi^L (\br)=0$, $
(\partial_y  -\gamma\partial_x)\phi^R(\br)=0$ 
in the entire strip. The eigenvalues of these modes
are  separated {}from those of the other modes by a
gap of order $L^2/W^2$ \cite{Yosef}. In other words, the  term 
$\int d^2 r\, {\rm tr}[(\partial_y z+\gamma\partial_x z) 
(\partial_y \zd -\gamma\partial_x \zd)]$ 
in the action can be ignored in the $1D$ limit:
\bea
\lim_{W/L \rightarrow  0}\int D[z,\zd]\, \zd_{ij} z_{kl} e^{S_0}&=&  
\int D [z,\zd]\, \zd_{ij} z_{kl} e^{\textstyle -\frac{\sx^0}{4}
(1+\gamma^2)W \int dx\, {\rm tr}[\partial_x z \partial_x \zd]}\non\\
&=&4\delta_{il}\delta_{jk}d^{1D}/\sigma_{xx}^0W,
\eea
where 
\be
d^{1D}(x,x')=\frac{4}{(1+\gamma^2)L}\sum_{m=1}^{\infty}
\frac{\sin(m\pi x/L)\sin(m\pi x'/L)}{(m\pi/L)^2}.
\label{d1D}
\ee
More generally, by symmetry, we obtain the NL$\sigma$M in
one dimension (where no topological term is possible) with 
$\sx^{0,1D}=\sx^0(1+\gamma^2)W=W/\rho_{xx}^0$ as the coefficient 
in the action.

\section{Two-probe Conductance}

\subsection{The boundary contribution}

{}From the nonlinear $\sigma$-model, the average conductance is
\be
\langle g \rangle =-\frac{1}{L^2}
\lim_{{\bf A}\rightarrow 0, n\rightarrow 0}
\int_0^L dx_1 \int_0^L dx'_1\, 
\langle \langle \frac{\dl S}{\dl A^{+-}_{x,11}(x_1)}
\frac{\dl S}{\dl A^{-+}_{x,11}(x'_1)}+ \frac{\dl^2 
S}{\dl A^{+-}_{x,11}(x_1) 
\dl A^{-+}_{x,11}(x_1')} \rangle\rangle,
\ee
where
\bea
\left. \frac{\dl S} 
{\dl A^{+-}_{x,11}(x_1)}\right|_{{\bf A}=0}&=&
\int_0^W dy_1\, \left[\frac{i\sx^0}{2}(\partial_x \zd_{11} 
 +\gamma \partial_y 
\zd_{11})+\frac{i\sx^0}{4} (\zd \partial_xz \zd)_{11}+O(z^5)\right],\non\\
\frac{\dl^2 S}{ 
\dl A^{+-}_{x,11}(x_1) \dl A^{-+}_{x,11}(x_1')}&=
&-\int_0^W dy_1\,\sx^0\dl(x_1-x_1')[1-
\frac{1}{4}(z\zd)_{11}
-\frac{1}{4}(\zd z)_{11} -
\frac{1}{16}(z\zd z\zd )_{11}\non\\
&&
 -\frac{1}{16}(\zd z\zd z)_{11}+\frac{1}{8}(z\zd)_{11}(\zd
z)_{11}+O(z^6)].
\eea
To leading order in $1/\sx^0$, we get 
\bea
g^0&=&\frac{1}{L^2}\int_A d^2 r\int_A d^2 r'\,\sx^0[\dl(\br-\brp)-
\frac{\sx^0}{4}(\partial_x+\gamma\partial_y)(\partial'_{x}-\gamma\partial'_{y})
\langle\langle\zd_{11}(\br)z_{11}(\brp)\rangle\rangle_0]
\non\\
&=&\frac{1}{L^2}\int d^2 r \int d^2 r'\, \sx^0 \left[\dl(\br-\brp)
-(\partial_x+\gamma \partial_y)(\partial'_{x}-\gamma 
\partial'_{y})d(\br,\brp)\right].
\eea
We have thus recovered the result {}from the diagrammatic 
expansion, and the other approaches described in Sec.\ I\@. 
Therefore the following results apply to any of these approaches.

The long-ranged term in the above expression comes {}from the ladder diagrams
(i.e. diffusion) and in the absence of magnetic field, it 
does not contribute to the two-probe conductance \cite{Kane}. 
In the presence of magnetic field, this is no longer the
case. One can show that the local term gives the Ohmic conductance $\sx^0 W/L$.
The long-ranged term, which involves volume integrals of total derivatives, 
gives the difference of boundary values of the diffusion propagator. Since the 
diffusion propagator goes to zero in the leads, the $\partial_x $ and 
$\partial'_x$ terms vanish upon volume integral.
We are left with the boundary difference at the upper and lower edges,
\bea
g^0&=&\sx^0\frac{W}{L}+\sx^0 \gamma^2 \frac{1}{L^2}
\int_0^L dx \int_0^L
dx'\,
\left[d(x,W;x',W)+
d(x,0;x',0)\right.\non\\
& &\left. -d(x,0;x',W)-d(x,W;x',0)\right]. 
\label{g(L)}
\eea
The $\sigma_{xy}^0$-dependent part is expressed as a boundary term 
and vanishes when the magnetic field is zero ($\gamma=0$), or if
the system is subject to periodic boundary condition 
in the transverse direction.

At one-loop level, which is the next order in $1/\sigma_{xx}^0$, 
we have verified that the interference correction to $\langle g\rangle$ 
vanishes in the limit $n\rightarrow 0$. Therefore the presence of 
edges does not change the conclusion of the previous perturbative calculations 
\cite{Brezin,Hikami,Chalker2} that there is no weak localization
correction to $\sx$ of relative order $1/\sx^0$. In general, we do not expect 
the presence of edges to have any effect on the renormalization flow of 
$\sx$ in the perturbative regime, since it is dominated by short distance
effects in the bulk. Whether $\sigma_{xy}$ is ever renormalized perturbatively 
when the system has edges is less clear to us.

\subsection{The small $\gamma$ correction for the two-probe 
conductance}

For small $\gamma$, we can make use of the propagator $d^0$ at $\gamma=0$ 
to obtain the leading correction to $g^0$. Plugging $d^0(\br,\brp)$ in the 
boundary term of the two-probe conductance, we get
\be
g^0(\gamma)
=\sx^0 W/L[1 +\gamma^2 f_1(L/W)]+O(\gamma^4),
\label{dg-ga}
\ee     
where
$$
f_1(L/W)=\frac{64}{\pi^4}\frac{L^2}{W^2}\sum_{m=1,odd}^\infty
\sum_{n=1,odd}^\infty\frac{1}{m^2}\frac{1}{m^2+\frac{L^2}{W^2}n^2}
\rightarrow \left\{\ba{ccc}
14\zeta(3)L/\pi^3W,&W\gg L;\\
1/2,&W=L;\\
1,&W\ll L.
\ea \right.   
$$
(Here $\zeta(s)=\sum_{m=1}^\infty m^{-s}$ is the Riemann zeta function, and we 
note that $\zeta(3)\simeq 1.202$.) Thus, for an extremely wide sample, 
$W\gg L$, the effect of the edges can be ignored.
In the 1D limit ($W/L \rightarrow 0$), we get $g^0\rightarrow(1+\gamma^2)
\sx^0 W/L$, which is consistent with our result that in the NL$\sigma$M,
$\sx^{0,1D}=(1+\gamma^2)\sx^0 W$. This is also consistent with other results,
as noted in Sec.\ IA, valid for arbitrary $\gamma$, which show that the 
mean conductance in the 1D limit can be obtained {}from this one-dimensional 
conductivity $\sx^{0,1D}=W/\rho_{xx}^0$.

\section{Variance of the Conductance}

In this section, we evaluate the variance of conductance to leading 
order in $1/\sx^0$. {}From the nonlinear $\sigma$-model, we get
\bea
\langle g^2\rangle-\langle g\rangle ^2  &=& \frac{1}{L^4}
\lim_{{\bf A}\rightarrow 0, 
n\rightarrow 0}
\int dx_1 \int dx'_1 \int dx_2 \int dx'_2  \non\\
&&\langle \langle 
\frac{\dl^2 S}{\dl A^{+-}_{x,11}(x_1) \dl A^{-+}_{x,11}(x_1')}  
\frac{\dl^2 S}{\dl A^{+-}_{x,22}(x_2) \dl A^{-+}_{x,22}(x_2')}  
\non\\
&& +\frac{\dl^2 S}{\dl A^{+-}_{x,11}(x_1) \dl A^{-+}_{x,22}(x_2')} 
\frac{\dl^2 S}{\dl A^{+-}_{x,22}(x_2) \dl A^{-+}_{x,11}(x_1')}\non\\    
&& +\frac{\dl^2 S}{\dl A^{+-}_{x,11}(x_1) \dl A^{+-}_{x,22}(x_2)} 
\frac{\dl^2 S}{\dl A^{-+}_{x,11}(x_1') \dl A^{-+}_{x,22}(x_2')}\non\\
&& +\frac{\dl^2 S}{\dl A^{+-}_{x,11}(x_1) \dl A^{-+}_{x,11}(x_1')}
\frac{\dl S}{\dl A^{+-}_{x,22}(x_2) }
\frac{\dl S}{\dl A^{-+}_{x,22}(x'_2)}+ 1\leftrightarrow 2 \non\\   
&&  +\frac{\dl^2 S}{\dl A^{+-}_{x,11}(x_1) \dl A^{-+}_{x,22}(x_2')}
\frac{\dl S}{\dl A^{+-}_{x,22}(x_2)}
\frac{\dl S}{\dl A^{-+}_{x,11}(x'_1)}+ 1\leftrightarrow 2   \non\\
&& +\frac{\dl^2 S}{\dl A^{+-}_{x,11}(x_1) \dl A^{+-}_{x,22}(x_2)}
\frac{\dl S}{\dl A^{-+}_{x,22}(x'_2)}
\frac{\dl S}{\dl A^{-+}_{x,11}(x'_1)}+ (+\leftrightarrow-)\non\\
&&+\frac{\dl S}{\dl A^{+-}_{x,11}(x_1)}\frac{\dl S}
{\dl A^{-+}_{x,11}(x'_1)}
\frac{\dl S}{\dl A^{+-}_{x,22}(x_2)}
\frac{\dl S}{\dl A^{-+}_{x,22}(x'_2)}\rangle\rangle_{\rm connected}.
\eea
where
\bea
\frac{ \dl^2 S}{ 
\dl A^{+-}_{x,11}(x_1) \dl A^{-+}_{x,22}(x_2')}&=&
-\int dy_1\, \dl(x_1-x'_2)
\frac{\sx^0}{8}(\zd
z)_{12}(z\zd)_{21}+O(z^6),  \non\\        
\frac{ \dl^2 S}{\dl A^{+-}_{x,11}(x_1) \dl A^{+-}_{x,22}(x_2)}&=&
\int_0^W dy_1\, \dl(x_1-
x_2)\frac{\sx^0}{2}\zd_{12}\zd_{21},
\eea 
and the other functional derivatives are all evaluated at ${\bf A}=0$.
The leading diagrams are shown in Figure~\ref{diagrams2}. 
Various vertices are denoted by polygons with the wavy tail indicating
$\frac{\textstyle \dl}{\textstyle \dl A}$, while the 
lines linking the vertices are the diffusion propagators. The diamond 
shaped vertex with no wavy tails comes {}from the 4-point interaction term in 
$S_1$. We have also obtained the same set of diagrams using the 
diagrammatic approach. The diagrammatic approach is complicated because 
various vertices need to be evaluated separatedly. In the presence of 
magnetic field, the vertices are dressed with non-vanishing $G^+ G^+$ and 
$G^-G^-$ ladders, although in the end they can all be expressed in terms of 
$\sx^0$ and $\sy^0$. For the NL$\sigma$M, the vertices can be obtained 
{}from the action. Figure~\ref{diagrams3} shows that one particular 
vertex {}from $\frac{\textstyle \dl^2 S}{\textstyle 
\delta A^{+-}_{x,11} \delta A^{-+}_{x,11}}$ is equal to the sum of 
four diagrams in the diagrammatic approach.

Figs.\ \ref{diagrams2}a and \ref{diagrams2}b 
are the only diagrams considered in previous UCF theories. The rest of the
diagrams have been considered by Kane et al.\ \cite{Kane} and it is known 
that, for $\gamma=0$, they give rise to the long-ranged  correlation in 
local current response but they do not contribute to the variance of 
conductance in the absence of magnetic field \cite{Kane}.  
One can show that these additional diagrams can all be written as 
boundary contributions and they vanish when $\gamma=0$
for the same reason as the ladder series vanish in the case of the 
average conductance, when written in the area-averaged form. However, in the 
presence of the magnetic field the additional diagrams give rise to
Hall-ratio--dependent contributions. 
The work of KY and ML discussed the effect of the tilted boundary condition 
on the diffusion propagator but did not consider the additional diagrams. 

{}From Figs.\ \ref{diagrams2}a and \ref{diagrams2}b alone, we get 
\be
 \langle \dl g^2 \rangle_{a, b}=
\frac{1}{L^4}
\left\{ 4 {\rm Tr}( dd^T)+
2  {\rm Tr}( dd)\right\}.
\label{varinc}
\ee       
Using the classical network model of Sec.\ IB, we 
have calculated the diffusion propagator $d$ for a range of values of $\gamma$
and $W/L$. Using this propagator, we find that ${\rm Tr}(dd)$ and ${\rm Tr} 
(dd^T)$ are smooth functions of $\gamma$ and $L/W$. The peaks reported in ML in
the variance of the conductance (as given by eq.\ ({\ref{varinc})) are not 
observed in our exact numerical calculation. The argument advanced by ML for the
existence of ``resonant'' peaks due to the tilted boundary condition effects is
not supported by this calculation. We emphasize again that, in any case,
eq.\ (\ref{varinc}) is not the full expression for the variance of the 
conductance, because there are other diagrams that were omitted by KY and ML.

\subsection{The recovery of UCF result in the 1D limit}

The importance of the additional diagrams can be best demonstrated 
in the quasi-1D limit ($W \ll L$), where $\sx^0$ and $\sy^0$ combine to 
form a single parameter, $\sx^{0,1D}$. This limit is well 
described by the random matrix theory of the unitary ensemble. For general
reasons given earlier, we expect the variance of the conductance to approach 
the well-known 1D UCF result, independent of the value of $\gamma$. 

Plugging in the $1D$ diffusion propagator of eq.\ (\ref{d1D}), we get   
\be
\langle \dl g^2 \rangle_{a, b, 1D}=\frac{6}{\pi^4}
\left (\sum_{m=1}^{\infty}\frac{1}{m^4}\right) 
\frac{1}{(1+\gamma^2)^2}
= \frac{1}{15}(1-2\gamma^2)+O(\gamma^4),\ee   
where we used
$$ \sum_{m=1}^{\infty}\frac{1}{m^4}=\frac{\pi^4}{90}. $$ 
This is essentially the argument used by KY and ML, except that they gave 
versions applicable at finite temperature. However, this result of these
authors, that the variance of the conductance depends on $\gamma$ even in the
1D limit, is incorrect.
      
Among the additional diagrams, 
Figs.\ \ref{diagrams2}c, d and the sum of ~\ref{diagrams2}i
and i$'$ vanish to order $\gamma^2$  for all $W/L$; 
Figs.\ \ref{diagrams2}g, h, i, i$'$, j and j$'$ vanish 
as $W/L\rightarrow 0$. 
The dominant contributions come {}from  Figs.\ \ref{diagrams2}e and f:
\be
 \langle \dl g^2 \rangle_{e, f, 1D}\simeq
\frac{6}{\pi^4}\left(\sum_{m=1}^{\infty}\frac{1}{m^4}\right) 
2\gamma^2f_1(\infty)= \frac{1}{15} (2\gamma^2).
\ee   
Figs.\ \ref{diagrams2}e and f thus cancel the $\gamma^2$  correction {}from 
Fig.\ \ref{diagrams2}a and b. We get in total, to order $\gamma^2$,
\be
\langle \dl g^2 \rangle_{W/L\rightarrow 0}=\frac{1}{15}.
\ee 
Thus, in 1D the UCF result of \cite{Alsh2,LSF,Random} is recovered, at least 
to order $\gamma^2$. We remind the reader that the 1D UCF result holds only
when the length $L$ is less than the 1D localization length, $\xi_{1D}$, and
that $\xi_{1D}$ is of order $\sx^{0,1D}=W/\rho_{xx}^0$, which is much larger
than the lower limit ($W$) on $L$ in the diffusive regime $\rho_{xx}^0\ll 1$. 

\subsection{The variance of the conductance in 2D}

For a wide sample with $W/L$ arbitrary, the variance of the conductance 
depends on the Hall ratio. We will calculate the correction to the usual 
result for the $\gamma=0$ unitary ensemble, to order $\gamma^2$. 
In the 2D limit, the individual diagrams, Figs.\ \ref{diagrams2}e--h, j and
j$'$, all have logarithmically-divergent parts, however, their logarithmic 
contributions cancel out. 
(The cancellation is guaranteed by the fact that $S$ is 
not renormalized at one-loop level.) There can be even more 
divergent diagrams containing $\dl(0)$, but these diagrams are canceled 
by diagrams generated by the measure $I[z,\zd]$. (Since they are at least of 
order $\gamma^4$, they are not explicitly calculated in this article.)
The total $\gamma^2$ correction is finite. Summing up the contributions 
{}from Figs.\ \ref{diagrams2}a--j$'$, we get for a square sample
\be
(\dl g)^2_{L=W}= \left[
9.06\frac{1}{\pi^4} + 2.40\gamma^2
\frac{256}{\pi^8}\right]+O(\gamma^4).       
\label{finalresult}
\ee  
The expression for the variance for arbitrary $W/L$  
is given in Appendix D.

\section{Conclusion}

In this paper, we considered the mesoscopic conductance and its fluctuations in 
the presence of a magnetic field for a realistic two-probe geometry. 
Our perturbation theory has a different structure {}from previous 
theories \cite{Alsh2,LSF,Xiong} because of the presence of two conductivity 
parameters, $\sx^0$ and $\sy^0$. We found that $\sy^0$ not only enters the 
boundary condition for diffusion, as was noted in Refs.\ 
\cite{Yosef,Maslov,Read}, but also appears in the current vertex and other 
vertices which govern the interference processes. As a result the two-probe 
conductance and its variance in the perturbative regime depend on the Hall 
ratio $\gamma=\sy^0/\sx^0$. Our calculations differ {}from the previous results 
\cite{Yosef,Maslov} since we have not only modified the boundary condition but 
also considered additional diagrams which vanish in the zero field limit or 
in an edgeless system. Our main result is that the UCF are modified in the 
presence of edges; the variance of the two-probe conductance, although it is 
still of order 1, increases with the Hall ratio, as shown in eq.\
(\ref{finalresult}). However, in the quasi-1D limit of a long sample, the usual
universal result is recovered.

The reflecting boundary condition at the ``hard'' wall (or ``edge'') is 
crucial for the dependence of the conductance on the Hall ratio 
$\gamma$ that we find. If this is replaced by a periodic transverse 
boundary condition (i.e.\ a system on the surface of a cylinder), the results 
of the usual unitary ensemble in 2D are obtained; the results of Xiong and 
Stone \cite{Xiong} are easily modified for this case, for which they are
correct. While a cylinder may seem hard to realize experimentally, it can be 
mapped to an annulus by a conformal mapping. The annulus is sometimes known as 
the Corbino disk, in which there are no edges, and a radial voltage drop is 
applied to induce a current flow. Thus for the disk, the conductance 
fluctuations should be a universal function of the ratio of the inner and 
outer radii, with no dependence on $\gamma$.

The experimental observation of the effects we find depends first on being in
the regime $L_{in}\gg L$, $W$, so the system is phase coherent, and on having
an elastic mean free path $l$ due to impurities such that $L$, $W\gg l$. Our
calculations only address the metallic regime of conductance fluctuations at
large diagonal conductivity $\sx^0$, where perturbation theory is valid. In 
principle, this approach is valid for any value of the Hall ratio 
$\gamma=\sy^0/\sx^0$, or of the Hall angle $\theta_H=\tan^{-1}\gamma$. For 
simplicity, we expanded most of our results also to first nontrivial order in 
$\gamma^2$. The terms in $(1/\sx^0)^2$ that are left out cannot be neglected 
if the system size $L$ or $W$ exceeds the order of $\xi_{pert}$, the crossover 
scale at which the renormalized conductivity becomes of order 1 or less. If 
$L$, $W$ are greater than $\xi_{pert}$, the system crosses over either to the 
localized regime where $\sy$ becomes quantized, or, for Fermi energies near 
the critical values that lie near the centers of the Landau bands, to the 
critical transition region between the plateaus; our theory does not apply 
to either of these. Therefore, one must use mesoscopic systems that are not 
too large. Fortunately, since $\xi_{pert}\sim l e^{(\sx^0)^2}$, this is not 
difficult if $\sx^0\gg1$. According to the SCBA results reviewed in Sec.\ IIB, 
$\sx^0$ will be large unless either the Landau level index $N$ of the highest 
partially-occupied Landau level is of order 1, or the Fermi energy lies in the 
tail of the density of states of the disorder-broadened Landau bands, when
$\omega_c\tau_0$ is large enough that these are well developed. Thus the 
magnetic field $B$ must be large enough to suppress the Cooperons, so the 
system is in the unitary (broken time-reversal symmetry) regime, but not too 
large. (We do {\em not} generally require $\omega_c\tau_0>1$, though this would
ensure that $\gamma \geq O(1)$.) In effect, for the observation of the effects 
found in our theory, ideal conditions would be that the system should exhibit 
Shubnikhov-de Haas (SdH) oscillations, but not well-developed quantized Hall 
plateaus, even for asymptotically low temperatures. As the Fermi energy or 
magnetic field varies through a Landau band, yielding such an oscillation in 
$\sx^0$, $\sy^0$ varies monotonically, which implies that the ratio 
$\gamma=\sy^0/\sx^0$ varies. There is therefore a lot of scope for varying 
$\gamma$ by varying either the field $B$ {}from low values ($\gamma\simeq 0$) 
to larger, or as the field sweeps through a single SdH oscillation. However, 
since the amplitude of the fluctuations depend on $\gamma$, it will be 
necessary to collect statistically-independent values of the conductance 
without changing $\gamma$ too much. Thus the simplest experimental method, 
which uses magnetic field as the ergodic parameter, will not work and some 
other technique must be used to vary the sample conductance at fixed $B$. 
Finally, as the quantized Hall plateas are reached, localization effects will 
suppress fluctuations strongly between the centers of the LLs, and our theory 
is not applicable (although such measurements would be interesting).

While the calculations in this paper have addressed only the weak-coupling 
regime at $\sx\gg 1$, it is interesting to speculate about the effects of 
$\sy$ on the conductance and its fluctuations in the critical regime of the 
integer quantum Hall effect, when the system has edges. The critical regime 
can be defined by the conditions $L_{in}\gg L$, $W$, and $L$ and $W$ between 
$\xi_{pert}$ and $\xi$, where $\xi\geq \xi_{pert}$ is the localization length, 
which diverges as $E_F$ approaches any of the critical values $E_{cN}$, $N=0$,
$1$, \ldots. We expect that the renormalized local conductivity parameters 
$\sx$, $\sy^I$, $\sy^{II}$ are still meaningful, and that $\sx$ and 
$\sy=\sy^I+\sy^{II}$ take on universal values ($\equiv 1/2$ (mod $1$), in the 
case of $\sy$) at the critical points. This raises the question of the 
renormalization of the two pieces $\sy^{I}$ and $\sy^{II}$, and whether the 
values of these are universal separately at the critical point. We note that 
at the localized fixed point, the behavior may be described by saying 
$\sy^I=0$, $\sy^{II}\equiv0$ (mod $1$), so that these parameters do approach 
universal values in this regime. For $\sx$, there is a widespread belief that 
it takes the universal value $1/2$ at the critical fixed point, though it is 
not always clear if the calculations done to support this are describing the 
local conductivity parameter $\sx$, rather than a mean conductance in a 
particular geometry. The relation of these is not known in the critical regime 
at the present time, and, as we have seen, is not simple even in the 
perturbative regime, if the system has edges. We expect that for a two-probe 
system with a periodic transverse boundary condition, $\sy$ should not 
contribute to the conductance in the critical regime, just as it does not in 
the perturbative theory in this paper. Even then, the mean conductance is not 
in general given by $\sx W/L$, since non-Ohmic behavior is expected at least 
for $L\gg W$ where the system approaches a localized quasi-1D limit. Thus, 
even in the case of a square sample with $W=L$ and periodic transverse 
boundary condition, it is not clear that $\langle g \rangle=\sx$. The effect 
of the edges in the critical region is nicely shown in a recent paper 
\cite{Cho}, which examined the mean, variance, and distribution of the 
conductance in a two-probe geometry like ours, with $W=L$, and for both 
reflecting and periodic transverse boundary conditions, i.e.\ with and without 
edges. The results show that the boundary conditions do make a difference
(however, finite size effects are significant, as shown for the periodic 
transverse boundary condition case in Ref.\ {\cite{Wang}). The 
authors tentatively attribute this to ``edge currents,'' but as we have seen 
in the perturbative regime, there are edge effects (described by $\sy$), that 
are not solely due to edge currents carried by edge states (which are described 
by $\sy^{II}$). The boundary effects make themselves felt throughout the 
system, due to the long-range correlations in the critical regime. They are 
relatively unimportant only when $W\gg L$. In fact, dependence on the boundary 
conditions, say on whether they are periodic or reflecting, would occur even 
in the absence of $\sy$, as it does in the weak coupling regime (see e.g.\
\cite{Xiong}). A further implication, suggested by our results, is that the
critical conductance properties, in a given geometry that possesses edges, 
may depend on the fixed point value of $\sy$, i.e.\ on which transition 
is being studied. While the structure of the critical field theory [including 
$\sy$ (mod $1$), $\sx$, and the critical exponents] should be universal, this 
may not be true for the conductance, because the edge brings in dependence on 
the integer part of $\sy$. A first example of this is the simple fact that the 
mean of the Hall conductance, that can be defined in a multiprobe geometry
(with edges), depends on which transition is being studied, thus violating 
universality to this extent. The same may be true of the critical conductance 
fluctuations in geometries with edges. On the other hand, for a Corbino disk, 
which has no edges, there should be full universality among the integer 
quantum Hall transitions. Clearly, it would be of interest to study this 
numerically or experimentally. One way to do so numerically would be using the 
Chalker-Coddington model \cite{Chalker1}, with additional co-moving edge 
channels coupling to the edges by hopping terms to obtain $|\sy^0|>1$.

Returning to the perturbative, metallic regime, $\sx\gg1$, we expect that
similar phenomena to those studied here in 2D should occur also in higher
dimensions, for example in 3D. No isotropic topological term containing only 
two gradients is possible in higher dimensions. However, the Hall conductivity
should make an appearance in the NL$\sigma$M effective action, since it is a
part of the measurable conductivity. It appears in a generalization of the 2D
action to 3D \cite{Kotliar},
\be
S=-\frac{1}{8}\sigma^0\int d^3r\, {\rm tr}\,[\partial_\mu Q\partial_\mu Q] 
-\frac{1}{8}\sigma_H^0 \int d^3r \epsilon_{\lambda\mu\nu}n_\lambda 
{\rm tr}\, [Q \partial_\mu Q \partial_\nu Q].
\ee
Here $\bf n$ is a unit vector in the direction of the magnetic field $\bf B$,
and $\sigma^0$ and $\sigma_H^0$ are the diagonal (dissipative) and Hall 
conductivities, respectively. Thus the action is anisotropic, because the 
$\bf B$ field specifies a direction. (However, for simplicity we neglected the 
possible anisotropy in the diagonal conductivity $\sigma^0$.) The action can be 
viewed as resulting directly {}from considering layers stacked perpendicular to
the magnetic field, each of which has a Hall conductivity and is descibed by 
the 2D NL$\sigma$M action, plus a transition amplitude for electrons hopping 
between the layers. Such models have recently been studied numerically 
\cite{Chalker3}. For systems with boundaries, the $\sigma_H^0$ term in the 
action now leads in perturbation theory to phenomena similar to those in 2D, 
such as a tilted boundary condition, a dependence of $\langle g \rangle$ and 
the conductance fluctuations on $\sigma_H^0/\sigma^0$, and so on. Thus in 3D, 
and also in still higher dimensions, the conductance fluctuations in general 
depend on the Hall ratio (or angle). However, for the localization transition 
in 3D, which would be expected to be in the unitary class since time-reversal 
symmetry is broken by the magnetic field, we suspect that the $\sigma_H^0$ 
term is irrelevant at the critical fixed point, so that the properties of the 
transition are universal, independent of the bare Hall ratio, at least to the 
same extent as in 2D, as discussed above. Similarly to 2D, the $\sigma_H^0$ 
term contributes to the action of configurations in which each layer has a 
non-zero instanton number (insofar as this number is well-defined, if the 
system has boundaries). In 3D, there also exist topologically-stable 
point-singular configurations of the $Q$ field (known as ``hedgehogs'' in the 
literature), which may be viewed as points at which the instanton number 
changes {}from one layer to the next. The $\sigma_H^0$ term counts the number 
of layers with each value of the instanton number, and thus is sensitive to 
the presence and location of the hedgehogs. However, in the case of 
NL$\sigma$M's studied in connection with antiferromagnets, it appears that the 
hedgehogs are irrelevant as far as the critical properties are concerned, even 
though they may affect the behavior in the phases on either side of the 
transition (see, e.g., Ref.\ \cite{readsach}). Therefore, we suspect that, 
while the $\sigma_H^0$ term plays a role in the metallic phase, and also 
(after renormalization) in the 3D quantized Hall phase of layered systems 
\cite{Chalker3}, it may have no effect on the critical properties, except 
perhaps for the conductance when edges are present. Clearly, these are 
questions that may repay further study. 

\acknowledgements  
This work was supported by NSF grants  nos.\ DMR-92-15065 and 
DMR-91-57484. S.~X. thanks L. I. Glazman, and N.~R. thanks S.-J.~Rey and 
S.~N.~Majumdar, for helpful conversations.

\appendix

\section{Boundary Condition at the Hard Walls for the White-Noise Model}

In this appendix we briefly derive the boundary condition on the diffusion
propagator within the SCBA, by using current conservation at the reflecting
walls.

Using the SCBA equation $(E-H_0-\Sigma^{\pm}(\br))\Gb^{\pm}(\br,\brp)=
\dl(\br-\brp)$, we can show that 
\bea
\bna\cdot {\bf J^{+-}}(\br,\brp)&=&\bna\cdot \left[\Gbm(\brp,\br)
\left(\frac{-i\hbar}{2 m_e}\DD\right)\Gbp(\br,\brp)\right]\non\\
&=&\frac{i}{\hbar}(\Sigma^+-\Sigma^-)\left[-\dl(\br-\brp)/u +
\Gbp(\br,\brp)\Gbm(\brp,\br)\right].
\eea 
Let us define the $G^+G^-$ ladder diagram with $n$ impurity lines as
 $S^{+,-,(n)}$ and the ladder diagram with one current vertex attached 
to the left as 
$${\bf v}^{L,(n)}(\br,\brp)=u\int d^2 r_1\,{\bf J^{+-}}(\br,\br_1)
S^{+,-,(n)}(\br_1,\brp).$$ 
Using the above property of ${\bf J}^{+-}$, we get the
following  recursive relation:
\be
\bna\cdot {\bf v}^{L,(n)}=\tau
\left[S^{+,-,(n)}-S^{+,-,(n+1)}\right].
\label{recursive}
\ee
Summing up all ladder diagrams, we get
\be
\bna\cdot {\bf v}^L(\br,\brp)=-\tau \dl(\br-\brp),
\ee
where ${\bf v}^L(\br,\brp)$ represents the ${\bf J^{+-}S^{+,-
}}(\br,\brp)$. This shows that, on the finest length-scale resolution, 
$\langle\sigma_{\mu\nu}(\br,\brp)\rangle_{\rm SCBA}$ obeys
$\nabla\cdot\sigma(\br,\brp)=0$, for $\br$ sufficently far {}from $\brp$. 

For $\br$ at the reflecting boundary, the normal component of $\bf j$ 
{\em outside} the sample is zero. In the presence of a boundary current, 
which, {}from a coarse-grained, large-scale point of view, can be treated as 
$\delta$-functions $\delta(y-W)$, $\delta(y)$ in the components tangential 
to the edge, the surface integral of the current emerging
{}from a small box centered on the top edge reduces to 
\be
\int_{W-0^+}^{W+0+}dy\,\partial_x\langle\sigma_{x\nu}(\br,\brp)\rangle_{\rm 
SCBA}
-\langle\sigma_{y\nu}(\br,\brp)\rangle_{\rm SCBA}|_{y=W} =0,
\ee
for $\br\neq\brp$. Thus any normal current (just inside the edge)
must be converted to a $\delta$-function tangential current at the edge. 
This condition was discussed for the current ${\bf j}'$ in Sec.\ IA. Within the 
SCBA, it leads (using (\ref{smunuSCBA})) to the conclusion that it is 
$\gamma =\sy^0/\sx^0$, not $\sy^{I,0}/\sx^0$, which appears in the boundary 
condition (\ref{propagator}) on the diffusion propagator $d$. A similar 
argument holds for the $\brp$ dependence. The extension of this discussion 
to include the situation $\br=\brp$ is given in Sec.\ IIIC.

\section{Proof of the Current Conservation
Identities within SCBA}

We will show that within SCBA, $\sigma_{\mu\nu}(\br,\brp)$
satisfies the constraints imposed by current conservation.
We start with the $\sigma^{+-}$ term. 
We can write the ladder diagrams  in the following fashion,
\be
\sigma^{+-}_{\rm ladder}(\br,\brp)=\frac{\hbar^4}{4m_e^2}u \int
d^2 r_2 \,v^{L}(\br,\br_2)J^{-,+}(\brp,\br_2).
\ee
Using the recursive relation (\ref{recursive}) for ${\bf v}^{L,(n)}$,
and denoting  $S^{+-,(n)}J^{-+}$  as ${\bf v}^{R,(n)}$, we get 
another recursive relation:
$$
 \bna\cdot{\bf \sigma}^{+-
,\;(n)}(\br,\brp)=\frac{2m_e}{\hbar^2}(\Sigma^{+}
-\Sigma^{-
})[ {\bf v}^{R, (n-1)}(\br,\brp)-{\bf v}^{R, (n)}(\br,\brp)].
$$ 
Summing up all the ladder diagrams, we get
$$
 \bna\cdot{\bf \sigma}^{+-
}(\br,\brp)=\frac{2m}{\hbar^2}\sum_N\sum_{N'}\left\{\Gbp_{N}(
\br,\brp)\DD'_jP_{N
'}(\brp,\br)-P_{N}(\br,\brp)\DD' \Gbm_{N'}(\brp,\br)
\right\},
$$
where $P_N(\br,\brp)$ is the projection operator onto the $N$th 
Landau level.
The right hand side is a short-ranged function
of $|\br-\brp|$ ,
which we can treat as a $\delta$-function. We can write
\be
\bna\cdot{\bf \sigma}^{+-}(\br,\brp)={\bf c}^{+-}\delta(\br-
\brp)
,\ee 
where
$$c_{j}^{+-}=-
i\hbar\sum_{n}\sum_{n'}(\Gbp_n v_{nn'}^j-
v_{nn'}^j\Gbm_{n'}), $$
where $v_{nn'}$  is the matrix elements of the velocity operator, and
$$ c_j^{+-}\left\{\begin{array}{cl}=0 &\mbox{for $B=0$;}\\
                               \neq 0 &\mbox{for $B\neq0$.}
                      \end{array}\right. 
$$
To evaluate $\sigma^{++}(\br,\brp)$ and $\sigma^{--}(\br,\brp)$, 
we use the following trick:
\bea
\sigma^{++}(\br,\brp)&=&\int_{-\infty}^E dE'\, 
f(E')\lim_{E_1=E_2\rightarrow
E'}\frac{\partial }{\partial E_1}\sigma^{++}(\br,\brp;E_1,E_2),\non\\
\sigma^{--}(\br,\brp)&=&\int_{-\infty}^E dE'\, 
f(E')\lim_{E_1=E_2\rightarrow
E'}\frac{\partial }{\partial E_2}\sigma^{--}(\br,\brp;E_1,E_2) ,
\eea          
where $\sigma^{aa}(E_1,E_2)$ involves ladder sum $S^{aa}$ ($a=+,-$).
Define ${\bf v}^{R,aa,(n)}=S^{aa,(n)}{\bf J}^{aa}$,
we can show that for the $n$th ladder diagram,
\bea
\nabla\cdot \sigma^{++,(n)}(\br,\brp;E_1,E_2)&=& 
\frac{2m_e}{\hbar^2}
[\Sigma^+(E_1)-\Sigma^{+}(E_2)]{\bf v}^{R,++,(n-1)}(\br,\brp)  \non\\
&&-\frac{2m_e}{\hbar^2}[\Sigma^+(E_1)-\Sigma^{+}(E_2)]{\bf v}^{R,++, 
(n) }
(\br,\brp)\non\\
&& +\frac{2m_e}{\hbar^2}
[E_1-E_2]{\bf v}^{R, ++, (n)}(\br,\brp) . 
\eea
One can see that there is cancellation between the $n$th ladder 
diagram and the $(n+1)$th ladder diagram. Similar relations can be derived for 
$ \nabla\cdot \sigma^{--,(n)}(\br,\brp;E_1,E_2)$. Summing up all the ladder 
diagrams, taking the derivative over energy and then the limit $E_1=E_2
\rightarrow E'$, we can show that
\bea
\nabla\cdot 
[\sigma^{++}(\br,\brp)+\sigma^{--}(\br,\brp)]&=&-c^{+-}\dl(\br-\brp)
\non\\
  &-& \frac{\hbar^2}{2m_e}\int dE'\,f(E') [{\bf
v}^{++,R}(\br,\brp,E')-{\bf v}^{--,R}(\br,\brp,E')].
\eea
We can see that $\nabla\cdot \sigma^{+-}(\br,\brp)$
is canceled by contributions {}from $\nabla\cdot \sigma^{++}(\br,\brp)$  and
$\nabla\cdot \sigma^{--}(\br,\brp)$,
\be
\bna\cdot {\bf \sigma}(\br,\brp)=- \frac{\hbar^2}{2 m_e}\int_{-\infty}^E dE'\,
f(E') [{\bf
v}^{++,R}(\br,\brp,E')-{\bf v}^{--,R}(\br,\brp,E')]       \label{divsigma}
\ee
 Since
$$\bna'\cdot {\bf v}^{R,++,(n)}(\br,\brp)=\frac{2m_e}{\hbar^2}[
S^{++,(n-1)}(\br,\brp)-S^{++,(n-1)}(\br,\brp)] =0, $$ 
and
$$ \bna'\cdot{\bf v}^{R,--,(n)}(\br,\brp)=\frac{2m_e}{\hbar^2}[
d^{--,(n-1)}(\br,\brp)-d^{--,(n-1)}(\br,\brp)] =0, $$ 
we get
\be
\stackrel{\rightarrow}{\nabla}\cdot\langle\sigma(\br,\brp)\rangle_{\rm SCBA}
\cdot \stackrel{\leftarrow}{\nabla'}=0. 
\label{divdivsigma}
\ee
Using eqs.\ \ref{divsigma} and \ref{divdivsigma} and the 
asymptotic property of the Green's function \cite{Baranger}
$$\int dS'|_{\brp=\infty} G^{\pm}(\br,\brp)\DD' G^{\pm}
(\brp,\br)=0,$$
we can show finally that
\be
\bna\cdot\int\langle\sigma(\br,\brp)\rangle_{\rm SCBA}\cdot d{\bf S}'=0.
\ee

\section{Remarks on Edge States and Quantization}

Here we return to the topological considerations of Sec.\ IIIB, and relate them
to edge states and the quantization of the Hall conductance in the localized
regime. The topological considerations of Sec.\ IIIB are closely related to 
the problem of setting up a path integral for a quantum spin, by which we mean 
an irreducible representation of the symmetry group, which is SU($2n$) here 
(for a review, see e.g.\ \cite{readsach}), and this connection is also 
utilised in the mapping {}from the Chalker-Coddington model (representing a 
network of edge states) to a quantum spin chain or the NL$\sigma$M 
\cite{Read2} (the connection between the latter two problems, and the relation 
to the quantum Hall effect, was discussed earlier \cite{affleck}). 
In the quantum spin problem, we would take imaginary 
time, with a periodic boundary condition in the time direction, and the 
action would contain only the $\sigma_{xy}^{II,0}$ terms {}from $S[{\bf A}]$,
eq.\ (\ref{S[A]}); the 
system would be taken to be a disk, with the single edge corresponding to the 
world line of the quantum spin with its periodic boundary condition. 
For the two-probe geometry, this corresponds to regarding
$x$ as imaginary time, and the two edge channels are then a pair of 
quantum spins, with the spins fixed at $Q=\Lambda$ at the initial and final
``times'' $x=0$, $L$. In the absence of the rest of the action, 
quantum-mechanical consistency requires in either geometry that the coefficient 
$\sigma_{xy}^{II,0}$ be quantized to integer values, for
reasons closely related to the properties of ``large'' (topologically
non-trivial) gauge transformations; for the case of SU(2), this corresponds 
to $2S=$ integer, as usual. Essentially, 
the argument says that, since the only degree of freedom in the problem is 
the value of $Q$ on the edge, then its continuation into the interior, needed
to write the topological term, is arbitrary, and the path integral should be 
invariant under a change in $Q$ in the interior that does not affect the edge; 
such changes are the ``large'' gauge transformations. Since the change in the
action under such a change is $2\pi i\sy^{II,0}q$ for some integer $q$, this 
implies that $\sy^{II,0}$ is an integer.
This is related to arguments for quantization of the Hall conductivity,
once localization sets in \cite{LLP,Pruiskenrev}. In this case, we may imagine 
that the localized system is described by the NL$\sigma$M but with 
$\sigma_{xx}^0$ replaced by a renormalized value $\sigma_{xx}$ equal to 
zero because of localization. Then a similar argument requires that the 
renormalized $\sigma_{xy}^{II}$ is quantized to integer values. Thus 
quantization of the Hall conductance and quantization of spin are closely 
connected \cite{kim}. This argument is also connected \cite{Pruiskenrev} with 
the gauge-invariance argument for quantization \cite{Halperin}. The edge
states, that are the only degrees of freedom able to transport current over
large distances in the localized regime, correspond to the quantum spin (in the
$n\rightarrow 0$ limit). We note that {}from this point of view of the edge
states, in which $x$ plays the role of imaginary time, $A_x$ plays the role of 
an external magnetic field, in the sense of the familiar Zeeman coupling in 
the SU(2) case. It is coupled to $Q$, which corresponds to the spin operator,
or the current operator for the edge state.
$Q$ corresponds for SU($2n$) to the three-component unit vector
$\bf \Omega$ that describes an SU(2) spin, which can be obtained explicitly by 
writing, for $n=1$, $Q={\bf\Omega}\cdot\mbox{\boldmath$\tau$}$, where 
{\boldmath $\tau$} is the vector of Pauli matrices. For SU(2), 
the corresponding operators in the quantum theory, after rescaling to absorb 
the coefficient analogous to our $\sigma_{xy}^{II,0}$, are the familiar 
operators $\bf S$, which generate SU(2) rotations and are conserved when 
the Hamiltonian is SU(2) invariant. In the presence of the vector potential 
$\bf A$, which enters multiplied by the magnitude of the spin, $S$, to give 
the Zeeman coupling, just 
as in our action, the equation of motion (or Ward identity) for a single
quantum spin describes the familiar precessional dynamics, which can 
therefore also be viewed as the covariant conservation of the spin.

By contrast, for the full action in the weak coupling regime $\sx\gg 1$,
where the rest of the action depends on the form of $Q$ in the interior, 
there is no reason why either $\sigma_{xy}^0$ or $\sigma_{xy}^{II,0}$ should 
be quantized, in accordance with our physical expectations. 
The same applies to derivations of $S[{\bf A}]$ starting {}from the network
model \cite{Read2}, which also represents only the Fermi energy response, 
except that in this case the splitting of $\sigma_{xy}^0$ into 
$\sigma_{xy}^{I,0}$ and $\sigma_{xy}^{II,0}$ is a matter of an arbitrary 
definition, as we discussed for the linearized model in Sec.\ IB. In the
network model, the links of the lattice can be viewed as quantized edge
channels, and these are the only degrees of freedom, so the coupling of 
$\bf A$ to these links is of the edge form discussed above, but with 
$\sigma_{xy}^{II,0}$ replaced by the quantized value $1$. Of course, the
coarse-grained values of $\sigma_{xx}^0$, $\sigma_{xy}^{I,0}$, and 
$\sigma_{xy}^{II,0}$ are determined by the parameters of the vertices in the
network model, and by the definition of the coarse-grained currents 
\cite{Ruzin}, so they are not quantized.

\section{Computation of the Variance Diagrams}

The conventional diagrams Figs.\ \ref{diagrams2}a and b
depend on $\gamma$ through the diffusion propagator. Using the 
boundary
perturbation expansion, we get
\bea
 {\rm Tr}(dd)&=& {\rm Tr }(d^0d^0)+{\rm Tr}(d^0\; d^0 {\cal B} d^0)
+{\rm Tr}(d^0 {\cal B} d^0\;d^0)\non\\
& &\mbox{}+ 2 {\rm Tr}(d^0\; d^0{\cal B}d^0 {\cal B}d^0)
+{\rm Tr}(d^0 {\cal B} d^0\; d^0 {\cal B} d^0)+O(\gamma^3),\\
 {\rm Tr}(dd^T)&=& 
 {\rm Tr}(d^0d^0)+{\rm Tr}(d^0\; d^0{\cal B}^Td^0)+{\rm Tr}(d^0 {\cal B}^T 
d^0\;d^0)\non\\
& &\mbox{}+ 2 {\rm Tr}(d^0\; d^0 {\cal B}^Td^0{\cal B}^Td^0)+
{\rm Tr}(d^0 {\cal B}^T d^0\; d^0 {\cal B}^T d^0)+O(\gamma^3),
\eea
where the matrix ${\cal B}$ has the following 
elements in the basis of $\phi^0_{nm}$:
$$
\langle n'm'| {\cal B} |nm\rangle =-\frac{\gamma}{WL}\frac{8mm'}{(m')^2-
m^2}\dl_{m+m',odd}\dl_{n+n',even} .
$$
The linear term in $\gamma$ is zero, because the matrix $ {\cal B} $ is 
anti-symmetric. We get
\be
(\dl g)^2_{a, b}=\frac{1}{\pi^4}[6f_2-\gamma^2\frac{64}{\pi^4}(12f_3-2f_4)],
\ee
where 
\be
f_2\left(\frac{L}{W}\right)=\sum_{n=0}\sum_{m=1}\frac{1}{(m^2+n^
2L^2/W^2)^2},
\ee        
\be
f_3\left(\frac{L}{W}\right)=\sum_{n=0,m=1}
\sum_{n'=0,m'=1}\frac{1}{(m^2+n^2 \frac{L^2}{W^2})^3}
 \frac{1}{m'^2+n'^2\frac{L^2}{W^2} } \frac{(mm')^2}{[(m')^2-m^2]^2}
\dl_{m+m',odd}\dl_{n+n',even},
\ee
\be
f_4\left(\frac{L}{W}\right)=
\sum_{n=0,m=1}\sum_{n'=0,m'=1}
\frac{1}{(m^2+n^2\frac{L^2}{W^2})^2}
 \frac{1}{(m'^2+n'^2 \frac{L^2}{W^2})^2} \frac{(mm')^2}{[(m')^2-m^2]^2}
\dl_{m+m',odd}\dl_{n+n',even}.
\ee
The upper bounds for $n$ and $m$ in all sums are $n_{max}\sim W/l$ 
and $m_{max}\sim L/l$. For a square sample with $L=W$, we have $f_2 
\simeq 1.51$,       
\be
(\dl g)^2_{a, b, L=W}=  \frac{1}{\pi^4}[ 9.06-
2.35\gamma^2]+O(\gamma^3).       
\ee        

Diagrams \ref{diagrams2}e and f give
\bea
(\dl 
g)^2_{e,f}&=&\frac{256}{\pi^8}\gamma^2\frac{L^2}{W^2}\sum_{n_1,m
_1=odd}\sum_{n_2,m_2=odd}\sum_{n_3=0,m_3=1}\frac{1}{m_1}\frac{1}{m_2}\non\\
&&\times\frac{1}{m_1^2+n_1^2L^2/W^2}\frac{1}{m_2^2+n_2^2L^2/W^2}
\frac{1}{(m_3^2+n_3^2L^2/W^2)^2}\non\\
&&\times\left\{3 \left(m_3^2 D_3D_2 + n_3^2 L^2/W^2 \;D_1 D_4\right) +3
\left( m_1 m_2 D_5 D_2 +n_1 n_2
L^2/W^2 D_1 D_6\right)\right.\non\\
&&\left.  \mbox{}+10 \left(m_1 m_3 D_7 D_2 + n_1 n_3 L^2/W^2 D_1 
D_8\right) \right\},
\eea          
where
\bea
D_1(m_1,m_2,m_3,m_3)&=& \dl_{m_1,m_2} -\frac{1}{2}\dl_{m_1, 
2m_3+m_2}-
\frac{1}{2}\dl_{m_1, -2m_3+m_2}+\frac{1}{2}\dl_{m_1, 2m_3-m_2},
\non\\    
D_2(n_1,n_2,n_3,n_3)&=& \dl_{n_1,n_2} +\frac{1}{2}\dl_{n_1, 
2n_3+n_2}+
\frac{1}{2}\dl_{n_1, -2n_3+n_2}+\frac{1}{2}\dl_{n_1, 2n_3-n_2},
\non\\      
D_3(m_1,m_2,m_3,m_3)&=& \dl_{m_1,m_2} +\frac{1}{2}\dl_{m_1, 
2m_3+m_2}-
\frac{1}{2}\dl_{m_1, -2m_3+m_2}-\frac{1}{2}\dl_{m_1, 2m_3-m_2},
\non\\     
D_4(n_1,n_2,n_3,n_3)&=& \dl_{n_1,n_2} -\frac{1}{2}\dl_{n_1, 
2n_3+n_2}+
\frac{1}{2}\dl_{n_1, -2n_3+n_2}-\frac{1}{2}\dl_{n_1, 2n_3-n_2},
\non\\     
D_5(m_1,m_2,m_3,m_3)&=& \dl_{m_1,m_2} -\frac{1}{2}\dl_{m_1, 
2m_3+m_2}-
\frac{1}{2}\dl_{m_1, -2m_3+m_2}-\frac{1}{2}\dl_{m_1, 2m_3-m_2},
\non\\     
D_6(n_1,n_2,n_3,n_3)&=& \dl_{n_1,n_2} +\frac{1}{2}\dl_{n_1, 
2n_3+n_2}+
\frac{1}{2}\dl_{n_1, -2n_3+n_2}-\frac{1}{2}\dl_{n_1, 2n_3-n_2},
\non\\     
D_7(m_1,m_2,m_3,m_3)&=& - \frac{1}{2}\dl_{m_1, 2m_3+m_2}
+\frac{1}{2}\dl_{m_1, -2m_3+m_2}+\frac{1}{2}\dl_{m_1, 2m_3-m_2},
\non\\ 
D_8(n_1,n_2,n_3,n_3)&=&  \frac{1}{2}\dl_{n_1, 2n_3+n_2}-
\frac{1}{2}\dl_{n_1, -2n_3+n_2}+\frac{1}{2}\dl_{n_1, 2n_3-n_2}.
\eea     
This term has a logarithmic part (it diverges with system size as 
$\log (L/l)$). It comes {}from the first term in the curly brackets, 
when the two derivatives of the 4-point interaction are applied to the closed
loop of two diffusion propagators. The second term results {}from  
applying the two derivatives to the two external propagators. The third term 
arises when one of the derivative is applied to the closed loop, 
one is applied to the external propagator. 

Diagram Figs.\ \ref{diagrams2}g and h are both logarithmic. They are of 
opposite signs, but the amplitude of diagram \ref{diagrams2}g, which is 
positive, is twice that of \ref{diagrams2}h. We get
\bea
(\dl 
g)^2_{g,h}&=&\frac{256}{\pi^8}\gamma^2\frac{L^2}{W^2}\sum_{n_1,
m_1=odd}
\sum_{n_2,m_2=odd}\sum_{n_3=0,m_3=1}\frac{1}{m_1}\frac{1}{m_
2}\non\\
&&\times\frac{1}{m_1^2+n_1^2L^2/W^2}\frac{1}{m_2^2+n_2^2L^2/W^2}
\frac{1}{m_3^2+n_3^2L^2/W^2}D_1D_2.
\eea
Diagram \ref{diagrams2}j and \ref{diagrams2}j$'$ also have logarithmic 
divergence. We get {}from diagram \ref{diagrams2}j
\bea
(\dl 
g)^2_{j}&=&\frac{256}{\pi^8}\gamma^2\frac{L^2}{W^2}\sum_{n_1,m_1=odd}
\sum_{n_2,m_2=odd}\sum_{n_3=0,m_3=1}\sum_{n_4=0,m_4=1}
\frac{1}{m_1}\frac{1}{m_2}\non\\
&&\times\frac{1}{m_1^2+n_1^2L^2/W^2}\frac{1}{m_2^2+n_2^2L^2/W^2}
\frac{m_3}{m_3^2+n_3^2L^2/W^2}\frac{m_4}{m_4^2+n_4^2L^2/W^2}
\non\\
&&\times[\dl_{m_1,m_4\pm m_3}-\dl_{m_1,m_3-m_4}]
[\dl_{m_2,m_3\pm m_4}-\dl_{m_2,m_4-m_3}]\non\\
&&\times[\dl_{n_1,n_4\pm n_3}+\dl_{n_1,n_3-n_4}]
[\dl_{n_2,n_4\pm n_3}+\dl_{n_2,n_3-n_4}].
\eea       
We get {}from \ref{diagrams2}j$'$
\bea
(\dl g)^2_{j'}&=&-
\frac{256}{\pi^8}\gamma^2\frac{L^2}{W^2}\sum_{n_1,m_1=odd}
\sum_{n_2,m_2=odd}\sum_{n_3=0,m_3=1}\sum_{n_4=0,m_4=1}
\frac{1}{m_1}\frac{1}{m_2}\non\\
&&\times\frac{1}{m_1^2+n_1^2L^2/W^2}\frac{1}{m_2^2+n_2^2L^2/W^2}
\frac{m_3^2}{m_3^2+n_3^2L^2/W^2}\frac{1}{m_4^2+n_4^2L^2/W^2}\non\\
&&\times[\dl_{m_1,m_4\pm m_3}-\dl_{m_1,m_3-m_4}]
 [\dl_{m_2,m_4\pm m_3}-\dl_{m_2,m_3-m_4}]\non\\
&&\times[\dl_{n_1,n_4\pm n_3}+\dl_{n_1,n_3-n_4}]
[\dl_{n_2,n_4\pm n_3}+\dl_{n_2,n_3-n_4}].
\eea       
Both diagrams \ref{diagrams2}j and j$'$ are negative. Their logarithmic parts 
combine to cancel those {}from diagram e, f, g and h. The variance, 
which is the sum of a--j$'$, is finite.



\begin{figure}
\caption{The two-probe geometry.}
\label{two probe geometry}
\end{figure}

\begin{figure}
\caption{The Chalker-Coddington network model. Each unit cell contains 
four distinct links, A, B, C, D. The tilted boundary condition arises 
{}from the fact that along each link, the random walk is along only 
one direction.} 
\label{network}
\end{figure}

\begin{figure}
\caption{(a) The SCBA single-particle Green's function. It sums up all the
non-crossing diagrams. The thin line denotes $G^0$, the Green's function in the
absence of disorder. The thick line denotes the SCBA Green's function $\Gb$. 
(b) The ladder sum for the SCBA two-particle Green's function. (c) The 
diagrams for the SCBA bilocal conductivity tensor.}
\label{diagrams1}
\end{figure}

\begin{figure}
\caption{The diagrams for the variance of conductance to leading 
order in $1/\sx^0$ and to order $\gamma^2$. The shaded polygons are 
vertices. The lines connecting the vertices are the diffusion propagators.}
\label{diagrams2}
\end{figure}

\begin{figure}
\caption{The equivalence between the diagrammatic approach and the 
NL$\sigma$M approach. One vertex {}from the NL$\sigma$M, $\delta^2 S/ 
\delta A^{+-}_{x,11} \delta A^{-+}_{x,11}$, is equal to the sum of four 
diagrams in the diagrammatic approach.}
\label{diagrams3}
\end{figure}
      

\begin{references}
\bibitem{Webb}S. Washburn and R. Webb,  Adv. in Phys. {\bf 
35}, 375  (1986).
\bibitem{Stone2}A. D. Stone,  Phys. Rev. Lett. {\bf 54}, 2692 (1985). 
\bibitem{LS}P. A. Lee and A. D. Stone,  Phys. Rev. Lett. {\bf 55}, 1622 
(1985).
\bibitem{Alsh2}B. L. Al'tshuler and B. L. Shklovskii,  Sov. Phys. JETP,
{\bf 64}, 127 (1986).
\bibitem{Skocpol}W. J. Skocpol, L. D. Jackel, R. E. Howard, P. M. 
Mankiewich, D. M. Tennant, A. E. White, and R. C. Dynes,  Surface Sci. 
{\bf 170}, 1 (1986); R. E. Howard, L. D. Jackel, P. M. Mankiewich, 
and W. J. Skocpol, Science {\bf 231}, 346 (1986).   
\bibitem{Oded}O. Millo et al., Phys. Rev. Lett. {\bf 65}, 1494 (1990).   
\bibitem{LSF}P. A. Lee, A. D. Stone and H. Fukuyama,  Phys. Rev. B {\bf 
35}, 1039 (1987).
\bibitem{gang4}E.  Abrahams, P. W. Anderson, D.C. Licciardello and T. 
V. Ramakrishnan,  Phys. Rev. Lett. {\bf 42}, 673 (1979).                   
\bibitem{AKL}B. L. Al'tshuler, V. E. Kravtsov and I. V. Lerner,  Sov.
Phys. JETP {\bf 64}, 1352 (1986).  
\bibitem{Imry}Y. Imry,  Europhys. Lett. {\bf 1}, 249 (1986). 
\bibitem{Mello} P. Mello and A. Stone, Phys. Rev. B {\bf 44}, 3559 
(1991).        
\bibitem{Random}A. D. Stone, P. A. Mello, K. A. Muttalib and J. 
Pichard, ``Random Matrix Theory and Maximum Entropy Models 
for Disordered Conductors", in {\it Mesoscopic Phenomena in Solids},
edited by B. L. Altshuler, P. A. Lee and R. A. Webb (Elsevier Science
Publishers B. V., 1991), Chapter 9.   
\bibitem{Beenakker} A. M. S. Mac\^{e}do and J.~T.~Chalker, Phys. Rev. B {\bf
46}, 14985 (1992); C.~W.~J.~Beenakker and B.~Rejaei, Phys. Rev. B {\bf 49},
7499 (1994).
\bibitem{Baranger2}
H. U. Baranger and P. A. Mello, Phys. Rev. Lett. {\bf 73}, 142 (1994).
\bibitem{Jalabert}
R. A. Jalabert, J.-L. Pichard and C. W. J. Beenakker, Europhys. Lett
{\bf 27}, 255 (1994).
\bibitem{Klitzing}K. von Klitzing, G. Dorda, and M. Pepper,  Phys. Rev. 
Lett. {\bf 45}, 494 (1980).            
\bibitem{Timp}G. Timp, A. M. Chang, R. E. Behringer, J. E. Cunningham, 
T. Y. Chang and R. E. Howard,  Phys. Rev Lett. {\bf 58}, 2814 (1987).
\bibitem{Geim}A. K. Geim el al.,  Phys. Rev. Lett. {\bf 67}, 3014 (1991);
{\bf 69}, 1248 (1992).
\bibitem{Kinaret}J. M. Kinaret and P. A. Lee, Phys. Rev. B {\bf 43},
3847 (1991).     
\bibitem{Bhatt2}Y. Huo, R. E. Hetzel, and R. N. Bhatt, Phys. Rev. Lett.
{\bf 70}, 481 (1993). 
\bibitem{Ando2}T. Ando,  Phys. Rev. B {\bf 49}, 4679 (1994).   
\bibitem{Wang}Z. Wang, B. Jovanovi\'{c}, and D.-H. Lee, Phys. Rev. Lett. {\bf
77}, 4426 (1996).
\bibitem{Cho}S. Cho and M. P. A. Fisher, LANL preprint no.\ 
cond-mat/9609048 (unpublished).
\bibitem{Ando1}T. Ando and Y. Uemura,  J. Phys. Soc. Japan,
 {\bf 36}, 959 (1974); T. Ando, Y. Matsumoto and Y. Uemura, J. 
Phys. Soc. Japan, {\bf 39}, 279 (1975).       
\bibitem{Brezin}E. Brezin, S. Hikami and J. Zinn-Justin,  Nucl. 
Phys. {\bf B165}, 528 (1980).                                         
\bibitem{Hikami}S. Hikami,  Prog. Theor. Phys. {\bf 72}, 722 (1984). 
\bibitem{Chalker2}P. Carra, J. T. Chalker and K. A. Benedict,  Ann. Phys. 
{\bf 194}, 1  (1989).
\bibitem{Pruisken1}A. M. M. Pruisken, Nucl. Phys. B {\bf 235}, 277 (1984). 
\bibitem{LLP}H. Levine, S. B. Libby, and A. M. M. Pruisken, Nucl. Phys. B {\bf
240}, 30, 49, 71 (1984).
\bibitem{Pruiskenrev}A. M. M. Pruisken, in {\it The Quantum Hall Effect}, 
edited by R.E. Prange and S.M. Girvin (Springer-Verlag, New York, 1990), 
Chap. 5.    
\bibitem{Pruisken2}A. M. M. Pruisken, Nucl. Phys. B {\bf 285}, 719 (1987); 
{\bf 290}, 61 (1987).
\bibitem{Xiong}S. Xiong and A. D. Stone,  Phys. Rev. Lett. {\bf 68}, 
3757 (1992).
\bibitem{Yosef}D. E. Khmel'nitskii and M. Yosefin, Surface Sci. {\bf 305}, 
507 (1994).
\bibitem{Maslov}D. L. Maslov and D. Loss,  Phys. Rev. Lett. {\bf 71}, 
4222 (1993). 
\bibitem{Read}N. Read, unpublished.        
\bibitem{Kane2}C. L. Kane, P. A. Lee and D. P. DiVincenzo, Phys. Rev. B 
{\bf 38}, 2995 (1988).
\bibitem{Hershfield}S. Hershfield, Ann. Phys. {\bf 196}, 12 (1989).  
\bibitem{Kane}C. L. Kane, R. A. Serota and P. A. Lee,  Phys. Rev. B {\bf 37},
6701 (1988).
\bibitem{Girvin}R. W. Rendell and S. M. Girvin, Phys. Rev. B {\bf 23}, 
6610 (1980).         
\bibitem{Halperin}B. I. Halperin,  Phys. Rev. B {\bf 25}, 2185 (1982). 
\bibitem{Buttiker}M. Buttiker,  Phys. Rev. Lett. {\bf 57}, 1761 (1986); 
M. Buttiker, IBM J. Res. Develop. {\bf 32}, 317  (1988).   
\bibitem{Baranger}H. U. Baranger and A. D. Stone, Phys. Rev. B {\bf  
40}, 8169 (1989).         
\bibitem{Chalker1}J. T. Chalker and P. D. Coddington,  J. Phys. C {\bf 21}, 
2665 (1988).
\bibitem{Fertig}
H. A. Fertig and B. I. Halperin, Phys. Rev. B {\bf 36}, 7969 (1987).
\bibitem{Kucera}
J. Kucera and P. Str\v{e}da, J. Phys. C {\bf 21}, 4357 (1988).
\bibitem{Szafer2}P. L. McEuen, A. Szafer, C. A. Richter, 
B. W. Alphenaar, J. K. Jain, A. D. Stone, and R. G. Wheeler,  Phys. Rev. Lett. 
{\bf 64}, 2062 (1990); A. Szafer, Ph.D.\ thesis, Yale University, 1991, 
(unpublished); 
A.~Szafer, A.~D.~Stone, P.~L.~McEuen, and B.~W.~Alphenaar, in {\it Granular
Nanoelectronics}, ed.\ by D.~K.~Ferry, J.~R.~ Barker, and C.~Jacoboni (Plenum,
New York, 1991).
\bibitem{Ruzin}
A. M. Dykhne and I. M. Ruzin, Phys. Rev. B {\bf 50}, 2369 (1994); I. M. Ruzin
and S. Feng, Phys. Rev. Lett. {\bf 74}, 154 (1995). 
\bibitem{Callan}A. Abouelsaood, C. G. Callan, C. R. Nappi, and S. A. Yost, Nucl.
Phys. B {\bf 280} [FS{\bf 18}],  599 (1987).
\bibitem{Read2} N. Read, unpublished; D.-H.~Lee, Phys. Rev. B {\bf 50}, 10788
(1994); M.~R.~Zirnbauer, Ann. Physik {\bf 3}, 513 (1994). 
\bibitem{Landauer}R. Landauer,  IBM J. Res. Develop. {\bf 1}, 
233 (1957); R. Landauer, Phil. Mag. {\bf 21}, 863 (1970).       
\bibitem{Laughlin}R. B. Laughlin, Phys. Rev. B {\bf 23}, 5632 (1981).  
\bibitem{Jackson}
See, e.g., J.~D.~Jackson, {\it Classical Electrodynamics} (Wiley, New York, 
2nd.\ Ed., 1975), Chapter 1, pp.\ 40--44; S.~L.~Sobolev, {\it Partial
Differential Equations of Mathematical Physics} (Dover, New York, 1989),
pp.~180--187.
\bibitem{Mahan}  G. D. Mahan, {\it Many-Particle  Physics} 
(Plenum, New York, 2nd.\ Ed., 1990), Chapter 3, Chapter 7.             
\bibitem{niu} Q. Niu and D. J. Thouless, Phys. Rev B {\bf 35}, 2188 (1987).
\bibitem{Streda}L. Smr\v{c}ka and P. Str\v{e}da,  J. Phys. C {\bf 10}, 2152  
(1977); P. Str\v{e}da, {\it ibid.} {\bf 15}, L717 (1982).     
\bibitem{Math}
P. M. Morse and H. Feshbach, 
{\it Methods of Theoretical Physics} (McGraw-Hill, New York, 1953), Chapter 7.
\bibitem{Amit}D. J. Amit, {\it Field Theory, the Renormalization Group, 
and Critical Phenomena} (World Scientific, Singapore, 1984, 2nd.\ Ed.), Part II,
Chapter 6.
\bibitem{Kotliar}M. Biafore, C. Castellani, and G. Kotliar, Nucl. Phys. B {\bf
340}, 617 (1990).
\bibitem{Chalker3} J. T. Chalker and A. Dohmen, Phys. Rev. Lett. {\bf 75}, 4496
(1995).
\bibitem{readsach}
N. Read and S. Sachdev, Nucl. Phys. B {\bf 316}, 609 (1989);
Phys. Rev. B {\bf 42}, 4568 (1990).                 
\bibitem{affleck}
I. Affleck, Nucl. Phys. B {\bf 257}, 397 (1985).
\bibitem{kim}
Y. B. Kim, Phys. Rev.  B {\bf 53}, 16420 (1996).



\end{references}
\end{document}